\documentclass[aps,pre,twocolumn,groupedaddress,showkeys]{revtex4-2}
\usepackage{graphicx,caption}
\usepackage{multirow}
\usepackage{dcolumn}
\usepackage{bm}
\usepackage{hyperref}
\usepackage{float}
\usepackage{placeins}
\DeclareUnicodeCharacter{200B}{\hskip 0pt}
\begin{document}
\title{Cation-DNA outer sphere coordination 
in DNA polymorphism}
\author{Elena A. Zubova}
\email{zubova@chph.ras.ru}
\author{Ivan A. Strelnikov}
\affiliation
{N.N. Semenov Federal Research Center for Chemical Physics, Russian Academy of Sciences, 4 Kosygin Street, Moscow 119991, Russia}
\begin{abstract}
\noindent
There are two approaches to describing DNA-ions interactions. The physical approach is an analysis of electrostatic interactions between ions and charges on the DNA molecule. The coordination chemistry approach is a search for modes of direct binding of ions to ionophores of DNA. We study both the inner and outer sphere coordination of ions by ionophores of the A and C forms of DNA in molecular dynamics simulations in two low-polarity solvents: in ethanol-water and methanol-water mixtures. We show that the counterion-DNA outer sphere coordination plays a key role in the experimentally observed conformational polymorphism of the DNA molecule: a transition to the A form in ethanol and to the C form in methanol. We identify the ionophores responsible for the existence of the A- and C-complexes. In both the complexes, the ions' inner sphere ligands are mostly water molecules, the ions reside in water clusters. In ethanol-water mixture, the water clusters are large, the major groove of the A-DNA is filled with water, and all ionophores are accessible to ions. In methanol-water mixture, the water clusters are small, and a large number of methanol clusters are present near DNA surface. They interfere with the coordination of ions in one of the ionophores of the major groove, and also with other ionophores near phosphates. Therefore, in methanol, the interaction energy of counterions with A-DNA cannot compensate for the repulsion between closely located phosphates. Therefore, the ions fill more accessible ionophores of the C-complex, converting DNA into the C form.
\end{abstract}
\maketitle
\section{Introduction}
\label{section-Intro}
\noindent
The interaction of a DNA polyion with ions and other molecules, as well as the change in the conformation of the molecule, is directly related to the performance of its biological functions. DNA locally transforms into the C form as a part of a nucleosome, and into the A form when interacting with polymerases and endonucleases (see review \cite{2023-DNA-review-we} and references therein). In the B form, the usual conformation of natural DNA in physiological (salt) solution, the angle of rotation of a base pair relative to the adjacent one (parameter $Twist$) is equal to about 36\textdegree. In the C form, the molecule is twisted more: $Twist\sim$38\textdegree, the minor groove is narrowed, and a large fraction of phosphates is in BII conformation ($>$40\%). The C-DNA is often considered as belonging to the 'B-family'. The A-DNA, on the contrary, is twisted less than the B-DNA, $Twist\sim$30\textdegree. In contrast to the C form, the minor groove of the A-DNA is wide, and the major groove is narrow, which is achieved by the transition of deoxyriboses from the south to the north (C3'$endo$) conformation.

The transition of the DNA to the A form in water can be caused by some trivalent ions: hexaamminecobalt(III) [Co(NH$_3$)$_6$]$^{3+}$ \cite{1995-CoNH36-A-DNA-X-ray}, aluminum and gallium \cite{1996-Al-Ga-A-DNA}. In the resulting A form, the ions are found between the phosphates of the opposite strands in the narrow major groove, and also near atom N7 of guanine ((G)N7). In the last case, a simultaneous interaction of the ion with an oxygen atom of a phosphate is detected. Mono- and divalent ions at low concentrations in water do not change the B conformation of DNA. 

When a large amount of a weakly polar solvent is added, a transition to the A- or C-form takes place (depending on the chemical nature of the solvent) in the presence of even monovalent ions \cite{1973-Ivanov-CD-A-B-C}. The authors of the work \cite{1973-Ivanov-CD-A-B-C} hypothesized that the transition of the DNA to the C-form is caused by the ions entering the minor groove, although there is no confirmation of this hypothesis to date. Molecular dynamics (MD) modeling in the framework of phenomenological force fields showed that during the transition to the A-form, ions concentrate near phosphates between the DNA strands in the major groove \cite{1997-B-to-A-hexaaminecobalt-MD-Chetham,2008-Mazur-Electrostatic-Origin-DNA-Polymorphism}. It is logical to assume that in solution the ions stabilize the C and A forms, narrowing the minor and major grooves, respectively. Therefore, it is more accurate to call the A-DNA and C-DNA in solution the A- and C-complexes.

Theoretical works (analytical and modeling in the framework of phenomenological force fields) normally use the physical approach to the analysis of DNA-ions interaction. The approach comes from the works by Manning (see, for example, \cite{2002-Manning-condesation}) and assumes that the structure of the DNA-ions complex is determined by electrostatic interactions and the entropy of the ions. Even in the case of all-atom modeling of the solvent, the localization of ions near the DNA surface is analyzed as the probability density of detection (see, for example, \cite{2015-MD-seq-dep-ions-distr-around-DNA}).

In contrast, experimental studies (X-ray, IR spectroscopy) focus on searching for ions primarily in direct contact with ionophores on the DNA surface (see reviews \cite{2001-flexible-ionophors,2008-DNA-ions-compl-Williams} and \cite{2001-Xray-ions-in-grooves-B-DNA,2011-Xray-Mg-many-places-B-DNA,2016-Xray-Mg-only-phosphates}). For the formation of direct contacts, the chemical nature of the ion plays a decisive role. But it is generally assumed that the outer sphere chelation of an ion by DNA atoms can be regarded as an 'electrostatic interaction', with the exception of ions that are very strongly bound to their inner sphere ligands and actually represent a complex, for example, [Mg(H$_2$O)$_6$]$^{2+}$ or [Co(NH$_3$)$_6$]$^{3+}$. 

In the present work, we show that outer sphere coordination of counterions by DNA ionophores is critical for the conformational dynamics of DNA. We use molecular dynamics modeling to study the interactions of ions with DNA in the process of formation of the A- and C-complexes. In experiment, upon addition of a low-polarity solvent to an aqueous solution, the DNA molecule changes its conformation from B to C or A depending on the solvent: upon addition of methanol - to C, and ethanol - to A \cite{1973-Ivanov-CD-A-B-C}.  Using the umbrella sampling scheme\cite{1977-Umbrella} combined with the Weighted Histogram Analysis Method (WHAM) \cite{1992-WHAM} we investigate the transition between the C- and A-complexes in both the solutions: the change in free and potential energy, as well as the structure of the solution around the DNA molecule. We find the binding modes of counterions to DNA in both the complexes and determine the modes whose blocking makes it impossible for the DNA molecule to exist in the A- or C-conformations. Accordingly, we determine the source of the DNA polymorphism in different solvents.
\section{A- and C-complexes: free energy, potential energy of the system componetns and helix parameters.}
\label{section-WHAM-U}
\noindent
When an organic solvent is added to DNA in water, the dielectric constant of the solution decreases, the electrostatic interactions of counterions with the DNA polyion increase, and the counterions approach the DNA molecule. The space available to the ions and, accordingly, their entropy decrease. The only way to decrease the free energy is to approach the charges on the DNA and to form an A- or a C-complex. Experimentally, at room temperature, DNA is in the B form in water, in the A form - in 80 vol.\% ethanol solution, and in the C form - in 80 vol.\% methanol solution \cite{1973-Ivanov-CD-A-B-C}. 

We carried out molecular dynamics (MD) modeling of decamer 5'-CCGGGCCCGG-3' (and the 20-mer composed of two such sequences) with Na$^+$ counterions in water and in 80 vol.\% solutions of methanol (molar fraction 0.643) and ethanol (molar fraction 0.552) with water. We used the all-atom CHARMM36 force field \cite{2012-CHARMM-for-BII} with NBFIX corrections for ion-phosphate interactions \cite{2012-parametrization-DNA-ions,2018-CUFIX-review} (see Section~\ref{section-Methods-MD} for simulation details).

\begin{figure}
\begin{center}
\includegraphics[width=0.85\linewidth]{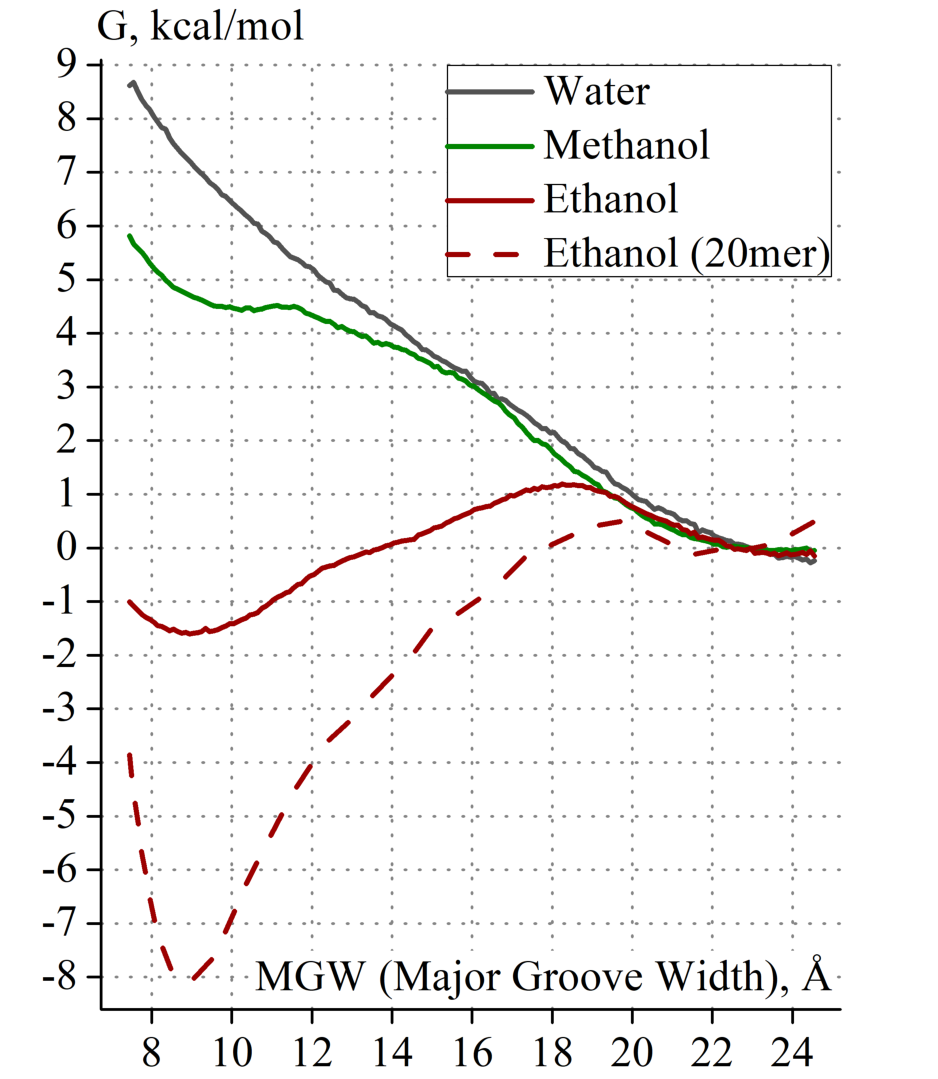}
\end{center}
\caption{Gibbs free energy G (NpT ensemble) profiles for decamer CCGGGCCCGG and 20-mer (CCGGGCCCGG)$_2$ in ethanol-water and methanol-water mixtures (the volume of solution is V; 80 vol.\% alcohol). For the decamer in the ethanol solution, the statistical error is less than $\sim$0.1kcal/mol, and the systematic error is $\sim$0.4kcal/mol (the minimum for the A form is lower, see Section~\ref{section-Methods-WHAM} for details).
\label{fig-WHAM-eth-meth}
}
\end{figure}

For the decamer in all the solutions, we calculated the free (Gibbs) energy G of the system in (NpT) ensemble using  the umbrella sampling scheme combined with the WHAM (see Section~\ref{section-Methods-WHAM} for details). The profiles are shown in  Figure~\ref{fig-WHAM-eth-meth}, the changes in free and potential energy, as well as in entropy after the C-A transition are listed in Table~\ref{table-deltaUGS-WHAM}. 

The reaction coordinate MGW is the average distance between phosphorus atoms on the opposite strands. In the region near A-DNA, MGW is the width of the major groove. This reaction coordinate was chosen to differentiate between the A and C forms in the ethanol and methanol solutions. The C and B forms correspond to the same value of MGW. 

We should note here that, in the CHARMM force field, the geometry of the C-DNA is poorly reproduced, and we cannot unambiguously identify the form of DNA (B or C) in the local minimum near MGW=23\AA{} (for details, see Section~\ref{section-Methods-C-form-CHARMM}). However, near MGW=23\AA{} the minor groove in water is wider than in ethanol and methanol, and parameter $Twist$ is smaller (see Table~\ref{table-AC-helix-parameters}). Therefore we will call the DNA form in the local minimum near MGW=23\AA{} in the ethanol and methanol solutions the C form, and in the water - the B form, as having an intermediate value of parameter $Twist$ between the A and C forms. The narrower minor groove and the larger $Twist$ of the DNA molecule in alcohol solutions in the C-complexes result from much greater number of ions in the minor groove. At the chosen large fraction of alcohol in the solvent, the balance of interactions in the CHARMM force field corresponds to the experiment: in the ethanol solution, the A-complex is more favorable, and in the methanol solution - the C-complex.  

We also carried out simulations of free DNA decamer in the A- and C-forms. The additional calculations confirmed both the instability of the A form in the water and in the methanol solution and the presence of two local minima, A and C, in the ethanol solution (see Section~\ref{section-Methods-WHAM} for details). As the difference in free energy between the A- and C-complexes of the decamer in the ethanol solution was rather small, we also calculated the free energy profile for the 20-mer composed of two decamers to weaken edge effects.
\begin{table}
\caption{Changes in the potential $\Delta U$ and free $\Delta G$ energies (kcal/mol) in the transition from the C-complex to the A-complex (fixed by reaction coordinates MGW=23\AA{} and MGW=9\AA{}, respectively) in DNA oligomers (see Figure~\ref{fig-WHAM-eth-meth}). The change in the entropy is calculated as $T\Delta S=\Delta U-\Delta G$.
\label{table-deltaUGS-WHAM}
}
\begin{tabular}{|l|c|c|c|}
\hline 
System                                          & $\Delta $U & $\Delta $G & $T\Delta S$ \\
\hline
\bf{Ethanol}                                 &              &             &        \\
decamer                                         &\bf 16.3 &\bf -1.6 &\bf 17.9   \\
20-mer                                           & 38.0     & -8.0      & 46.0   \\
decamer, no P-cross                        & 22.5     & 4.9        & 17.6   \\
decamer, no O4'-cross                     & -3.4     & -8.5      & 5.1    \\
\hline
\bf{Methanol}                               &              &            &        \\
decamer                                       &\bf 22.2  &\bf 4.7  &\bf 17.5 \\
decamer, no P-cross                       & 23.5      & 9.2      & 14.3   \\
decamer, no O4'-cross                   & 4.7         & -4.8     & 9.5    \\
\hline
\bf{Water}                                   &        &        &          \\
decamer                                        &\bf 10.2&\bf 7.2 &\bf 3.0 \\
\hline 
\end{tabular}
\end{table}

For a detailed study of the systems in the A- and C- (or B-) minima (data in all the Figures and Tables except for Figure~\ref{fig-WHAM-eth-meth} and Table~\ref{table-deltaUGS-WHAM}), we performed additional MD simulations 100ns long (or even 1$\mu$s in some cases) at fixed reaction coordinates MGW=9\AA{} for the A-DNA and MGW=23\AA{} for the B- and C-DNA (see Section~\ref{section-Methods-MD} for details and Table~\ref{table-boxes} for the list of the systems).
The helix parameters are summarized in Table~\ref{table-AC-helix-parameters}.
\begin{table}
\caption{Helix parameters for decamer (CCGGGCCCGG) -  simulation with the fixed reaction coordinate: MGW=9~\AA{} for the A-complex, MGW=23~\AA{} for the B-, C- and 'C-no O4'-cross' complexes. Two terminal base steps are excluded from the calculation of parameters $Twist$, $Roll$ and $Slide$.
\label{table-AC-helix-parameters}
}
\begin{tabular}{|l|c|c|c|c|}
\hline 
                       & Exp.\footnotemark[1] & Water    & Methanol        & Ethanol    \\
\hline
\%BII($\varepsilon$-$\zeta >$0),            &                                 &                   &                     &                  \\           
A                     &                                 & 10$\pm$8  & 5$\pm$5     & 5$\pm$5   \\
B                     & 40\footnotemark[2]& 43$\pm$11 &                     &                  \\
C                     &$>$40                          &                    & 28$\pm$11 & 31$\pm$11 \\
E                      &  0                            &                   &                      &                  \\
C(B)-noO4'cr.  &                                & 40$\pm$11     & 12$\pm$9 & 11$\pm$9  \\
\hline
\%north. sug.  &    &               &               &               \\
A                      &    & 67$\pm$12     & 75$\pm$12     & 84$\pm$7      \\
B                      &    & 7$\pm$6       &     &        \\
C                      &    &                      & 7$\pm$6       & 11$\pm$10        \\
E                      &100 &               &               &                 \\
C(B)-noO4'cr.  &    & 9$\pm$7       & 41$\pm$11     & 58$\pm$15       \\
\hline
Min. gr.  &                        &                   &              &                  \\    
width, \AA  &                        &                   &              &                  \\           
A                     & 16.9                 & 16.00$\pm$0.31         & 16.18$\pm$0.34   & 16.28$\pm$0.27   \\
B                     &  12.6                & 14.87$\pm$1.13        &                     &                  \\
C                     &  9.5                  &                    & 14.11$\pm$0.99  & 14.10$\pm$0.97     \\
E                      & 15.3                &                   &                      &                  \\
C(B)-noO4'cr. &                        &  14.98$\pm$0.91        &  15.55$\pm$0.62  & 15.64$\pm$0.40     \\
\hline
$Twist$, \textdegree     &    &               &               &               \\  
A                      & 30 & 30.9$\pm$1.3  & 31.1$\pm$1.2  & 31.1$\pm$1.1  \\
B                      & 36 & 32.2$\pm$3.1    &               &               \\ 
C                      & 38 &  & 33.9$\pm$1.8  & 34.0$\pm$1.8  \\
E                      & 32 &               &               &               \\
C(B)-noO4'cr. &       & 32.1$\pm$1.9  & 30.4$\pm$1.7  & 29.9$\pm$1.9  \\
\hline
$Slide$, \AA             &    &               &               &               \\
A                      &-1.4&-1.31$\pm$0.27 &-1.32$\pm$0.25 &-1.40$\pm$0.18 \\
B                      & 0.5& 0.48$\pm$0.42 &               &               \\ 
C                      & 1.8&               & 0.33$\pm$0.31 & 0.36$\pm$0.33 \\
E                      &-1.6&               &               &               \\
C(B)-noO4'cr. &       & 0.34$\pm$0.37 &-0.69$\pm$0.48 &-0.95$\pm$0.58 \\
\hline
$Roll$, \textdegree      &    &               &               &               \\
A                      & 12 & 12.7$\pm$1.8  & 13.8$\pm$2.0  & 13.4$\pm$1.9  \\
B                      & 2  &  6.1$\pm$2.1   &               &               \\ 
C                      &-12 &   &  5.4$\pm$2.2  &  5.4$\pm$2.1  \\  
E                      &2.5 &               &               &               \\  
C(B)-noO4'cr. &      &  6.2$\pm$2.1  &  6.6$\pm$2.6  &  6.0$\pm$2.1  \\
\hline 
\end{tabular}
\footnotetext[1]{The experimental parameters for the A, B and C forms are taken from review~\cite{2023-DNA-review-we} (Table~3 and Figure~10); for the E form - from article~\cite{2000-Dickerson-stable-intermediate-CAT}.}
\footnotetext[2]{In the experiment, the phosphates on steps CpC, GpG and CpG adopt the BII conformation with a very high probability - unless the DNA molecule is in the A form, see Fig.~10 of the review~\cite{2023-DNA-review-we}.}
\end{table}

We have analyzed the difference in the potential energy of the system components between the C-complex and the A-complex (fixed by reaction coordinates MGW=23\AA{} and MGW=9\AA) (see Section~\ref{section-Methods-U} for details). We calculated separately the internal energy of the DNA molecule (DNA-DNA), the interaction of ions with each other (Na-Na), ions with water (Na-Water), and so on. The results are presented in Table~\ref{table-energy-components-Delta-U-eth-meth-wat}.
\begin{table}
\caption{Change in the components of the potential energy (in kcal/mol) in transition from the C-complex to the A-complex (fixed by MGW=23\AA{} and MGW=9\AA{}) in the water (W) and in the 80 vol\% methanol (M) and ethanol (E) solutions. T=300K. The standard errors of the estimated means (calculated by block averaging; block sizes are 10-20ns; trajectory length $\sim$1$\mu$s) are given as an example in the case of the decamer in the ethanol solution. The large errors in measurement of the energy components reflect slow energy transfer between them.
\label{table-energy-components-Delta-U-eth-meth-wat}
}
\resizebox{\linewidth}{!}{%
\begin{tabular}{|l|ccc|ccc|}
\hline 
                                & E      & M     & W     & E20  & M20 & W20 \\
\hline
Wat-Wat                  & 52.9$\pm$2.8      & 34.5     & -2.1    & 66      & 52      & -266 \\
Wat-Alc                   &  2.6$\pm$5.0     & 22.2     &         & -88     & -55     &   \\
Alc-Alc                   & -49.6$\pm$3.5    & -60.1    &         & -190    &   -238  &       \\
$\Sigma$:\bf{Solvent-}&                         &             &         &            &             &       \\
\bf{-Solvent}              & \bf{5.9}           & \bf{-3.4}&\bf{-2.1}&\bf{-211}&\bf{-241}&\bf{-266} \\
\hline
DNA-Wat                    & -139.8$\pm$7.1     & -141.6     & -89    & -86    & -242    & -245 \\ 
DNA-Alc                    & 51.6$\pm$6.6       & 47.7       &        & 270    & 147     &      \\
Na-Wat                     & 26.8$\pm$6.5       & 39.5     & 129.9    & 22     & 147     & 857 \\
Na-Alc                     & 104.8$\pm$2.9      & 114.2    &          & 391    & 540     &     \\
$\Sigma$:\bf{Solute-}&          &                &         &            &             &       \\
\bf{Solvent}                & \bf{43.4}          &\bf{59.8} &\bf{40.9} &\bf{597}&\bf{591} &\bf{612} \\
\hline
DNA-DNA                      & 230.4$\pm$2.6      & 205.4    & 194.3    & 1554   & 1521    & 1475 \\
Na-Na                        & 129.9$\pm$8.1      & 81.1     & 92.8     & 1483   & 1148    & 945 \\
DNA-Na                       &-392.9$\pm$13.3      &-320.4    &-315.7    & -3385  & -2999   &-2757 \\
$\Sigma$:\bf{Complex}        &\bf{-32.6}          &\bf{-33.9}&\bf{-28.6}&\bf{-349}&\bf{-330}&\bf{-337}\\
\hline
$\Delta$U                    &16.6$\pm$0.8        & 22.4     & 10.3     &     37  &     21  & 10 \\
\hline 
\end{tabular}
}
\end{table}

From Table~\ref{table-deltaUGS-WHAM} one can see that the change in the entropy of the system in the C-A transition for the decamer is approximately the same for the ethanol and methanol solutions. The disadvantage of the A form in the case of methanol occurs because of a higher increase in the potential energy of the system $\Delta$U. In the water, the role of entropy in the transition is small, the A-complex is unfavorable also due to its much higher potential energy. From Table~\ref{table-energy-components-Delta-U-eth-meth-wat} one can see that the main contribution to the decrease in the potential energy of the system in C-A transition is made by the interaction of DNA with counterions, and this term is the lowest of all for DNA in the ethanol solution. In the next Section we will anylize the character of the ion-DNA interaction in both the A- and C-complexes. 
\section{Ions as inherent part of A- and C-complexes in low polarity solvents: ionophores P-cross and O4'-cross}
\label{section-ion-DNA-coordination-in-stability}
\begin{figure*}
\begin{center}
\includegraphics[width=0.9\linewidth]{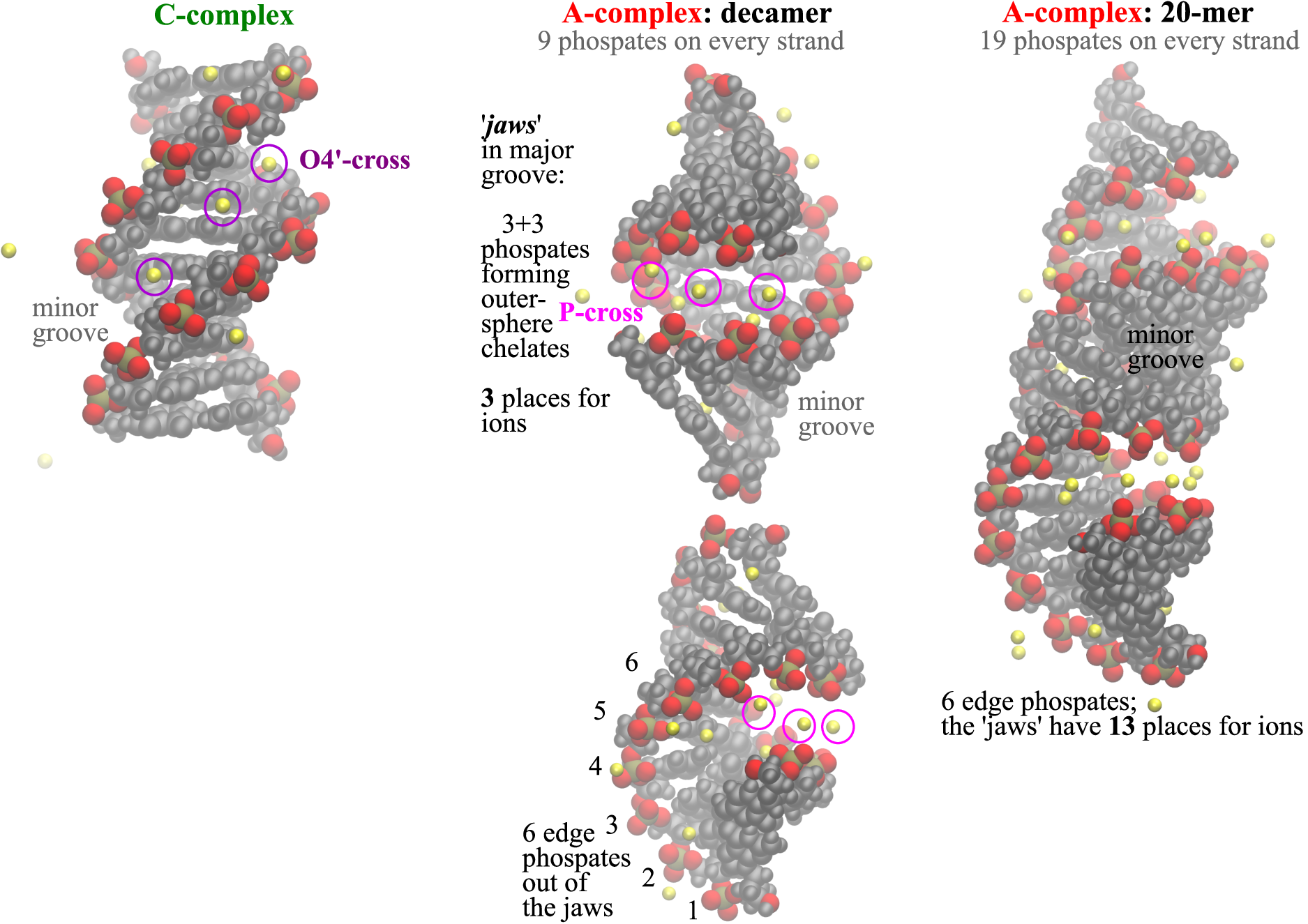}
\caption{Ions providing stability of the C- and A-complexes in low-polarity solvents: ionophores O4'-cross and P-cross. We show snapshots from MD trajectories of the DNA decamer and the 20-mer in the ethanol solution. Coordination of ions by these and other DNA ionophores is detailed in Section~\ref{section-coord-compl}. Visualization (here and in the following): VMD program \cite{1996-VMD,VMD-site}.
\label{fig-ions-in-major-minor-grooves}
}
\end{center}
\end{figure*}

The A and C forms of DNA have specific sets of ionophores ('pockets') attracting positively charged atoms of proteins - and (counter)ions. In an ionophore, an ion is coordinated by several polar DNA atoms, which can be both in the first and in the second coordination sphere (CS) of the ion. In the first case, we will speak of a direct contact between the ion and the DNA, and in the second case, of an outer sphere contact (see~Section~\ref{section-Methods-Contacts}). A simple visualization of the MD trajectories (see~Figure~\ref{fig-ions-in-major-minor-grooves}) readily reveals the main ionophores P-cross and O4'-cross.
In Section~\ref{section-coord-compl} we consider the coordination of ions in these and the other ionophores.

To check if the possibility for ions to coordinate with ionophores is critical for stability of the A- and C-complexes, we have obtained the free energy profiles for DNA decamers with ionophores P-cross and O4'-cross blocked. If we forbid the ions to form (inner and) outer sphere contacts with atoms OP1 and OP2 belonging to the four phosphates coordinating the ions in the 'jaws'{} of the A-complex (see Figure~\ref{fig-ions-in-major-minor-grooves}), then, in both the alcohol solutions, the local minimum for the A form disappears (variant 'no P-cross'; see Table~\ref{table-deltaUGS-WHAM}). 

If we forbid Na$^+$ ions to form (inner and) outer sphere contacts with O4' oxygen atoms in the minor groove, the local minimum for the C form on the free energy curves disappears not only in the ethanol solution, but also in the methanol solution (variant 'no O4'-cross'; see Table~\ref{table-deltaUGS-WHAM}). The only minimum in the case 'no O4'-cross' in the alcohol solutions is the A-complex. In the water, the DNA remains in the B form, because, in such a polar solution, the ions have no need to approach the DNA surface and to fill in the ionophores of the major groove. 

So we have seen that the A-complex does not exist without ions in ionophores P-cross. Analogously, the C-complex does not exist without ions in ionophores O4'-cross. These ions are an inherent part of the complexes in solution.  As we will see later, the ions in other ionophores may also be crutial for the stability of the A-complex.    
\section{Counterion-DNA coordination in the A- and C- complexes}
\label{section-coord-compl}
\subsection{Modeling results}
\label{section-pockets-results}
\begin{figure*}
\begin{center}
\includegraphics[width=0.9\linewidth]{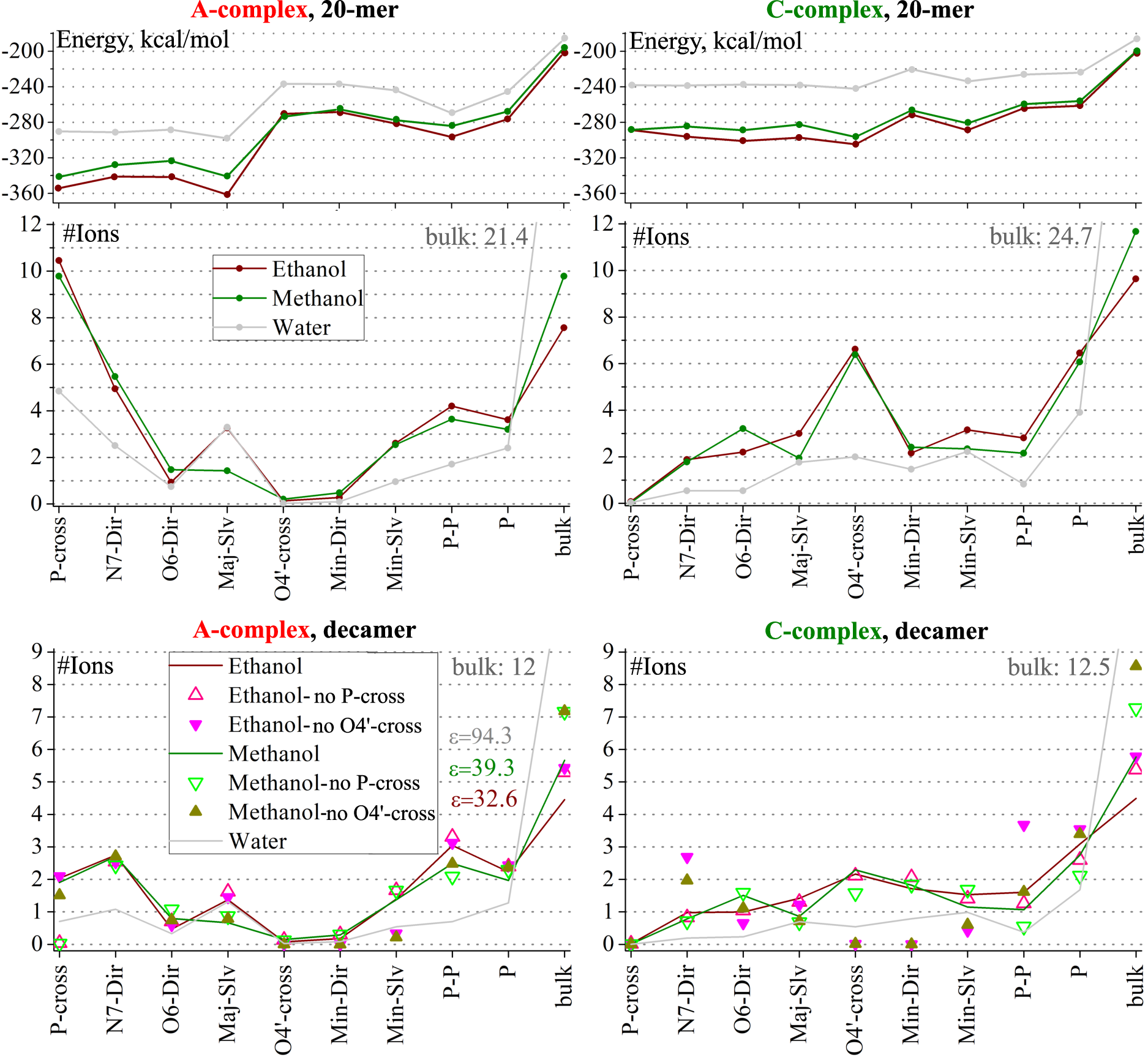}
\caption{Number of counterions in ionophores of  DNA. For the 20-mer, we also show the energy of an ion in the ionophores. Variants 'no P-cross' and 'no O4'-cross': results in cases of blocking outer shell contacts with phosphate oxygen atoms and  atoms O4' (see Section~\ref{section-ion-DNA-coordination-in-stability}). The total number of counterions for the decamer is 18, for the 20-mer - 38. The standard deviation of the number of counterions in an ionophore depends on the ionophore and lies within the range 0.5-1.6 for the decamer and 1.3-2.1 for the 20-mer. The standard deviation of the energy of an ion in an ionophore is 9-27~kcal/mol. The standard errors of the estimated mean values for the energy in the 20-mer (calculated by block averaging; block sizes are 15-20ns) vary from 1.1 to 3.5~kcal/mol.
\label{fig-pockets-eth-meth}
}
\end{center}
\end{figure*}

\begin{figure*}
\begin{center}
\includegraphics[width=0.85\linewidth]{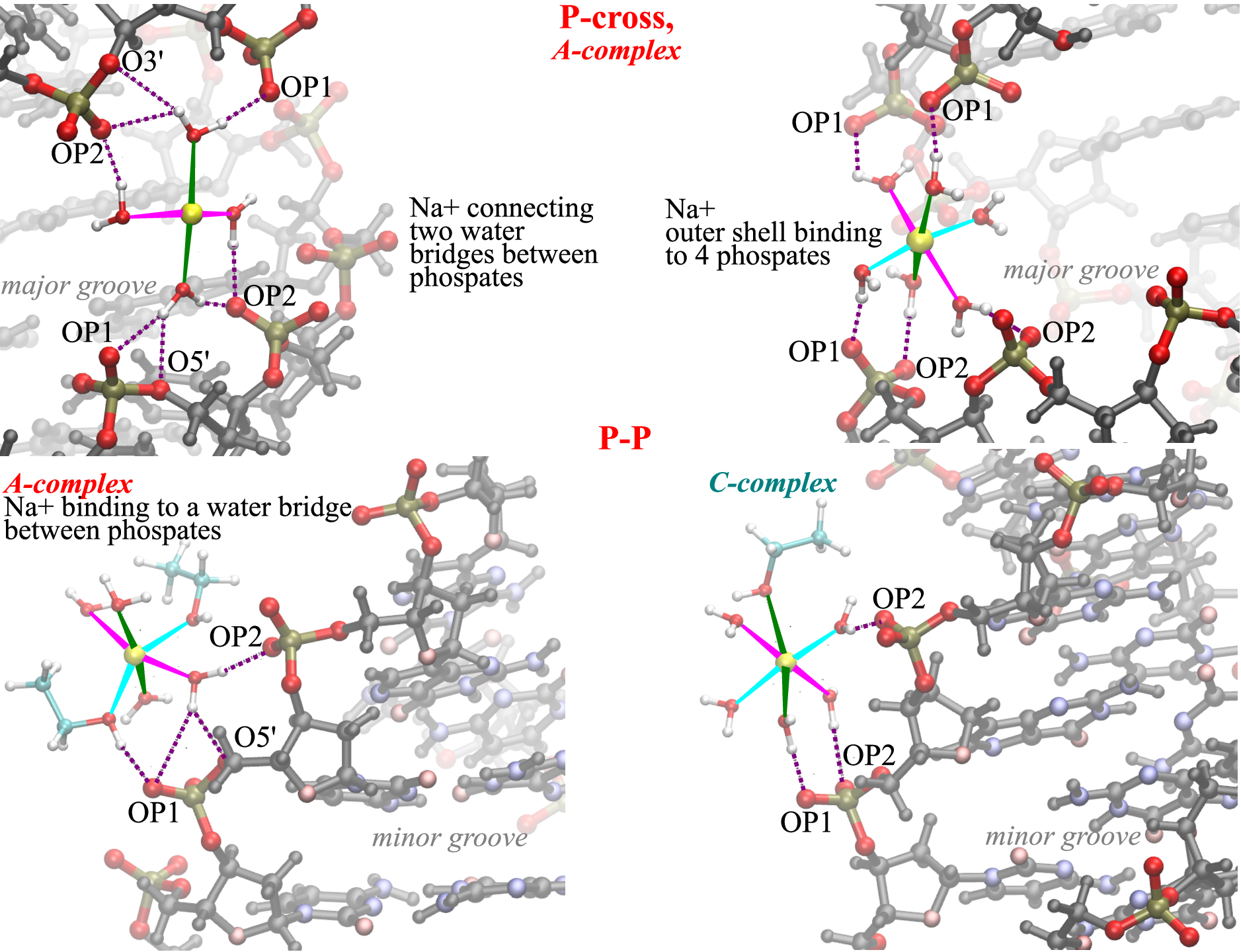}
\caption{Ionophores near phosphates (see Table~\ref{table-pockets-description}). The main ionophore of the A-complex P-cross: two typical binding modes (with a water bridge in the first CS ($\sim$70\%) and without it). Ionophore P-P in the A- and C-complexes. Of the six ligands of the ion, only those necessary for binding to DNA may be shown. Different colors are used for orthogonal directions.
\label{fig-coord-compl-P}
}
\end{center}
\end{figure*}

\begin{figure*}
\begin{center}
\includegraphics[width=0.85\linewidth]{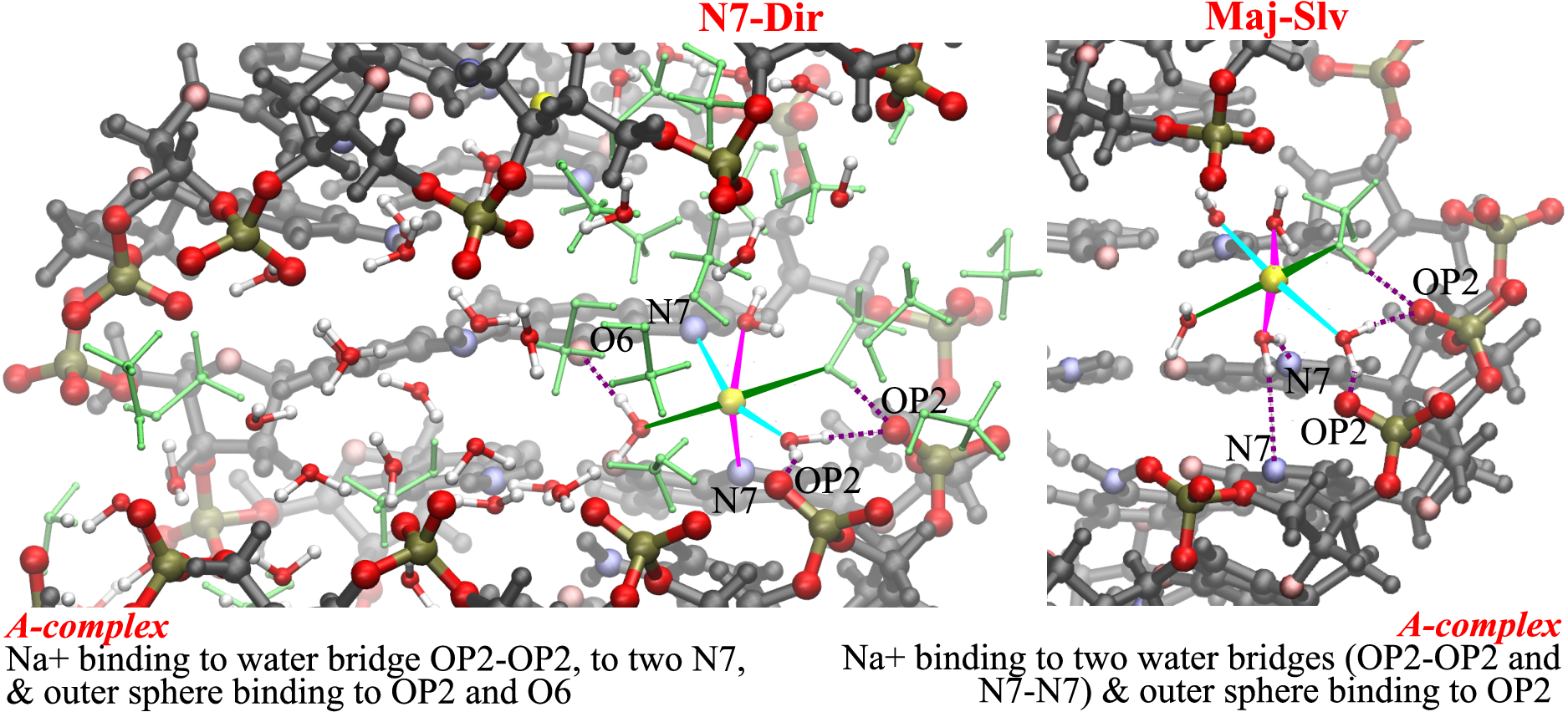}
\caption{Ionophores N7-Dir and Maj-Slv (see Table~\ref{table-pockets-description}) in the major groove of the A-complex in the methanol solution. For ionophore N7-Dir, all solvent molecules in the groove are shown.
\label{fig-coord-compl-Maj}
}
\end{center}
\end{figure*}

\begin{figure*}
\begin{center}
\includegraphics[width=0.85\linewidth]{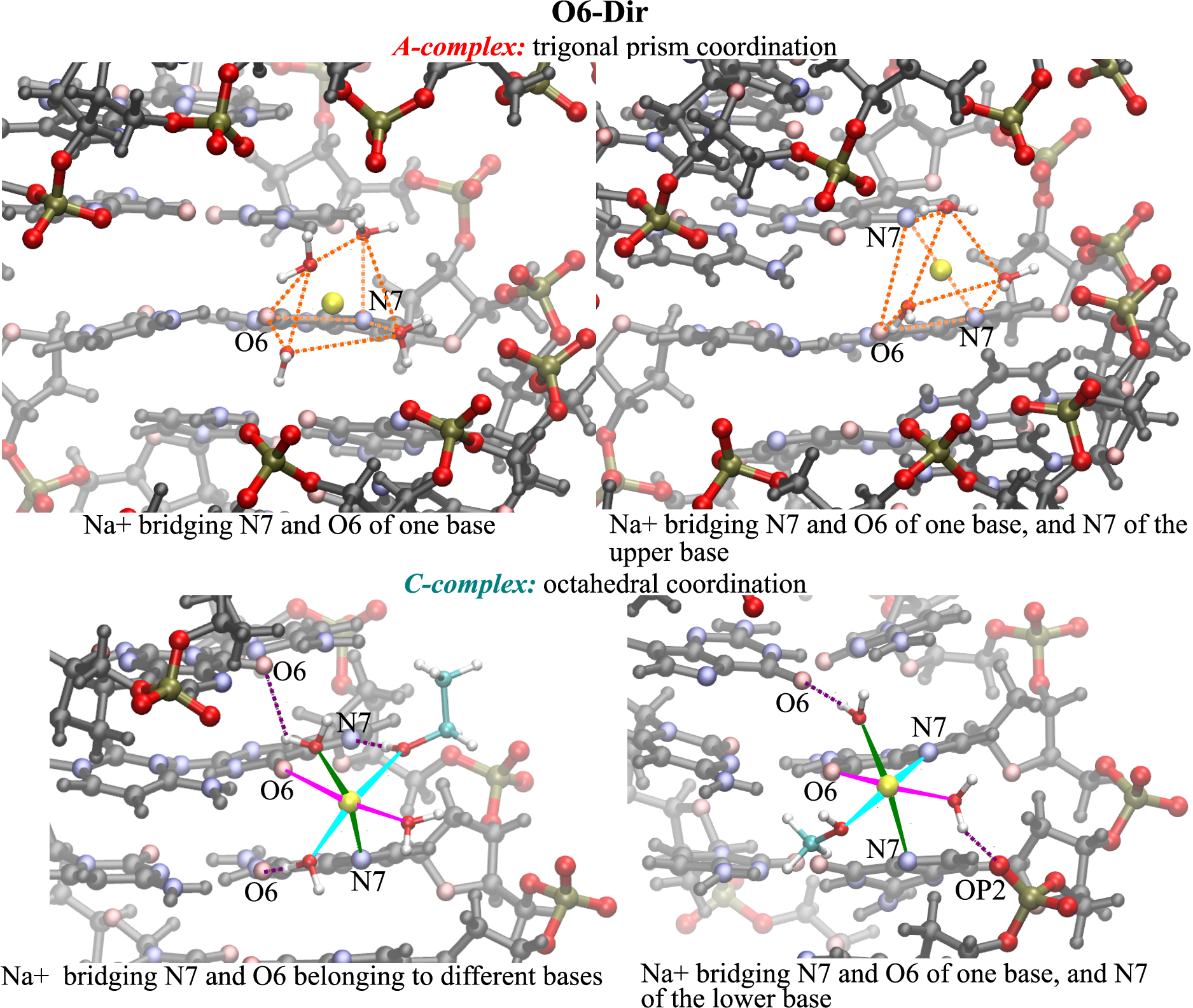}
\caption{Ionophore O6-Dir (see Table~\ref{table-pockets-description}) in the A- and C-complexes. In the A-complex, instead of the directions to ligands, we depicted trigonal prisms, at the vertices of which the ligands are located. 
\label{fig-coord-compl-O6-Dir}
}
\end{center}
\end{figure*}

\begin{figure}
\begin{center}
\includegraphics[width=0.85\linewidth]{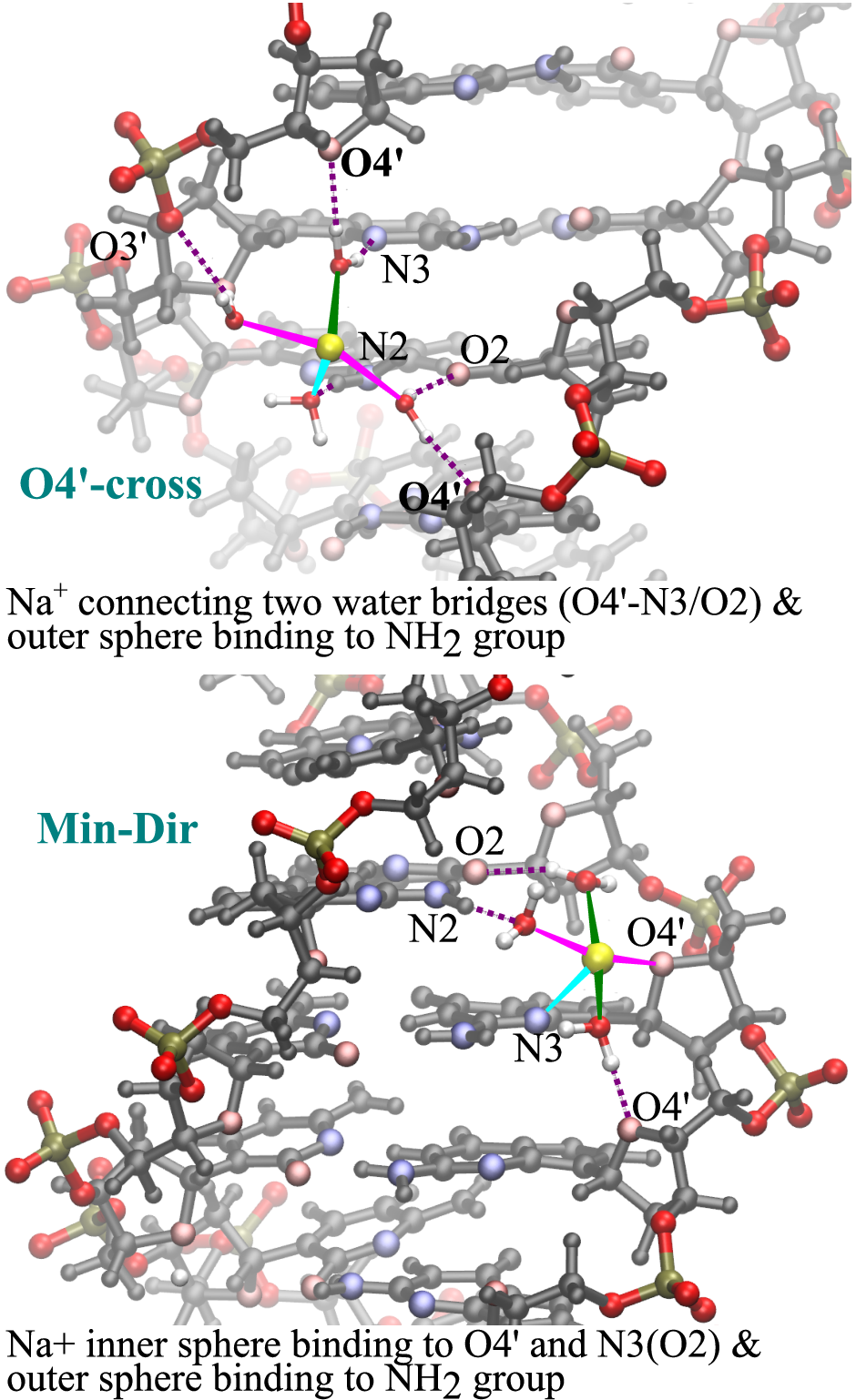}
\caption{Ionophores in the minor groove of the C-complex: the main ionophore O4'-cross and its replacement in the case of a wide minor groove Min-Dir (see Table~\ref{table-pockets-description}).  
\label{fig-coord-compl-O4cross}
}
\end{center}
\end{figure}

\noindent
We observed several ionophores (coordination sites, 'pockets') in the A- and C-complexes. They are listed in Table~\ref{table-pockets-description}.
\begin{table*}
\caption{Coordination of ions by DNA atoms in ionophores. The number after the ionophore name indicates its priority in assignment in case the ion's location corresponds to more than one ionophore. For the definitions of direct and outer sphere contacts, see Section~\ref{section-Methods-Contacts}. 
\label{table-pockets-description}
}
\begin{tabular}{|l|l|}
\hline
Ionophore             &  Description   \\
\hline
{\bf P-cross} (1)   & ion is inner or outer sphere coordinated by two (or more) oxygen atoms \\
                             & OP1 or OP2 belonging to different DNA strands (Fig.~\ref{fig-coord-compl-P})\\
N7-Dir (3)           & ion is in direct contact with one (or more) atom (G)N7 \\
                             & of the major groove (no direct contacts with atoms (G)O6) (Fig.~\ref{fig-coord-compl-Maj}) \\    
O6-Dir (4)            &  ion is in direct contact with one (or more) atom (G)O6 of the major groove, \\
                             & inner sphere coordination by (G)N7 is possible (Fig.~\ref{fig-coord-compl-O6-Dir}) \\                          
Maj-Slv (6)          & ion is in outer sphere contact with (G)N7, (G)O6, H41/H42\\
                            & (hydrogen atoms of (C)N4) in the major groove (Fig.~\ref{fig-coord-compl-Maj})\\
\hline
{\bf O4'-cross} (2) & ion is inner or outer sphere coordinated by two atoms O4' of different strands (Fig.~\ref{fig-coord-compl-O4cross})\\
Min-Dir (5)         & ion is in direct contact with one (or more) polar atom of the minor groove:\\
                           & (G)N3, (C)O2, O4' (Fig.~\ref{fig-coord-compl-O4cross})\\
Min-Slv (7)         &ion is in outer sphere contact with one (or more) polar atom of the minor groove:\\
                           & (G)N3, H21/H22 (hydrogen atoms of (G)N2), (C)O2, O4' \\
\hline
P-P (8)             & ion is inner or outer sphere coordinated by two oxygen atoms (OP1, OP2, \\
                    & O3', O5') of different phospates from one strand (Fig.~\ref{fig-coord-compl-P})\\
P (9)               & ion is inner or outer sphere coordinated by one (or more) oxygen atom \\
                    & (OP1, OP2, O3', O5') of one phosphate\\
Bulk (10)           & ion has no direct or outer sphere contacts with DNA atoms\\
\hline
\end{tabular}
\end{table*}
The potential energy of an ion in a given ionophore (the total energy of interaction of the ion with the rest of the system) and the number of ions in the ionophores are presented in Figure~\ref{fig-pockets-eth-meth} (for more details, see Tables \ref{table-pockets-energies-A} and \ref{table-pockets-energies-C} in Section~\ref{section-Methods-U}). The average lifetime of an ion in the indicated positions is listed in Table~\ref{table-pockets-lifetime}. The large standard deviations indicated in Table~\ref{table-pockets-lifetime} are due to the fact that the probability of a given lifetime of an ion in an ionophore has a distribution that is well approximated by the sum of three exponential distributions. The first two times characterize frequent and short-term events when an ion leaves an ionophore to replace solvent molecules in the second and the first CS. The ions in each groove also change their coordination with DNA remaining in approximately the same place. The average lifetimes in the grooves (labeled $Major$ and $Minor$) are also presented in the Table.
\begin{table}
\caption{Average time (in ps) of ion residence in the designated position (see table~\ref{table-pockets-description}), in the DNA grooves (\it{Major} \rm and \it{Minor}\rm) (see details in Section~\ref{section-Methods-Contacts}), and in direct contact with the atoms of the major groove (\it{N7+O6}\rm). The reported data are for the A- and C-complexes of the 20-mer, at the fixed reaction coordinate MGW=9\AA{} and 23\AA{}, respectively. For comparison: the lifetime of hydrogen bonds in solutions ranges from 0.65 to 1.5 picoseconds, the residence time of a solvent molecule in the first CS of Na$^+$ ranges from 7.7 to 12 picoseconds. 
\label{table-pockets-lifetime}
}
\begin{tabular}{|c|ccc|}
\hline
Ionophore           &  Ethanol          & Methanol & Water        \\
\hline
\hline
{\bf P-cross}, A    &  133\textpm 250   & 223\textpm 502 & 77\textpm 116 \\
\hline
N7-Dir, A   &  124\textpm 175   & 153\textpm 283 & 87\textpm 103 \\
N7-Dir, C   &  54\textpm 83     & 57\textpm 104  & 38\textpm 41 \\
\hline
O6-Dir, A         & 37\textpm 62      & 52\textpm 129  & 40\textpm 54 \\
O6-Dir, C         & 81\textpm 133     & 123\textpm 210 & 46\textpm 52 \\
\hline
Maj-Slv, A        & 70\textpm 109     & 68\textpm 150 & 65\textpm 86 \\
Maj-Slv, C        & 67\textpm 95      & 67\textpm 102 & 51\textpm 58 \\
\hline
\it N7+O6, A      &\it 244\textpm 344 &\it 609\textpm 1132 &\it 147\textpm 186 \\
\it N7+O6, C      &\it 222\textpm 377 &\it 397\textpm 852  &\it 95\textpm 116 \\
\it Major, A      &\it 343\textpm 871 &\it 723\textpm 3050 & \it 213\textpm 424 \\
\it Major, C      & \it 239\textpm 715&\it 378\textpm 1324 & \it 118\textpm 222 \\
\hline
\hline
{\bf O4'-cross}, C & 69\textpm 103     & 86\textpm 134 & 44\textpm 48 \\
\hline
Min-Dir, A        & 62\textpm 463     & 110\textpm 528 & 53\textpm 138\\
Min-Dir, C        & 138\textpm 754    & 181\textpm 1260& 176\textpm 798\\
\hline
Min-Slv, A        & 115\textpm 187    & 100\textpm 182 & 54\textpm 72 \\
Min-Slv, C        & 31\textpm 63      & 30\textpm 66   & 34\textpm 49 \\
\hline
\it Minor, A      & \it 228\textpm 521 &\it 228\textpm 814    & \it 72\textpm 124 \\
\it Minor, C      & \it 872\textpm 2228 &\it 1249\textpm 3568 & \it 281\textpm 708 \\
\hline
\hline
P-P, A            & 37\textpm 51       & 51\textpm 81 & 21\textpm 21 \\
P-P, C            & 37\textpm 42       & 39\textpm 46 & 20\textpm 16 \\
\hline
P, A              & 33\textpm 46       & 42\textpm 60 & 23\textpm 22 \\
P, C              & 46\textpm 63       & 57\textpm 78 & 28\textpm 26 \\
\hline
Bulk, A           & 154\textpm 1107    & 237\textpm 1319 & 302\textpm 1098 \\
Bulk, C           & 145\textpm 1033    & 212\textpm 1075 & 219\textpm 817 \\
\hline
\end{tabular}
\end{table}
Figures \ref{fig-coord-compl-P}, \ref{fig-coord-compl-Maj}, \ref{fig-coord-compl-O6-Dir}, and \ref{fig-coord-compl-O4cross} visualize typical DNA binding modes for some ionophores.

From the middle part of Figure~\ref{fig-pockets-eth-meth} (and Tables \ref{table-pockets-energies-A} and \ref{table-pockets-energies-C} in Section~\ref{section-Methods-U}) one can see that the main contribution to the formation of the complexes is made by the ions located between DNA strands. Among them, the main ionophores are P-cross for the A-complex and  O4'-cross for the C-complex. These ionophores appear only when the DNA molecule takes the corresponding form: A or C. In both the cases, narrowing of the respective groove is necessary: the major groove for the A complex and the minor groove for the C complex. As we have already seen (see Section~\ref{section-ion-DNA-coordination-in-stability}), without ions at these positions, the corresponding complexes in the alcohol solutions do not exist.

In addition, sequence-specific ionophores located near guanine bases N7-Dir and Maj-Slv (Figure~\ref{fig-coord-compl-Maj}) may also play a role in stabilizing the A-complex. The potential energy of the ion in these positions is lower in the A-complex than in the C-complex, due to the proximity of the phosphates to the ion in the narrow major groove in the A form. The existence of the ionophores N7-Dir and Maj-Slv may be responsible for the well-known 'A-phility'{} of the DNA sequences with a high content of G:C pairs. Ions in these ionophores are almost always in outer sphere contact with phosphates, usually through a water molecule bridging neighboring phosphates of one strand in the A-complex (see Figure~\ref{fig-coord-compl-Maj}).

The potential energy of an ion in ionophore O6-Dir (which is in the middle of the major groove) is almost as low as in ionophore N7-Dir in both the forms (see Figure~\ref{fig-pockets-eth-meth} and Tables~\ref{table-pockets-energies-A} and \ref{table-pockets-energies-C}). However, in the A-complex, ionophore O6-Dir is much less populated because of the inconvenience of the ion coordination with the A form of DNA. In $\sim$15-35\% of cases, the ion in ionophore O6-Dir is in direct contact with one or two atoms (G)O6. The most frequent coordination ($\sim$40\% of cases) also includes a direct contact with atom (G)N7: located on the same base in the case of the A form, and on the adjacent base in the case of the C form (see Figure~\ref{fig-coord-compl-O6-Dir}). At the GpG step, a third direct contact with atom N7 is also often formed (in $\sim$20-35\% of cases). In the C form, due to the positive parameter $Slide$, the ion can be octahedrally coordinated by the ionophore. In the A form, due to the negative parameter $Slide$, in both the cases, O6+N7 and O6+2*N7, octahedral coordination is impossible. Trigonal-prismatic coordination turns out to be unstable: solvent molecules tend to shift from the vertices of the prism for better interaction with the DNA atoms. As a result, the number of ions in ionophore O6-Dir in the A-complex is two times lower than in the C-complex, in both methanol and ethanol solutions.

The ion in ionophore P-P in the C-complex binds to two neighboring phosphates (of one strand) by two different solvent molecules, and in the A-complex - in most cases by one water bridge between the phosphates (see Figure~\ref{fig-coord-compl-P}). In general, water bridges between atom OP2 and neighboring polar atoms (OP1, O5', (G)N7) are a characteristic feature of the A-complex; they appear due to the short distances between neighboring phosphates and between phosphates and the major groove in the A form.  

In the C-complex, ions appear in ionophores Min-Dir in places where the minor groove widens (in the decamer, mainly near the ends of the molecule), and the ion coordination by the main ionophore O4'-cross is impossible. The ion can be coordinated by ionophore Min-Dir only if the sugar adopts the south conformation, therefore the ions leave this pocket after the C-A transition.

From Figure~\ref{fig-pockets-eth-meth} (and Tables \ref{table-pockets-energies-A} and \ref{table-pockets-energies-C}) one can estimate the contribution of one or another ionophore to the stabilization of the A- and C-complexes. Ions in $Bulk$ expectedly have the highest potential energy; they are there only if they failed to occupy the ionophores at the DNA surface. In the water, 'pocket' $Bulk$ makes the main contribution to the potential energy because the main part of the free energy is the entropy $-T*S$. In the alcohol solutions, the A-complex is stabilized mainly by ions in ionophores P-cross, P-P and in the ionophores of the major groove. The contribution of ions located in ionophore P-P compared to ions located in ionophore P-cross is much more noticeable for the decamer. The ions are distributed more evenly around the C-DNA, although the contribution of the minor groove (O4'-cross, Min-Dir and Min-Slv) is the main one.

An ion in the major groove of the A-complex often 'moves' between ionophores N7-Dir, O6-Dir and Maj-Slv, changing the composition and the structure of its solvation shell, but shifting little relative to DNA. As we have already noted, in ionophore Maj-Slv, as well as in ionophores N7-Dir and O6-Dir, most of the ions are outer sphere coordinated by oxygens of phosphates. The ions in these three ionophores are located at approximately the same distance from the nearest phosphate.

In the minor groove of the C-complex, an ion also easily 'moves' between ionophores O4'-cross, Min-Dir, and Min-Slv, with little real movement relative to DNA. An ion in ionophore Min-Slv appears to approach, both in its location relative to DNA and in its coordination, either the position in ionophore O4'-cross or the position in ionophore Min-Dir. The distance from an ion in ionophore Min-Slv to the nearest phosphate is equal to the distance from ions in ionophores O4'-cross and Min-Dir to the nearest phosphate. This distance ($\sim$6\AA) is greater than the distance from ions in ionophores N7-Dir and Maj-Slv to the nearest phosphate ($\sim$5.25\AA), so the potential energy of an ion in the major groove is always lower than the potential energy of an ion in the minor groove, even in the C-complex (except for ionophore O4'-cross). The energy of an ion in the main ionophore of the A-complex P-cross is lower than the energy of an ion in the main ionophore of the C-complex O4'-cross. And in general, in all solvents the potential energy of the ions in the A-complex is noticeably lower than in the C-complex (see Tables \ref{table-pockets-energies-A} and \ref{table-pockets-energies-C}).

The force field we used is not perfect and thus might fail to reproduce some existent ionophores or create non-existent ones. Therefore, in the next Section, we gathered and analyzed experimental data on the location of the ionophores on the DNA surface for both the A- and C-complexes. We will see that our results are well supported by the experimental data.
\subsection{Ions near DNA surface: experiment and modeling in water and in alcohol-water solvents}  
\noindent
Ions Na$^+$ (as well as NH$_{4}^+$ and, at partial occupancy, K$^+$) scatter X-rays as water molecules \cite{2008-DNA-groupI-ions-X-ray-book1}. These ions can be identified only indirectly by their coordination number and coordination distances \cite{2016-Na-K}. However, ionophores on DNA surface that attract sodium ions can also coordinate other ions, and the resulting chelates can be more stable with divalent and trivalent ions. Ion Mg$^{2+}$, isoelectronic to ion Na$^+$, in water forms complexes [Mg(H$_2$O)$_6$]$^{2+}$ with the same octahedral coordination as ion Na$^+$, but with a longer lifetime. Localization of divalent or trivalent ions (Mg$^{2+}$, Ca$^{2+}$, Al$^{3+}$) at the DNA surface may indicate sodium ionophores that require outer sphere contacts. Other, more easily observed, ions of larger size and with less stable water shell (K$^+$, Tl$^+$, Rb$^+$) prefer direct contacts with DNA atoms and can be used to localize the corresponding ionophores. IR spectroscopy also allows detecting the attachment of ions to DNA atoms by shifts of bands and changes in their intensities \cite{1993-La-Eu-Tb-no-A-DNA,1996-B-to-A-in-solution-Hexaamminecobalt-exp,2005-DNA-Mg-phosphate-groups-compaction-Blagoi,2009-IR-DNA-ion-binding-Andrushchenko}. 
\subsubsection{Ionophores of the major groove of DNA in experiment}
\noindent
In ionophore P-cross of the A-complex, even the ions [Mg(H$_2$O)$_6$]$^{2+}$ were not reliably identified due to the diffuse electron density \cite{2000-Mg-A-DNA-X-ray}. The authors of the cited work believe that the reason is the high mobility of ions in this location.
They insist that this binding mode must exist for the ions [Mg(H$_2$O)$_6$]$^{2+}$, because it was found for hexaamminecobalt(III) ions [Co(NH$_3$)$_6$]$^{3+}$ (see \cite{1995-CoNH36-A-DNA-X-ray} and 212D (PDB)). Trivalent ions interact with DNA phosphates more strongly than divalent and monovalent ions. Therefore, hexaamminecobalt (III), aluminum, and gallium salts induce the B to A transition even in water \cite{1996-B-to-A-in-solution-Hexaamminecobalt-exp,1996-Al-Ga-A-DNA}. MD modeling of this transition in the presence of hexaamminecobalt ions \cite{1997-B-to-A-hexaaminecobalt-MD-Chetham} confirmed that the ions do reside between phosphates of opposite strands in the DNA major groove. Not all trivalent ions induce the B-A transition \cite{1993-La-Eu-Tb-no-A-DNA}, which shows 'chemical'{}, rather than 'electrostatic'{}, nature of ion binding to ionophore P-cross. For RNA molecule (which is always in the A form), MD modeling indicates the presence of ions Mg$^{2+}$ in ionophore P-cross \cite{2021-RNA-P-cross}. 

One case resembling the position in ionophore P-cross was reported for ion Mg$^{2+}$, but the ion was located outside the major groove and contacts two other DNA duplexes (PDB: 2B1B). In other ionophores near phosphates (P-P and P) even sodium ions are detected, but they also rather bridge phosphates of different DNA duplexes, maintaining the crystal structure of the sample. Probably, the location of ions between DNA duplexes in crystals is energetically more favorable for most ions than the position in ionophore P-cross.

In water, divalent ions Cu$^{2+}$, Zn$^{2+}$, Mn$^{2+}$, and Ca$^{2+}$ cause not the transition of DNA to the A form, but condensation of DNA into particles of about 40~nm in size \cite{2005-DNA-Mg-phosphate-groups-compaction-Blagoi}. With increasing salt concentration, these ions fill the ionophores of the major groove (especially N7-Dir), and then bind different duplexes - instead of filling ionophores P-cross and converting DNA to the A form. Magnesium ions, which do not bind to ionophore N7-Dir due to their strong bond with their solvation shell, cannot provide sufficient neutralization of DNA charges and do not cause the DNA condensation \cite{2005-DNA-Mg-phosphate-groups-compaction-Blagoi}. The absence of the transition to the A form in this case is explained by insufficient electrostatic energy of the interaction between divalent ions and DNA phosphates in a too polar solvent. For the ions, it is more advantageous to remain in water, which reduces the free energy due to entropy.

The presence of metal ions in direct contact with the major groove (ionophores N7-Dir and O6-Dir) is better documented than in ionophore P-cross because of the lower ion mobility in these ionophores and because these ionophores also function in the B form, which is studied more often than the A form. In both ionophores N7-Dir and O6-Dir, Tl$^+$ ions were detected \cite{2001-Xray-ions-in-grooves-B-DNA}. In ionophore O6-Dir, even Na$^+$ ions were found, both in native DNA (PDB: 1ZF6 and 1ZF9 \cite{2005-A-sequences}) and in DNA complexed with proteins (PDB: 7L35, 3G0R and 4RPX). In ionophore N7-Dir, a large number of ions misidentified as Mg$^{2+}$ have also been reported \cite{2016-Xray-Mg-only-phosphates}. The cases of outer sphere coordination of ions by ionophore Maj-Slv observed in our simulations were also found experimentally in the A-DNA: for the ions Mg$^{2+}$ \cite{2000-Mg-A-DNA-X-ray} and Ca$^{2+}$ (PDB: 1ZFA, \cite{2005-A-sequences}). The binding of magnesium ions to RNA molecules is much better known, including in the ribosome. Both inner and outer sphere coordination by atoms (G)N7 and (G)O6, as well as by phosphate oxygen atoms, are reported \cite{2011-ions-binding-RNA-Auffinger}.
\subsubsection{Ionophores of the minor groove of DNA in experiment}
\label{section-min-groove-ionophores}
\noindent
As for the ionophores of the B (or C) form of DNA, attention was focused on A-tracts in connection with the analysis of the 'spine' or 'ribbon' of hydration in the very narrow minor groove of these sequences. In the A-tracts, water molecules in the minor groove bridge polar atoms O2/N3 belonging to the bases of different strands (on the adjacent base pairs), thus stabilizing the B form \cite{1998-A-B-hydration-X-ray}. These water bridges are linked by water molecules into a whole continuous ribbon of hydration. Monovalent cations, for example, Rb$^+$, can take the place of the water bridges O2/N3-O2/N3 in the A-tracts \cite{1999-Rb-A-tract-X-ray}.

In the wider minor groove formed by G:C pairs, O2/N3-O2/N3 water or water-ion bridges between the strands cannot form because of (G)N2 amino groups. Instead, water bridges O4'-O2/N3 appear, both the atoms belonging to the same strand. One can see the bridges O4'-O2/N3 in the structures 1DPN and 460D (PDB) at the ends of the Dickerson-Drew dodecamer \cite{1998-A-B-hydration-X-ray,1999-Rb-A-tract-X-ray}. Water bridges O4'-O2/N3, as well as bridges O2/N3-O2/N3, can be connected by one or two water molecules.

In ionophore O4'-cross, ion Na$^+$ takes the place of a water molecule, linking two O4'-O2/N3 water bridges of different DNA strands (Figure~\ref{fig-coord-compl-O4cross}). Almost always, this ion is also coordinated by one (G)N2 amino group through a water molecule. The second CS of this ion may contain O4' oxygens of other sugars and O3' oxygens of phosphates. The second CS of the ion in the O4'-cross ionophore includes at least 5 polar atoms of different strands, which stabilizes the south sugars of the C-complex.

The binding mode of ionophore O4'-cross was first observed in crystallized native DNA (oligomers) for ion Mg$^{2+}$ (1EN3 (PDB) \cite{2000-O4-cross-exp}). In DNA complexed with proteins, even Na$^+$ ions were detected at locations close to the position in this ionophore (PDB: 6V8U, 1T9I, 5BOM; see also other structures from \cite{2015-O4cross-polymerasa}). This indicates that in crystals ions spend much more time in ionophore O4'-cross of the B-DNA than in ionophore P-cross of the A-DNA. This coincides with what we observed in solutions in MD simulations. The ion spends a very long time in the minor groove of the C-complex, often remaining in approximately the same place, but changing coordination between the ionophores O4'-cross, Min-Dir and Min-Slv (see~Table~\ref{table-pockets-lifetime}).  

In the wide minor groove away from the A-tract of the Dickerson-Drew dodecamer, in ionophore Min-Dir, ions Tl$^+$ (PDB: 1JGR, \cite{2001-Xray-ions-in-grooves-B-DNA}) and Na$^+$ (PDB: 3OPI, \cite{2011-Na-in-Min-Dir-accidently}) have been reported. The ionophore Min-Dir also exists near A:T pairs, although in this case the coordination of the ion with DNA is slightly different (see PDB: 5J0Y). In most cases in our simulations, the atoms O4' and O2/N3 forming the bidentate complex with the ion belong to the same nucleotide, while in the case of 5J0Y they belong to the different nucleotides. In addition, near A:T pairs, it is more likely to find an ion directly coordinated by two O2 atoms of thymines from adjacent base pairs, together with outer sphere contacts with two O4' atoms from the opposite strands (PDB: 7RVA, 1VE8).
\subsubsection{Ionophores on DNA surface in MD simulations}
\noindent
One can find a lot of works modeling the interactions between DNA and ions, but most of their authors did not aim to precisely localize (from the point of view of coordination chemistry) ionophores on the DNA surface and to determine their role in the conformational dynamics of DNA. A pleasant exception is the paper \cite{1999-MD-A-B-Pettitt}, in which the distribution of ions around DNA was considered an important factor determining the A- or B-conformation of the DNA molecule. Even in the framework of the early force fields used in this work, the authors found most of ionophores of DNA, although the role of each of them in stabilizing the A- and C-complexes could not be determined at that time. In work \cite{2004-MD-Na-K-DNA}, in the framework of the AMBER force field, it was shown that the interaction of DNA with counterions depends on their chemical nature (actually, the size): ions Na$^+$ and K$^+$ prefer different ionophores of the B-DNA: Na$^+$ more often is in the minor groove, while K$^+$ - in the center of the major groove (in ionophores O6-Dir and Maj-Slv). Pasi et al \cite{2015-MD-seq-dep-ions-distr-around-DNA} proposed a method for analyzing the ion distribution density around DNA. The method was applied to search for sequence-specific ionophores of DNA (B-DNA in the framework of the AMBER force field). The method allowed to localize the ionophores in the major groove near atoms N7 and O6 of guanine and to demonstrate that there is no ionophore near the atom N7 of adenine, at least for the ion K$^+$.    
\section{Energy of A- and C-complexes and filling of ionophores with ions}
\label{section-energy-constituents}
\noindent
The first step to understanding why a free DNA molecule can not take the A form in the methanol solution is to consider the relationship between the energies $\Delta$U$_{DNA-DNA}$, $\Delta$U$_{DNA-Na}$, and $\Delta$U$_{Na-Na}$ in different solutions (Table~\ref{table-energy-components-Delta-U-eth-meth-wat}). The change in the internal energy of the DNA molecule $\Delta$U$_{DNA-DNA}$ is mainly an increase in the Coulomb energy of repulsion between phosphates more closely located to each other in the A-complex. The value of $\Delta$U$_{Na-Na}$, similarly, increases when the ions approach each other, filling the major groove of the A-complex (the closely situated ionophores P-cross, N7-Dir, Maj-Slv). Both of these terms are compensated by a decrease in the interaction energy of the ions with the phosphates in the A-complex $\Delta$U$_{DNA-Na}$. Along with this,  the higher the value $\Delta$U$_{DNA-DNA}$, the lower is the sum $\Delta$U$_{DNA-Na}$+$\Delta$U$_{Na-Na}$. All six points for the systems presented in Table~\ref{table-energy-components-Delta-U-eth-meth-wat} are collinear. The highest energy $\Delta$U$_{DNA-DNA}$ is in the ethanol solution, because in this case the DNA molecule has the maximum proportion of north sugars (see Table~\ref{table-AC-helix-parameters}), so the molecule takes the real A form. In the methanol solution and in the water, even with the major groove width fixed by reaction coordinate MGW, the DNA molecule is not able to take the A form.

Comparing Tables~\ref{table-pockets-energies-A} and \ref{table-pockets-energies-C}, one can see that the potential energy of ions in the methanol solution is higher than in the ethanol solution, both in the A- and C-complexes. In the ethanol solution, ions fill up most of the ionophores better, and there are less ions left in $Bulk$.
In the methanol solution, ions occupy ionophores with direct contact with DNA slightly more often than in the ethanol solution, but the ionophores with outer sphere coordination (especially Maj-Slv) are occupied noticeably less often (see Figure~\ref{fig-pockets-eth-meth}). Although this is true for both the complexes, the disadvantage of the A-complex is greater than the disadvantage of the C-complex. Thus, the inability of counterions in the methanol solution to fill up the ionophores of the A form with outer sphere coordination results in the transition to this form not taking place in this medium.
\section{Ionophores and chemical nature of the solvent}
\noindent
The number of counterions situated in solution further than in the second solvation shell of DNA (in 'pocket' $Bulk$) is determined by the balance of the entropy, the electrostatic attraction to phosphates and the specificity of interaction between the ion and the solution. In the first approximation, this balance can be described by the static permittivity of the solution $\varepsilon$. In the water, the number of ions far from the DNA is maximum, in ethanol - minimum (see ~Figure~\ref{fig-pockets-eth-meth}). The more compact A form attracts ions more strongly. It was believed \cite{1973-Ivanov-CD-A-B-C} that adding a solvent with a higher permittivity (e.g. methanol, $\varepsilon_{exp}\sim33$) to water transforms DNA into the C form, and with a lower permittivity (e.g. ethanol, $\varepsilon_{exp}\sim25$) - into the A form. Indeed, the ionophores of the A-DNA require more ions, providing a lower energy of their interaction with DNA. In a polar solvent, condensation of ions on DNA may be unfavorable due to the loss in entropy of both the ions coordinated by ionophores and the solvent near these ions. 

However, the value of $\varepsilon$ does not determine the form into which DNA transforms upon addition of a low-polarity solvent to water. Indeed, let us consider the process of transformation of the DNA into the A form in ethanol solution in more detail \cite{1973-methanol-ethanol-C-noA-condensation}. When ethanol is added to an aqueous solution of NaDNA, the DNA molecule, according to the circular dichroism spectrum, indeed begins the transformation not into the A form, but into the C form. This continues up to 60\% ethanol, where 0.4 of the path to the C form has already been passed. However, at 70\% ethanol, the transformation into the A form begins, which ends at 80\% ethanol. When methanol is added to water, the transformation to the C form successfully ends at 80-95\% alcohol. When the value of $\varepsilon$ is the same as at the end of the B-A transition in the ethanol-water mixture, the DNA does not transform into the A form in the methanol-water mixture. The authors of work \cite{1973-methanol-ethanol-C-noA-condensation} even thought that the DNA in a mixture of water and ethanol does not transform into the A form, but condenses into an ordered structure - so unlikely did the transitions of the DNA into different conformations seem in solvents as close in chemical composition as methanol and ethanol. As was determined later, neither in \cite{1973-methanol-ethanol-C-noA-condensation} nor in \cite{1973-Ivanov-CD-A-B-C} did DNA condensation occur, because the circular dichroism spectrum was zero at wavelengths longer than 300 nm (for more details on the shape of the spectra of condensed DNA, see \cite{2009-Kypr-CD-review}). Thus, methanol, in contrast to ethanol, does not allow DNA to transform into the A form even with a sufficient number of ions near the molecule.

\begin{figure}
\begin{center}
\includegraphics[width=0.9\linewidth]{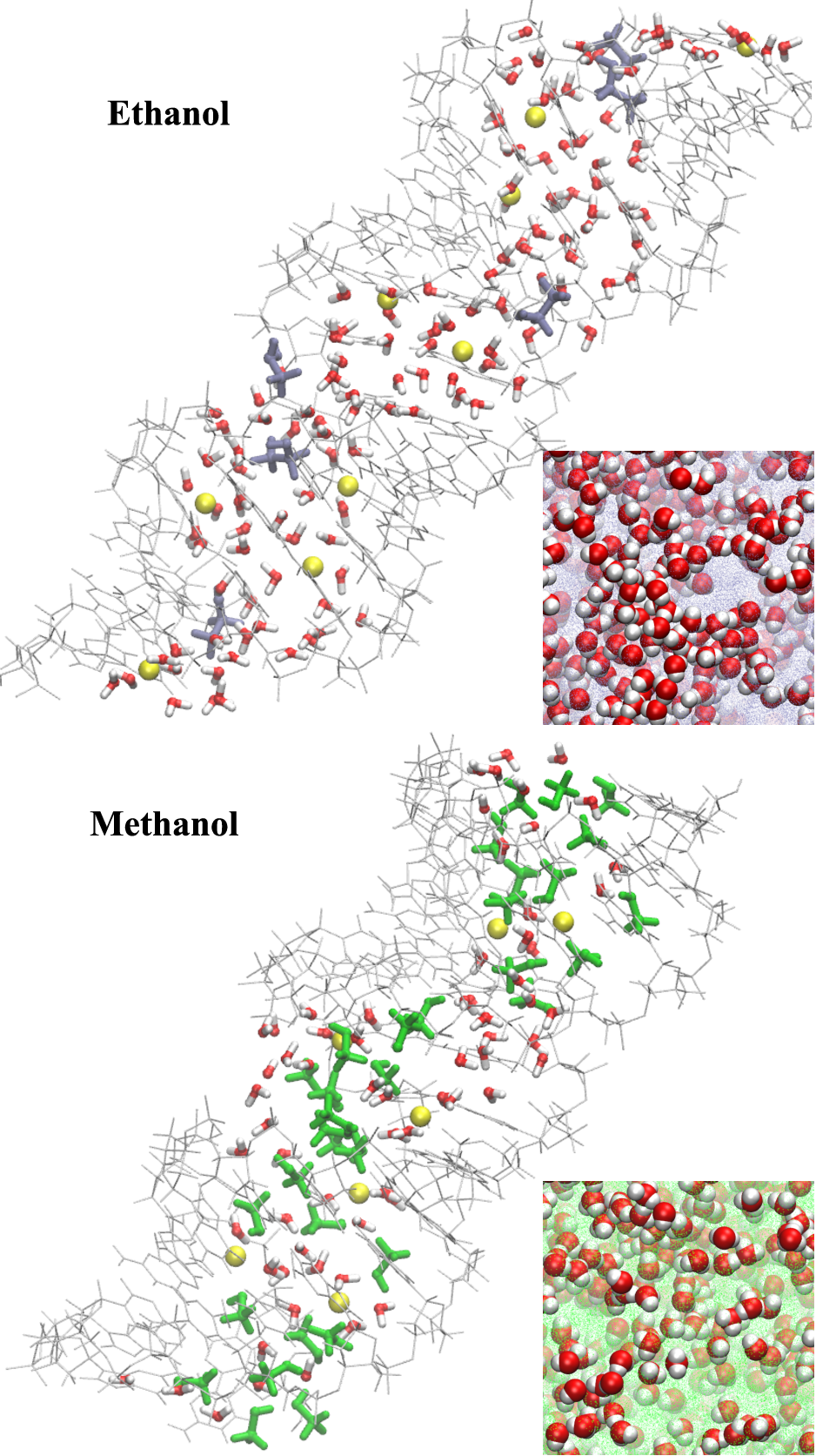}
\caption{Solvent inside the major groove of the A-complexes in 80 v/v\% ethanol and methanol solutions. The insets show water clusters in the same solutions without DNA and ions, alcohol molecules are not visualized.
\label{fig-wat-clusters}
}
\end{center}
\end{figure}

Let us investigate the DNA solvation by binary mixtures of water with alcohols. It is known that these mixtures are microheterogeneous: broken up into water and alcohol clusters \cite{2017-alcohol-water-clusters-bi-percolation,2022-exp-ethanol-water-clusters} of complex architecture \cite{2016-wat-alc-binary-microheterogeneity}. The insets in Figure~\ref{fig-wat-clusters} show the structure of water clusters in 80\% mixtures of ethanol and methanol in the CHARMM force field.
Although the proportion of the heavy atoms of alcohol is the same in the two solutions (0.79), the ethanol solution consists of much larger alcohol and water clusters. Small methanol molecules often integrate into chains of hydrogen bonds between water molecules, and water is much better 'dissolved' in methanol. Large alcohol clusters in ethanol solution are formed mainly via hydrophobic interactions between CH$_2$/CH$_3$ groups.

\begin{figure}
\begin{center}
\includegraphics[width=0.85\linewidth]{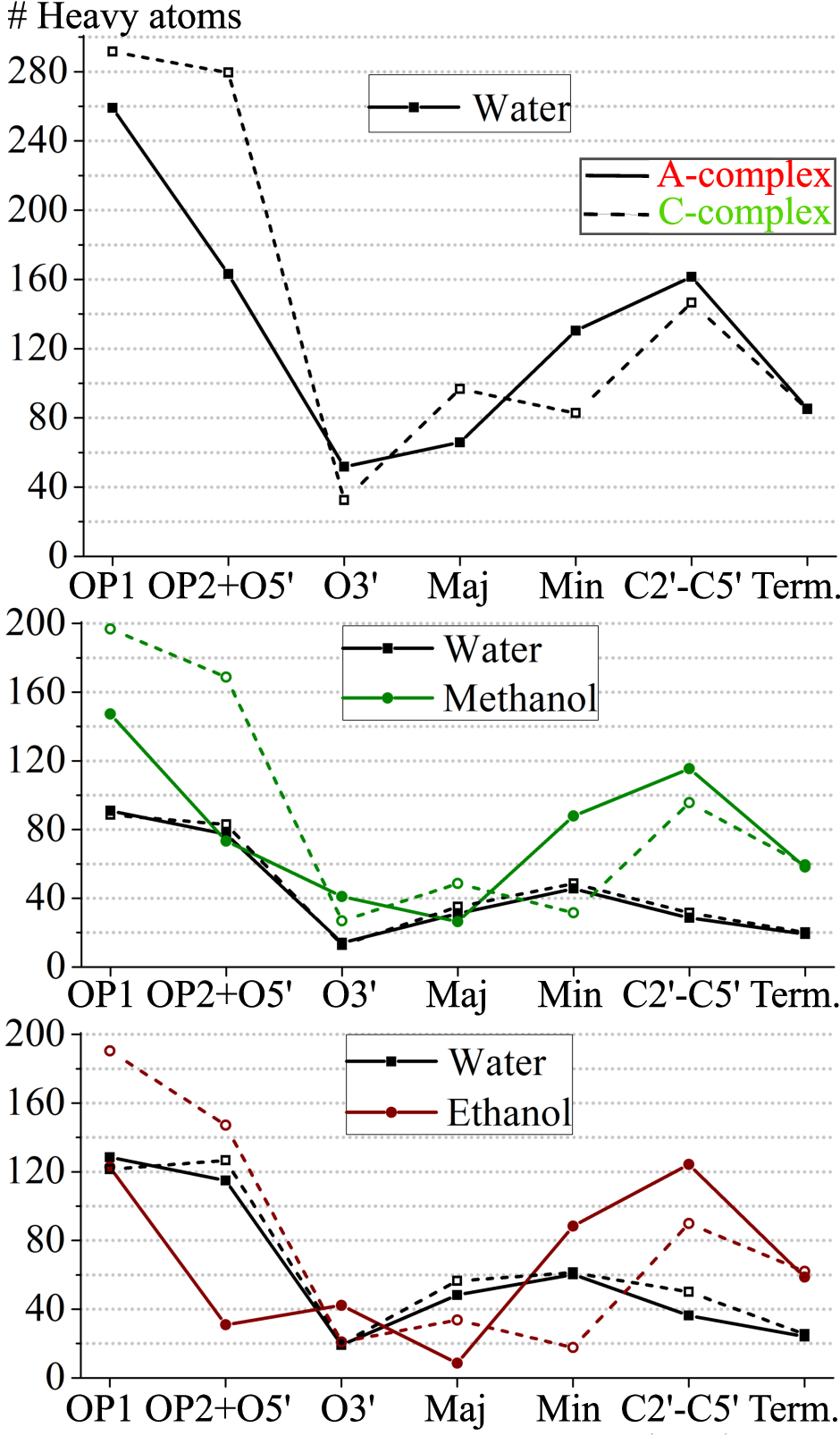}
\caption{The number of heavy atoms of alcohol and water at a distance less than 5.5\AA{} (the far boundary of the second CS) from the selected heavy atoms of the DNA 20-mer. $Maj$ and $Min$ are the atoms of the major and minor grooves, $Term$ are the atoms of the terminal base pairs. For the pure water (the top graph), the standard deviations  $\sigma_i$ of the presented mean numbers lie between 5 and 10. For the atoms of alcohols and water on the other two graphs $\sigma_i$ $\sim$3-15.
\label{fig-solvent-near-DNA}
}
\end{center}
\end{figure}

The alcohol-water mixture tends to retain its structure also when it solvates a DNA molecule, although polar atoms of DNA interact more readily with water molecules. The proportion of heavy atoms of alcohol at a distance closer than 5.5\AA{} from DNA heavy atoms (the far boundary of the second CS) is almost the same in the A- and C-complexes and is equal (in the case of the decamer) to 0.67(A)/0.68(C) in the methanol solution and 0.58(A)/0.59(C) in the ethanol solution. Larger water clusters in ethanol solution interact more readily with the surface of the DNA molecule than with ethanol clusters. The number of hydrophobic contacts of DNA with alcohol molecules is approximately the same in both the solutions, but methanol molecules form hydrogen bonds with the surface of the DNA more often. The number of hydrogen bonds between alcohol and A-DNA is 1.74 times higher in the methanol solution than in the ethanol solution. These hydrogen bonds are formed predominantly with oxygen atoms of phosphates OP1 and OP2 (see Figure~\ref{fig-solvent-near-DNA}).  

The greatest difference between the alcohols is in the solvation of the major groove and atoms OP2 and O5' of the A-complex. In the ethanol solution, there are few alcohol molecules there (see Figure~\ref{fig-wat-clusters} and Figure~\ref{fig-solvent-near-DNA}), while in the methanol solution the number of heavy atoms of water and alcohol is the same: there are two water molecules per one methanol molecule; water and alcohol occupying approximately the same volume. This effect is due to both the small size of the water clusters in the methanol solution and the small size of the methanol molecule. In Figure~\ref{fig-wat-clusters}, one can see that the major groove of the A-complex is filled with alternating water and methanol clusters, with the water clusters being linked to phosphates and atoms N7 and O6. The ions always reside in these water clusters. The same can be said about other ions in ionophores of DNA.

The effect of preferential solvation of ions by water or alcohol has been known for a long time \cite{1998-NaCl-water-methanol-MD-Hawlicka,2013-MgCl2-water-ethanol-Tembe,2015-NaCl-water-methanol-MD-Tembe}. The proportion of alcohol molecules in the first CS of an ion in MD simulation depends on the chosen force field (compare, for example, the results of \cite{1998-NaCl-water-methanol-MD-Hawlicka} and \cite{2015-NaCl-water-methanol-MD-Tembe}). In the CHARMM force field, in both the alcohols (in 80 vol.\% solutions of ethanol and methanol) the coordination number of $Na^+$ is $\sim$5.5, of which alcohol molecules occupy $\sim$1.5 in the ethanol solution and $\sim$2.5 – in the methanol solution (see Table~\ref{table-ions-DNA-contacts-A}, 'pocket'{} $Bulk$).    
\begin{table*}
\caption{Ligands in the first coordination sphere of ions $Na^+$ in some ionophores of the A-complex of DNA 20-mer. 
\label{table-ions-DNA-contacts-A}
}
\begin{tabular}{|l|l|ccccccc|}
\hline
\multicolumn{2}{|c|}{ } & P-cross      & N7-Dir      & O6-Dir    & Maj-Slv     & P-P         & P         & Bulk \\
\hline
number of       &    &            &            &           &           &          &           &       \\    
wat./alc.          &Ethanol& 5.33/0.39    & 4.48/0.19  & 3.67/0.18  & 5.71/0.21   & 4.78/0.7    & 4.48/0.98 &\ \ 4.09/1.34\ \ \\
molecules in    &Methanol& 4.14/1.47    & 3.38/1.02  & 2.63/0.94  & 4.47/1.33   & 3.61/1.79   & 3.24/2.26 & 2.99/2.63 \\
the 1st CS       &Water& 5.8          & 4.77       & 4.03       & 5.99        & 5.7         & 5.85      & 5.9    \\
                        &    &            &            &           &           &          &           &       \\    
\hline
                        &\multirow{2}{*}{Ethanol}& 0.9          & 1.09       & 0.36       & 1.15        & 0.58       & 0.02      &       \\
number of        &     &   32/48/19/1\footnotemark[1] &  24/53/23/0&  78/21/1/0 & 25/42/27/6  & 44/54/2/0  & 98/2/0/0  & \\ 
water bridges   &    &            &            &           &           &          &           &       \\
between           &\multirow{2}{*}{Methanol}& 0.94         & 1.09       & 0.3        & 1.14        & 0.57       & 0.02      &       \\
DNA atoms     &     &\ \ 29/50/21/0 \ \ &\ \ 25/58/17/0 \ \ &\ \ 82/18/0/0 \ \ &\ \ 23/45/27/5 \ \ &\ \ 44/54/2/0 \ \ &\ \ 98/2/0/0 \ \ & \\ 
in the 1st CS   &    &            &            &           &           &          &           &       \\
of an ion\footnotemark[1]&\multirow{2}{*}{Water}& 0.69         & 0.8        & 0.33       & 0.92        & 0.51       & 0.01      &      \\
                       &        & 44/43/12/1   &  39/46/15/0& 75/24/1/0  & 33/44/20/3  & 51/47/2/0  & 99/1/0/0  &   \\
\hline
number of      &    &            &            &           &           &          &           &        \\
hydrogen        &Ethanol& 4.17/0.17  & 4.31/0.09  & 2.47/0.06 & 4.1/0.06  & 2.4/0.16 & 1.02/0.13 &        \\
bonds with     &Methanol& 3.67/0.82  & 3.66/0.54  & 1.94/0.4  & 3.67/0.5  & 2.08/0.48& 0.8/0.35  &       \\ 
DNA via        &Water& 3.97       & 4.12       & 2.6       & 3.84      & 2.47     & 1.15      &       \\
wat./alc.         &    &            &            &           &           &          &           &       \\
\hline
\end{tabular}
\footnotetext[1]{The second line gives the percentage of ions with 0/1/2/3+ water bridges in the first CS.} 
\end{table*}
\begin{table*}
\caption{Ligands in the first coordination sphere of ions $Na^+$ in some ionophores of the C-complex of DNA 20-mer.
\label{table-ions-DNA-contacts-C}
}
\begin{tabular}{|l|l|ccccccc|}
\hline
\multicolumn{2}{|c|}{ } & O6-Dir       & Maj-Slv    &  O4'-cross &  Min-Dir    & Min-Slv    & P-P       & P \\
\hline
number of       &    &            &            &           &           &          &           &       \\    
wat./alc.          &Ethanol& 3.29/0.46    & 5.17/0.56  & 5.15/0.55  & 3.6/0.32   & 5.05/0.64  & 4.41/0.99 & 4.37/1.07\ \ \\
molecules in    &Methanol& 2.32/1.22    & 3.96/1.73  & 4.15/1.58  & 2.62/1.21  & 3.98/1.73  & 3.25/2.16 &\ \ 3.15/2.36\ \  \\
the 1st CS       &Water&  4.07         & 5.94       & 5.87       & 4.04       & 5.91       & 5.83      & 5.87      \\
                       &    &               &            &            &            &            &           &           \\
\hline
                     &\multirow{2}{*}{Ethanol}& 0.19         & 0.63       & 2.22       & 0.63        & 1.24       & 0.21      & 0.01     \\
number of        &     &\ \ 85/14/1/0\footnotemark[1]  &  49/40/10/1& 2/13/51/34 & 51/45/4/0   & 16/50/29/5 & 80/20/0/0 & 99/1/0/0 \\ 
water bridges   &    &            &            &           &           &          &           &       \\
between           &\multirow{2}{*}{Methanol}& 0.13         & 0.62       & 2.15       & 0.54        & 1.22       & 0.19      & 0.01     \\
DNA atoms   &     &\ \ \ 90/9/1/0\ \ \ &\ \ \ \ 50/40/9/1\ \ \ \ &\ \ 2/15/52/31\ \ &\ \ \ 57/40/3/0\ \ \ &\ \ 15/53/28/4\ \ &\ \ 81/19/0/0\ \ &\ \ 99/1/0/0\ \ \\ 
in the 1st CS   &    &            &            &           &           &          &           &       \\
of an ion\footnotemark[1]&\multirow{2}{*}{Water}& 0.26         & 0.57       & 2.1        & 0.5         & 1.03       & 0.14      & 0.01     \\
                       &        & 79/19/2/0    & 53/37/9/1  & 2/16/55/27 & 57/42/2/0   & 23/53/21/3 & 86/14/0/0 & 99/1/0/0 \\
\hline
number of      &    &               &            &            &             &            &           &           \\
hydrogen       &Ethanol& 2.08/0.24     & 3.12/0.14  & 5.76/0.14  & 2.88/0.11   & 3.88/0.11  & 2.15/0.29 & 1.06/0.17 \\
bonds with     &Methanol& 1.46/0.56     & 2.68/0.54  & 5.33/0.66  & 2.25/0.6    & 3.52/0.49  & 1.68/0.74 & 0.79/0.4  \\ 
DNA via        &Water& 2.49          & 2.97       & 5.6        & 2.93        & 3.48       & 2.36      & 1.19     \\
wat./alc.         &    &               &            &            &             &            &           &       \\
\hline
\end{tabular}
\footnotetext[1]{The second line gives the percentage of ions with 0/1/2/3+ water bridges in the first CS.} 
\end{table*}
In ionophores of DNA (except for P and P-P) the number of alcohol molecules in the first CS of Na$^+$ drops, especially in the major groove of the A-complex, to $\sim$0 in the ethanol solution and to $\sim$1 in the methanol solution. 

Table~\ref{table-ions-DNA-contacts-A} allows one to determine a typical structure of the solvation shell of an ion in the ionophores of the major groove of the A-complex. The ion is always six-coordinated. In ionophore Maj-Slv, in the water and in the ethanol solution all ligands are water molecules, of which one molecule is a bridge between DNA atoms and two water molecules have one hydrogen bond with the DNA each. The remaining three water molecules are not hydrogen bonded with the DNA, but they are still most likely connected with it through one or two other solvent molecules because of the closed geometry of the major groove in the A form. In the methanol solution one ligand of the ion in ionophore Maj-Slv is alcohol. The outer sphere coordination with DNA also is through a water bridge and two water molecules. The remaining molecules from the first CS of the ion (two water molecules and one methanol molecule) are not in direct contact with DNA. In ionophore N7-Dir, the (outer sphere) coordination with DNA remains the same: a water bridge and two water molecules. But now the ion has one direct contact with DNA in the water and in the ethanol solution, and two direct contacts in the methanol solution. Accordingly, in the fist CS of the ion, without direct contacts with DNA, there are two water molecules in the water and in the ethanol solution and one methanol molecule in the methanol solution. Thus, in both the alcohols and in both the ionophores Maj-Slv and N7-Dir of the A-complex, the ions actually reside in water clusters.   

Theoretically, an ion may form outer sphere contacts with DNA through alcohol molecules. However, this rarely happens even in the methanol solution, and almost never in the ethanol solution. We hypothesize that the ions prefer the coordination with DNA through water molecules because alcohol molecules tend to cluster, forming hydrophobic bonds between carbon atoms. These clusters prevent the second (and third, fourth and so on) sphere coordination of the ion with DNA by the remaining ligands. Figure~\ref{fig-coord-compl-Maj} shows a variant of ion coordination in ionophore N7-Dir near a methanol cluster, hydrophobic contacts between methanol molecules are clearly visible. 

Probably, because of the large number of methanol molecules near the DNA surface, the ions cannot occupy completely those ionophores in which the coordination with DNA requires several second or third sphere contacts made through water molecules. First of all, this is ionophore Maj-Slv of the A-complex. Ionophore Maj-Slv requires at least three water molecules for the second sphere contacts with DNA. In addition, the other two ligands are also water molecules, because they also participate (through 1+ solvent molecule) in fixing the ion near the polar DNA atoms. The situation is better for the other ionophores with outer sphere coordination: P-cross, P-P, P and Min-Slv, but in all these cases the number of ions in them is lower in the methanol solution than in the ethanol solution. Ionophores directly coordinating ions with DNA (N7-Dir in the A-complex and O6-Dir in the C-complex), on the contrary, win in the methanol solution (see Fig.~\ref{fig-pockets-eth-meth}), because they require less water, which is insufficient near the DNA in the methanol solution.
\section{Conclusion}
\noindent
In the present work, we have modeled the formation of DNA-counterions (Na$^+$) complexes in a low-polarity medium, at conditions close to the conditions in a living cell. In such a medium, the counterions approach the surface of the DNA molecule, and it changes its (common in water) B conformation to A or C (see~Figure~\ref {fig-ions-in-major-minor-grooves}). The choice of the conformation depends on the chemical nature of the solvent. This polymorphism has been known for 50 years, but until now it was not understood. 

We considered two binary mixtures, ethanol-water and methanol-water, in both the cases the volume fraction of alcohol was 80\%. We simulated the C-A transition in both the mixtures using the umbrella sampling scheme combined with the Weighted Histogram Method and obtained the profiles of the free energy. In the case of the ethanol solution, the A-complex has a lower free energy, while in the case of the methanol solution the C-complex is more favorable. In both the local minima of the free energy, we investigated the structure of the solution and the location of ions near the DNA surface.

It turned out that the possibility for the ions to occupy certain DNA ionophores is critical for the choice of the conformation, A or C, by the molecule. The geometry of the C form of DNA is close to that of the B form, only the molecule is slightly more twisted, and, correspondingly, the minor groove is a little more narrow. This geometry is provided by the counterions situated in the ionophores of the minor groove. The main ionophore of the minor groove is O4'-cross (see Table~\ref{table-pockets-description} and Figure~\ref{fig-coord-compl-O4cross}). The ion in this ionophore generally has no direct contacts with the DNA, but links two O4'-N3/O2 water bridges of opposite DNA strands, thereby narrowing the minor groove and (over)twisting the DNA helix. If we forbid ions to form outer sphere contacts with atoms O4', DNA in the methanol solution takes the A form instead the C form. Similarly, in the A-complex of DNA the main ionophore is P-cross (see Table~\ref{table-pockets-description} and Figure~\ref{fig-coord-compl-P}), the ion in which also has no direct contacts with DNA atoms, but outer sphere coordinated by the oxygen atoms of the phosphates from the opposite DNA strands. If we forbid ions to form such outer sphere contacts, DNA in the ethanol solution takes the C form instead of the A form.

However, other ionophores of the major groove are also important for stabilization of the A-complex. It is often supposed that direct contacts of ions with atoms N7 of guanine are responsible for the 'A-philicity' of the DNA sequences with a large number of G:C base pairs. But there are more ions in ionophore N7-Dir in the methanol solution than in the ethanol solution. Methanol, in contrast to ethanol, impedes the outer sphere coordination with the DNA atoms rather than the inner sphere coordination. The ionophore Maj-Slv in the A-complex suffers the most. In this ionophore, there are two times fewer ions in the methanol solution than in the ethanol solution.

The difference between the two binary mixtures ethanol-water and methanol-water is in the size and the structure of water and alcohol clusters. The alcohol clusters form largely by hydrophobic contacts between CH$_2$/CH$_3$ groups. In the methanol solution, the water clusters are small, and methanol molecules often integrate into chains of hydrogen bonds between water molecules (see ~Figure~\ref{fig-wat-clusters}). In the ethanol solution, both water and alcohol clusters are large. This difference between the two binary mixtures persists after DNA solvation. In the A-complex, inside the major groove and near atoms OP2 and O5', there are as many heavy atoms of methanol as heavy atoms of water (see Figure~\ref{fig-solvent-near-DNA}), while large water clusters of the ethanol solution prefer to interact with polar atoms of the DNA major groove rather than with the surface of ethanol clusters, and the major groove of the A-complex in the ethanol solution is almost entirely filled with water (see Figure~\ref{fig-wat-clusters}). 

Ions in binary alcohol-water mixtures are also preferentially solvated by water molecules. In the CHARMM force field, the coordination number of ion $Na^+$ in a binary mixture with 80 vol.\% alcohol is $\sim$5.5, of which alcohol molecules occupy $\sim$1.5 in the ethanol solution and $\sim$2.5 - in the methanol solution. In all the ionophores of DNA, the proportion of alcohol in the first CS of the ions is less, and the ions actually reside in water clusters, while in the case of the methanol solution there are not many water clusters in the major groove of the A-complex (see~Figure~\ref{fig-wat-clusters}). The ions in ionophores are coordinated by the DNA molecule not only in the first and the second coordination spheres. More distant DNA atoms, through a (water) hydrogen bond network, also contribute to stabilizing the position of the ion. Theoretically, an outer sphere contact of the ion with DNA may also be provided by the OH group of an alcohol molecule. But alcohol molecules are clustered. We hypothesize that the alcohol clusters in the major groove of the A-complex prevent the required coordination of the ion with DNA. This primarily applies to ionophore Maj-Slv, but ionophores P-cross, P-P, and P of the A-complex are also affected.

Since methanol clusters near the DNA surface prevent ions from filling certain ionophores of the A-complex, the attraction between counterions and DNA cannot compensate for the repulsion between phosphates, which are located closer to each other in the A form than in the C (or B) form. Therefore, the counterions fill the more accessible (requiring less water for coordination with DNA) ionophores of the C-complex or remain in solution near DNA, and the DNA molecule takes the C rather than the A form.     
\section{Methods}
\subsection{Details of MD simulations}
\label{section-Methods-MD}
\noindent
We performed molecular dynamics (MD) simulations of DNA oligomers in the framework of the phenomenological all-atom CHARMM36 force field \cite{2012-CHARMM-for-BII} in the package OpenMM \cite{2017-openmm}. In the CHARMM force field, the DNA molecule has a 'C-philic' backbone and an 'A-philic' base stacking \cite{2023-CBA-test-we}, and it can take (unfortunately, not very accurately) the A, B, and C forms of DNA, although the geometry of the C form differs from experiment (see Section~\ref{section-Methods-C-form-CHARMM}). 

For the ions, we used the potentials by Joung\& Cheatham \cite{2008-Young-Cheatham-ions} adjusted to reproduce the osmotic pressure of the sodium dimethyl phosphate solution in the CHARMM force field \cite{2012-parametrization-DNA-ions,2018-CUFIX-review} (NBFIX corrections). For self-consistency of the force fields, we chose the TIP3P water (used in the CHARMM force field to tune its parameters for DNA) and the standard CHARMM parameters for alcohols. The main physical properties of the studied alcohol-water solutions are listed in Table~\ref{table-list-of-alcohols}. One can see that the chosen models poorly reproduce the dielectirc constant of water, as well as the diffusion coefficients. The used force fields significantly accelerate the dynamics of the solutions as compared to experiment, but they well reproduce electrical interactions in the binary mixtures.
\begin{table}
\caption{The studied binary mixtures of alcohols (80 vol.\%) with water: comparison of experimental characteristics with simulation in the framework of the CHARMM force field \cite{2008-alcohols-CHARMM}. We calculated (for temperature T=300K and pressure p=1atm in NpT ensemble) the density $\rho$, static permittivity $\varepsilon$, dipole moments $\mu$, and diffusion coefficients of water $D_W$ and alcohol $D_A$ in the solutions. 
\label{table-list-of-alcohols}
}
\begin{tabular}{|l|l|l|l|l|l|}
\hline 
20vol.\%of wat.                & $\rho$,                               & \multirow{2}{*}{$\varepsilon$}      & $\mu$,                                                 & $D_W$/$D_A$, \\
+80vol.\% of                                       & kg/l                                  &                                       & Debye                                                  & 10$^{-9}$m$^2$/s \\ 
\hline
eth. sim.     & 0.846                                 & 34.3                                  & 2.41                                                   & 1.55/1.04       \\
eth. exp.                              & 0.859\cite{2004-MethEthDens}          & 34.6\cite{1932-eth-meth-epsilon-exp}  & 2.56\cite{2022-eth-meth-dip-mom-NOT-exp}\footnotemark[1] & 0.92/0.77\cite{2003-meth-eth-SD-exp} \\
\hline 
meth. sim.     & 0.828                                 & 41.0                                  & 2.39                                                     & 2.29/1.99\\
meth. exp.                             & 0.849\cite{2004-MethEthDens}          & 41.7\cite{1932-eth-meth-epsilon-exp}  & 2.61\cite{2022-eth-meth-dip-mom-NOT-exp}\footnotemark[1] & 1.26/1.45\cite{2003-meth-eth-SD-exp}\\
                                             &                                       &                                       &                                                  & 1.52/1.48\cite{1983-meth-SD-exp-alternative}\\
\hline
wat.TIP3P\cite{1985-water-TIP3P-parameters}  & 1.01                                 & 104.8                                  & 2.35                                                     & 4.77\\
wat. exp. \cite{2000-mu-liquid-water}       & 0.997                                & 77.8                                  & 2.9$\pm$0.6\cite{2000-water-dip-moment-exp}              & 2.35 \\
\hline 
\end{tabular}
\footnotetext[1]{The dipole moments of alcohol molecules in the liquid phase have not been measured experimentally; we give theoretical estimates obtained by the density functional method for pure alcohols.}
\end{table}       

For electrostatic interactions we used the PME (particle-mesh Ewald) method with the accuracy parameter ('ewaldErrorTolerance') of 10$^{-4}$. Non-bonded interactions were cut off at a distance of 12\AA, van der Waals interactions were smoothly zeroed with a switching function applied between 10 and 12\AA. We used the Langevin thermostat with a friction coefficient of 1~ps$^{-1}$, and the Monte Carlo barostat \cite{1995-MC-barostat1,2004-MC-barostat2}. The valence bonds X-H (X is any atom) and the angle H-O-H in water molecules were rigidly fixed. The integration step was equal to 2 fs. To calculate the free energy profiles (Figure~\ref{fig-WHAM-eth-meth} and $\Delta$G from Table~\ref{table-deltaUGS-WHAM}), only the coordinates of the DNA and the ions were output every 10~ps. 
 
In long-term simulations, the terminal base pairs of DNA open too frequently and for too long periods of time as compared to experiment \cite{2014-modeling-fraying,2015-modeling-against-fraying}. To avoid this, on the terminal base pairs we replaced the energy of the hydrogen bond N3-H1 with a parabolic potential. In the region of the minimum of the bond energy (2.2-2.3\AA), this potential approximates the sum of their van der Waals and Coulomb interactions. 

We imposed periodic boundary conditions and therefore used a sufficiently large computational cell (a cube with an edge $\sim60$\AA{} for the decamer). In it, the A-complex of DNA may be unfavorable due to the large space available for counterions \cite{2017-our-CG-DNA-model}. To stabilize the A-complex we chose the 'A-philic' sequence CCGGGCCCGG. This decamer crystallizes in the A form in experiment \cite{2005-A-sequences}, and in simulation in the framework of the CHARMM36 force field takes the A form in ethanol solution with high ethanol content \cite{2019-CCGGGCCCGG-MD-ethanol}. We also modeled a 20-mer composed of two decamers CCGGGCCCGG. 

For systems listed in Table~\ref{table-boxes}, we have obtained free energy profiles along reaction coordinate MGW (see the next Section) and carried out long simulations at two values of the reaction coordinate. To analyze the A form of DNA, we fixed MGW=9\AA{}, to analyze the B (in the water) and C forms - MGW=23\AA{}. The sample preparation procedure was similar to that described in  \cite{2023-CBA-test-we}. The simulation box contained no additional salt.    
\begin{table}
\caption{Parameters of the simulated systems for the detailed study of the forms of DNA (for all the Figures and Tables except for Fig.~\ref{fig-WHAM-eth-meth} and Table~\ref{table-deltaUGS-WHAM}). Decamer - CCGGGCCCGG; 20-mer is composed of two decamers; ethanol - 80 vol.\% ethanol solution; methanol - 80 vol.\% ethanol solution; N$_{wat}$ - number of water molecules, N$_{alc}$ - number of alcohol molecules, N$_{ions}$ - number of counterions, $\tau _{sim}$ - simulation time. (NpT) ensemble: p=1atm, T=300K. \it{Free energy profiles were not calculated for 20-mer in water and in methanol.} \rm
\label{table-boxes}
}
\resizebox{\linewidth}{!}{%
\begin{tabular}{|l|c|c|c|c|}
\hline
System                                      & N$_{wat}$                             & N$_{alc}$ & N$_{ions}$       & $\tau _{sim}$\\
\hline
decamer in ethanol &  \multirow{3}{*}{1379}       &  \multirow{3}{*}{1699}   & \multirow{7}{*}{18} & 1$\mu$s \\
decamer in ehtanol, no O'-cross  &                        &                       &  & 100ns \\
decamer in ehtanol, no P-cross   &                        &                       &  & 100ns \\
\cline{1-3}
\cline{1-3}
decamer in mehtanol    &  \multirow{3}{*}{1322}       &  \multirow{3}{*}{2417}   &                                   &  1$\mu$s  \\
decamer in mehtanol, no O'-cross  &         &     &  & 100ns  \\
decamer in mehtanol, no P-cross   & &  &  & 100ns \\
\cline{1-3}
decamer in water         &  6295       &      0       &                                   &  1$\mu$s  \\
\hline
20-mer in ethanol        & 4149        & 5113& \multirow{3}{*}{38} & \multirow{3}{*}{100ns} \\
\it{20-mer in methanol} \rm             &  3978       &7274  &                                    &                       \\
\it{20-mer in water} \rm                   &  18885     &      0      &                                    &    \\
\hline
\end{tabular}
}
\end{table}

To keep ions off the atoms O4' in the case 'no O4'-cross' and off the atoms OP1 and OP2 of two first phospates on every strand in the case 'no P-cross', we replaced the parameters of the Lennard-Jones potential between these particles with $\varepsilon$=0.01kcal/mol and R$_{min}$=7.7\AA.

Blocking ionophore O4'-cross at the fixed reaction coordinate MGW=23\AA{} in the alcohol solutions creates a high-energy state in which the DNA molecule constantly switches between the E and C forms. The E-DNA has been reported in experiment \cite{2000-E-DNA-Ho-1,2000-Dickerson-stable-intermediate-CAT}. This is a form intermediate between A and B, with north sugars (typical of the A form), negative parameter $Slide$ (as in the A form) and parameter $Roll$ close to zero (as in the B form) (see Table~\ref{table-AC-helix-parameters}).
\subsection{Free energy profiles}
\label{section-Methods-WHAM}
\noindent
The calculation of the free energy profile (Gibbs free energy G) was carried out by the umbrella sampling scheme \cite{1977-Umbrella} combined with the WHAM \cite{1992-WHAM}. As the reaction coordinate MGW (Major Groove Width) we chose the average distance $R(i1,i2)$ between the phosphorus atoms belonging to opposite strands and closest to each other in the A form. The distance $R(i1,i2)$ determines the local width of the major groove in the A form. In the decamer, there are two such distances, between the phosphorus atoms of the 2-nd and 3-rd bases of each strand (distances $R(2,3)$ and $R(3,2)$, if we number the bases from the 5' end to the 3' end). In the case of the 20-mer, we used the distances between the phosphorus atoms from the 2-nd base to the 13-th base of the first strand and the phosphorus atoms from the 13-th base to the 2-nd base of the second strand ($R(2,13)$, $R(3,12)$, ..., $R(13,2)$). The biasing potential of the reaction coordinate was parabolic: 
$$
U(MGW)=K\cdot\left(\frac{1}{n}\sum\limits_{i=1}^nR(i1,i2)-MGW\right)^2,
$$
where K=5~kcal/mol/\AA$^2$. 

The value of MGW in the region of B and C forms (MGW$>$20\AA) corresponds not to the width of the major groove, but rather to the elongation of the duplex along the helix axis. Therefore MGW poorly distinguishes between the B and C forms. The further increase in the reaction coordinate leads to conformations of the elongated helix. 

Reaction coordinate MGW took values from 8 to 24\AA{} with a step of 1\AA{}. For all the free energy curves, we took as zero the energy at the point 23\AA. The bin size of the histograms was 0.1\AA{}. Doubling the bin size changes the free energies profiles greatly, halving the bin size leaves the profiles practically unchanged. We obtained long trajectories in every window to ensure sufficient sampling of solvent and ion configurations near DNA. To ensure the continuity of the transition along the reaction coordinate, we did not include in the histograms the parts of the trajectories where the system was not in equilibrium (immediately after turning on the reaction coordinate potential). In addition, we excluded the periods with non-canonical duplex conformations (such as 90\textdegree{} bends or base openings). If such a conformation stabilized, we repeated the simulation starting from the final state of the adjacent (along the reaction coordinate) simulation. We considered the free energy profile converged if adding 20ns to every trajectory (i.e. 17*20=340ns simulation time overall) did not alter the profile in any bin by more than 0.1kcal/mol. The relevant parts of individual trajectories used to build histograms for the WHAM equations ranged from 55ns to 715ns, $\sim$3$\mu$s in sum for one profile. The histograms were scaled to represent the effective number of uncorrelated configurations estimated according to procedure described in \cite{2007-statistical-inefficiency}. The tolerance for the maximal change in the free energy shift of any biased system after WHAM iteration was 10$^{-10}$kcal/mol. 

Even on long trajectories, we could not collect sufficient statistics in some transition regions, especially for the case of the ethanol solution, where there are two local minima. The reason is that the C-A transition is of the first order, the conformational dynamics of DNA is slow, and there is a hysteresis in the space orthogonal to the reaction coordinate (resulting in a  systematic error, see \cite{2012-systematic-error-in-WHAM}). Therefore, we started from the C (B) form and generated initial states in the windows
by gradually decreasing the value of MGW, allowing for short equilibration at each window before further reduction of the reaction coordinate. Such a protocol resulted in undersampling of the A-like conformations. The statistical error was small (less than 0.1kcal/mol for the decamer in all the solutions estimated by bootstrapping \cite{Grossfield-bootstrap}, the autocorrelation time being estimated as in \cite{2007-statistical-inefficiency}), because we got long trajectories. We expected this protocol to result in an erroneous increase in the energy of the A-minimum, which, even so, for the decamer in the ethanol solution still proved to be lower than the B-minimum. 

\begin{figure}
\begin{center}
\includegraphics[width=0.85\linewidth]{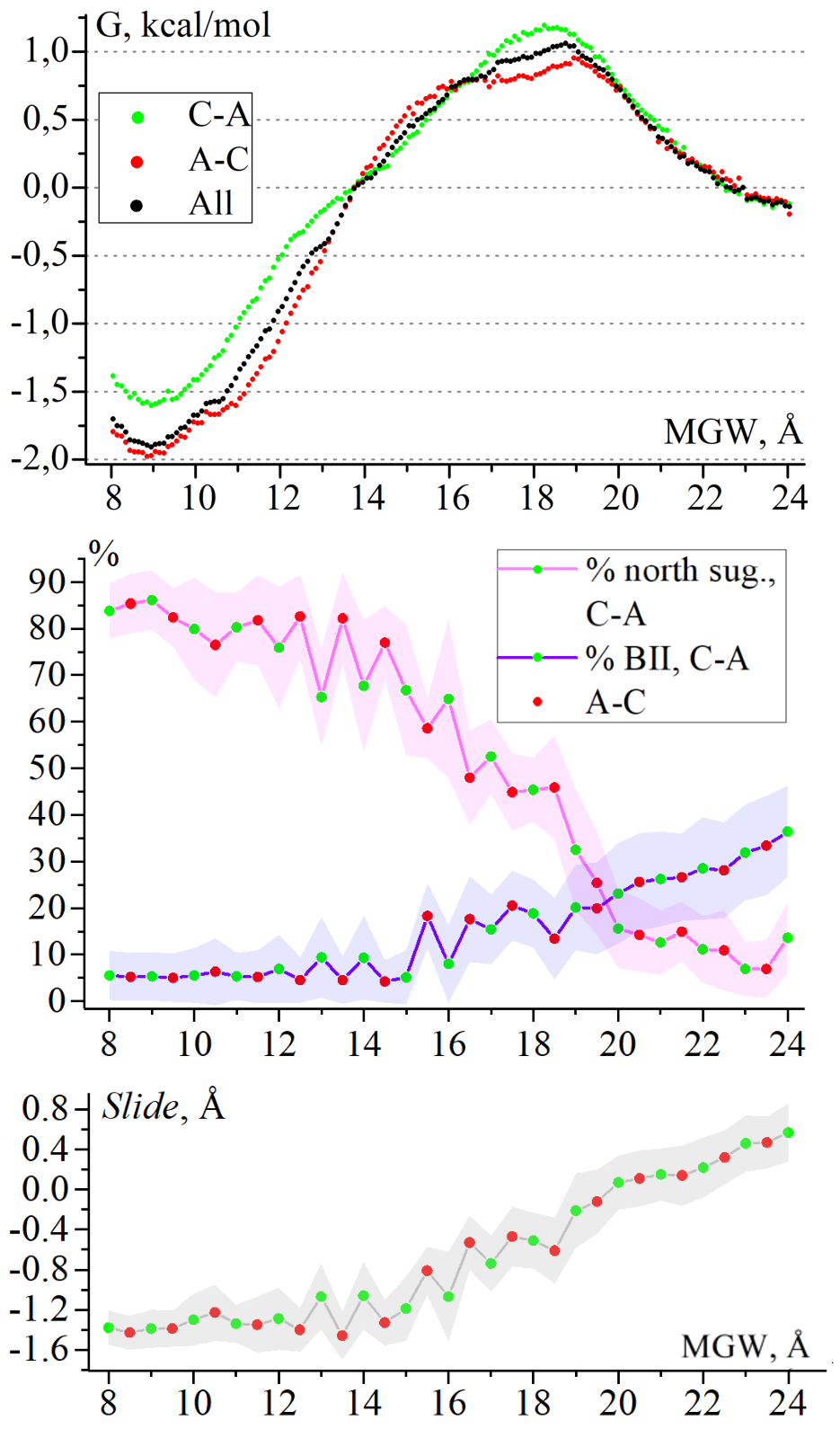}
\end{center}
\caption{Estimation of the systematic error in calculating the free energy difference between the A and C forms (10-mer in the ethanol solution). Green curve: the free energy profile was obtained moving from the C to the A form (the initial state of the system in the current window was the final state in the previous one), window centers are at points 24, 23, 22, ... \AA. Red curve: the opposite transition, from the A to the C form, windows at points 8.5, 9.5, ... \AA. Black curve: all data combined. The graphs at the middle and on the bottom show the percentages of north sugars and BII conformations, and also parameter $Slide$ with standard deviations.
\label{fig-A-C-G}
}
\end{figure}

To estimate the systematic error resulting from undersampling of the A-like conformations, we performed additional umbrella sampling, placing new windows between the old ones. We went through these windows in the opposite direction, from the A to the C form. As expected, the new profile yielded a minimum of the A form lower by 0.4 kcal/mol. The profile constructed using the full number of windows is closer to the new A-C profile (see Fig.~\ref{fig-A-C-G}).

We also performed simulations of a free DNA decamer, starting with conformations at MGW=9\AA{} (A form) in the cases of the water and the methanol solution, and with conformations at MGW=9\AA{} at MGW=23\AA{} (C form) in the case of the ethanol solution. In the water and in the methanol solution, the transition to the C form began immediately and completed in 6ns in water and in 80ns in methanol. In an ethanol solution, the transition from the A form to the C form began only after 670ns and completed in 33ns. A successful transition from the C to the A form began after $\sim$1.8 $\mu$s and completed in 120ns. Before the successful transition, one unsuccessful transition attempt was observed, which began $\sim$900ns after the start of the simulation and was not completed in the next 200ns, after which the DNA returned to the C form in approximately the same 200ns. Thus, both the instability of the A form in the water and in the methanol solution and the presence of two local minima, A and C, in the ethanol solution were confirmed.
\subsection{C-DNA in the CHARMM force field and in  experiment}
\label{section-Methods-C-form-CHARMM}
From Table~\ref{table-AC-helix-parameters} we see that, in the local minimum near MGW=23~\AA{}, only the DNA in the water has the number of BII conformations of phosphates sufficient to classify the molecule as the C form ($>$40\%), although the DNA should be in the B form in the water. But even in the water in our 'A-philic' sequence the parameters of the helix are shifted to the parameters of the A form due to the 'A-philicity' of stacking in the CHARMM force field \cite{2023-CBA-test-we}.

In experiment, changing the geometry of the DNA helix to the C form (narrowing the minor groove and increasing parameter $Twist$) is accompanied by the appearance of BII conformations of phosphates. In the CHARMM force field, the appearance of a BII conformation leads to an excessively strong decrease in angle $\beta$ (adjacent to angle $\zeta$ responsible for the transition to BII). This does not correspond to the X-ray data and leads to artifacts in the geometry of DNA helix. In our 'A-philic' sequence, with a large number of BII conformations in the water (see Table~\ref{table-AC-helix-parameters}), parameter $Twist$ is small, parameter $Roll$ is large, and, as a consequence, the minor groove is wide. When ions enter the minor groove and contract it (in the methanol and ethanol solutions), the number of BII conformations in the DNA in the CHARMM force field decreases - contrary to experiment. Because of these artifacts, a DNA molecule in the CHARMM force field may satisfy the definition of the C form by the helix parameters \cite{1958-MARVIN-Twist-C-DNA} (narrow minor groove, large $Twist$ and small $Roll$), while contradicting the definition of the C form by the backbone conformation ($>$40\% BII conformations \cite{2000-C-DNA-Levitt}).

After Ivanov \cite{1973-Ivanov-CD-A-B-C}, we think that the main role in the formation of the C-form is played by the narrowing of the minor groove when ions enter it. Therefore, we define the C form as the conformation of the C-complex, the form with the minimum width of the minor groove (and the maximum value of parameter $Twist$). Accordingly, we consider the DNA at MGW=23~\AA{} in the water to be in the B form, and in the ethanol and methanol solutions - in the C form.
\subsection{Potential energy of system components and of ions in ionophores}
\label{section-Methods-U}
\noindent
We estimated the energies of the system components (Table~\ref{table-energy-components-Delta-U-eth-meth-wat}) and the energies of ions in ionophores (Tables~\ref{table-pockets-energies-A} and \ref{table-pockets-energies-C}, and Figure~\ref{fig-pockets-eth-meth}) on CPU using the LAMMPS package \cite{1995-LAMMPS,LAMMPS-site}.
\begin{table*}
\caption{Average potential energy (kcal/mol) of ions in ionophores of the A-complex. Data format: E$_j$*N$_j$\ \ E$_{Sum}$, E$_j$ is energy of an ion in ionophore $j$, N$_j$ is number of ions, and E$_{Sum}$ is the total energy of all ions in this ionophore. 
\label{table-pockets-energies-A}
}
\begin{minipage}{\textwidth}
\centering
\begin{tabular}{|c|lr|lr|lr|r|}
\multicolumn{8}{c}{\bf{Decamer} \rm } \\
\hline
Ionophore& Water, A  &      & Eth., A     &        & Meth., A     &        & Eth.-Meth., A \\
\hline
P-cross  & -224*\ 0.7   & -157 & -272*\ 2.02 & -549   & -260*\ 1.89  & -492 & -57  \\
N7-Dir   & -220*\ 1.08 & -239 & -257*\ 2.74 & -705  & -248*\ 2.71  & -670 & -35  \\
O6-Dir   & -217*\ 0.33 & -72  & -260*\ 0.47 & -122   & -246*\ 0.79  & -195 & 73  \\
Maj-Slv  & -223*\ 1.3  & -290 & -274*\ 1.37 & -376   & -254*\ 0.66  & -167 & -210  \\
\hline
O4'-cross& -201*\ 0.02 & -5   & -226*\ 0.07 &   -17    & -226*\ 0.14  & -32 & 16   \\
Min-Dir  & -198*\ 0.08 & -16  & -225*\ 0.17 &   -38   & -223*\ 0.29  & -65 & 27   \\
Min-Slv  & -197*\ 0.53 & -105 & -222*\ 1.39 &  -310 & -220*\ 1.37  & -302 &-7    \\
\hline
P-P         & -206*\ 0.69 & -143 & -230*\ 3.02 & -694   & -225*\ 2.47  & -555 & -138  \\
P            & -197*\ 1.28 & -253 & -219*\ 2.27 & -495    & -213*\ 1.99  & -424 & -71  \\
Bulk      & -177*11.98  & -2116& -184*\ 4.48 & -822   & -182*\ 5.69  & -1036 &214   \\
\hline
Total     &                      &\ \ -3395&                 &\ \ \ -4127&              &\ \ -3939& -188  \\
\hline
\end{tabular}
\end{minipage}
\begin{minipage}{\textwidth}
\centering
\begin{tabular}{|c|lr|lr|lr|r|}
\multicolumn{8}{c}{\bf{20-mer} \rm } \\
\hline
Ionophore& Water, A  &        & Eth., A&          & Meth., A  &        & Eth.-Meth., A \\
\hline
P-cross  & -290*\ 4.82  & -1398  & -354*10.45  & -3703 & -342*\ 9.78  & -3341  & -362   \\
N7-Dir   & -291*\ 2.53  & -736   & -341*\ 4.95 & -1688 & -328*\ 5.47  & -1795  &  106   \\
O6-Dir   & -288*\ 0.74  & -214   & -342*\ 0.93 & -319  & -324*\ 1.47  & -477   &  158   \\
Maj-Slv  & -298*\ 3.28  & -979   & -362*\ 3.26 & -1179 & -341*\ 1.43  & -487   & -692   \\
\hline
O4'-cross& -237*\ 0.03  & -6     & -271*\ 0.14 &  -37  & -274*\ 0.21  & -57    &   20   \\
Min-Dir  & -237*\ 0.1   & -23    & -268*\ 0.28 &  -75  & -265*\ 0.48  & -128   &   53   \\
Min-Slv  & -244*\ 0.95 & -232   & -282*\ 2.61 &  -735 & -278*\ 2.55  & -708   &  -27   \\
\hline
P-P         & -270*\ 1.72  & -465   & -297*\ 4.2  &  -1246 & -284*\ 3.64  & -1034  &  -213   \\
P             & -246*\ 2.4   & -591   & -276*\ 3.62 &  -999  & -268*\ 3.2   & -858   &  -141   \\
Bulk       & -187*21.43  & -3997 & -202*\ 7.57 &  -1527 & -196*\ 9.78  & -1920  &   393   \\
\hline
Total      &                      &\ \ -8641  &                &\ -11508   &              & -10804  & -705 \\
\hline
\end{tabular}
\end{minipage}
\end{table*}
\begin{table*}
\caption{Average potential energy (in kcal/mol) of ions in ionophores of the C-complex in different solvents.  The data format is the same as in Table~\ref{table-pockets-energies-A}. 
\label{table-pockets-energies-C}
}
\begin{minipage}{\textwidth}
\centering
\begin{tabular}{|c|lr|lr|lr|r|}
\multicolumn{8}{c}{\bf{Decamer} \rm } \\
\hline
Ionophore& Water, C     &      & Eth., C    &        & Meth., C     &        & Eth.-Meth., C \\
\hline
P-cross  & -206*\ 0.005 & -1   & -246*\ 0.02 & -4     & -239*\ 0.01  & -2     & -3   \\
N7-Dir   & -203*\ 0.2   & -40  & -242*\ 0.98 & -237   & -233*\ 0.78  & -182   & -69  \\
O6-Dir   & -201*\ 0.23  & -46  & -250*\ 0.99 & -248   & -239*\ 1.51  & -360   & 124  \\
Maj-Slv  & -202*\ 0.69  & -139 & -243*\ 1.39 & -339   & -231*\ 0.84  & -195   & -141  \\
\hline
O4'-cross& -205*\ 0.53  & -108 & -251*\ 2.13 & -534   & -245*\ 2.25  & -552   & 16   \\
Min-Dir  & -194*\ 0.79  & -153 & -233*\ 1.72 & -400   & -227*\ 1.82  & -414   & 4    \\
Min-Slv  & -199*\ 0.99  & -196 & -238*\ 1.55 & -369   & -231*\ 1.18  & -272   & -110  \\
\hline
P-P      & -195*\ 0.36  & -70  & -222*\ 1.57 & -348   & -217*\ 1.05  & -228   & -115 \\
P        & -194*\ 1.68  & -325 & -218*\ 3.12 & -681   & -213*\ 2.76  & -590   & -77  \\
Bulk     & -177*12.54   & -2223& -185*\ 4.52 & -835   & -183*\ 5.79  & -1058  & 228  \\
\hline
Total    &              & \ -3302&           &\ \ -3996&             &\ \ -3853& -143 \\
\hline
\end{tabular}
\end{minipage}
\begin{minipage}{\textwidth}
\centering
\begin{tabular}{|c|lr|lr|lr|r|}
\multicolumn{8}{c}{\bf{20-mer} \rm } \\
\hline
Ionophore& Water, C    &      & Eth., C     &        & Meth., C     &        & Eth.-Meth., C \\
\hline
P-cross  & -238*\ 0.03 & -8   & -289*\ 0.07 & -21    & -289*\ 0.03  & -8     & -13  \\
N7-Dir   & -239*\ 0.54 & -130 & -296*\ 1.88 & -557   & -285*\ 1.8   & -512   & -45  \\
O6-Dir   & -238*\ 0.55 & -130 & -301*\ 2.2  & -662   & -289*\ 3.19  & -920   & 258  \\
Maj-Slv  & -238*\ 1.76 & -419 & -297*\ 3    & -891   & -283*\ 1.96  & -555   & -336 \\
\hline
O4'-cross& -242*\ 1.98 & -479 & -305*\ 6.62 & -2018  & -296*\ 6.46  & -1914  & -104 \\
Min-Dir  & -220*\ 1.5  & -331 & -271*\ 2.17 & -587   & -267*\ 2.41  & -643   & 56   \\
Min-Slv  & -234*\ 2.22 & -519 & -289*\ 3.16 & -912   & -281*\ 2.35  & -661   & -252 \\
\hline
P-P      & -226*\ 0.82 & -186 & -264*\ 2.81 & -743   & -260*\ 2.11  & -548   & -195 \\
P        & -224*\ 3.91 & -876 & -261*\ 6.46 & -1687  & -256*\ 5.99  & -1534  & -154 \\
Bulk     & -187*24.69  & -4609& -201*\ 9.64 & -1940  & -200*11.71   & -2347  & 407  \\
\hline
Total    &             & \ \ -7686&             & \ -10018 &              & \ -9640  & -378 \\
\hline
\end{tabular}
\end{minipage}
\end{table*}  
In order to do this, we used the trajectories obtained in the OpenMM package as the input to the LAMMPS package. The calculations were performed by the commands "compute group/group"{} with the option $kspace$ enabled for non-bonded interactions and "compute pe/atom"{} for intramolecular interactions. We had to modify the LAMMPS package in order to correctly take into account the 1-4 nonbonded energy.

The difference between the full energies of the C- and A-complexes calculated in the package LAMMPS slightly differs from value $\Delta U$ in Table~\ref{table-deltaUGS-WHAM}, because the packages OpenMM and LAMMPS  implement smooth zeroing of nonbonded interactions differently, and these packages use different methods for approximating the Ewald sum (in the case of LAMMPS, this is the PPPM (particle-particle particle-mesh Ewald) method \cite{2021-computer-book-Hockney}).
\subsection{Contacts between ions, solvent and DNA}
\label{section-Methods-Contacts}
\noindent
We considered two particles to be in direct contact (the first coordination sphere (CS), sometimes also called contact ion pair) if the distance between them was less than the distance $R_{min}$ at which the first minimum of their radial distribution function is located. For ion $Na^+$, the distances $R_{min}$ are: with the oxygen atom of water or alcohol 3.2\AA, with the oxygen atom of the PO$_4$ group of DNA - 3\AA, with (G)N7 - 3.4\AA, with (G)O6 - 3.1\AA, with (G)N3 - 3.25\AA, with (C)O2 and O4' - 3.3\AA. 

Direct contact of DNA with water or alcohol was determined by the presence of a hydrogen bond between a nitrogen or oxygen atom of one molecule and a hydrogen atom of the other molecule. We considered a hydrogen bond between atoms to be formed if the distance between them was less than 2.5\AA. We considered two particles to be in outer (second) sphere contact (the second CS, sometimes also called solvent-shared ion
pair) if they were in direct contact with the same solvent molecule (water or alcohol).

When we estimated the lifetimes of ions in ionophores (Table~\ref{table-pockets-lifetime}), we did not analyze contacts of the ions with particular atoms. Only the transition of an ion to another ionophore was considered as the exit of the ion from the ionophore. To determine the lifetime of contacts between ions Na$^+$ and water and alcohol molecules, we calculated short trajectories with the output of the coordinates of all particles every 0.1~ps. 
\begin{acknowledgments}
\noindent
This work was supported by the Program of Fundamental Research of the Russian Academy of Sciences.
\end{acknowledgments}
\vspace{\baselineskip}
\noindent
The input files for the calculations described in Section~\ref{section-Methods-WHAM} with coordinates and parameters are available at https://github.com/strlnkv/DNA-AB-FreeEnergy.
\FloatBarrier
\bibliography{DNA-coord-chem}

\begin{thebibliography}{73}%
\makeatletter
\providecommand \@ifxundefined [1]{%
 \@ifx{#1\undefined}
}%
\providecommand \@ifnum [1]{%
 \ifnum #1\expandafter \@firstoftwo
 \else \expandafter \@secondoftwo
 \fi
}%
\providecommand \@ifx [1]{%
 \ifx #1\expandafter \@firstoftwo
 \else \expandafter \@secondoftwo
 \fi
}%
\providecommand \natexlab [1]{#1}%
\providecommand \enquote  [1]{``#1''}%
\providecommand \bibnamefont  [1]{#1}%
\providecommand \bibfnamefont [1]{#1}%
\providecommand \citenamefont [1]{#1}%
\providecommand \href@noop [0]{\@secondoftwo}%
\providecommand \href [0]{\begingroup \@sanitize@url \@href}%
\providecommand \@href[1]{\@@startlink{#1}\@@href}%
\providecommand \@@href[1]{\endgroup#1\@@endlink}%
\providecommand \@sanitize@url [0]{\catcode `\\12\catcode `\$12\catcode
  `\&12\catcode `\#12\catcode `\^12\catcode `\_12\catcode `\%12\relax}%
\providecommand \@@startlink[1]{}%
\providecommand \@@endlink[0]{}%
\providecommand \url  [0]{\begingroup\@sanitize@url \@url }%
\providecommand \@url [1]{\endgroup\@href {#1}{\urlprefix }}%
\providecommand \urlprefix  [0]{URL }%
\providecommand \Eprint [0]{\href }%
\providecommand \doibase [0]{https://doi.org/}%
\providecommand \selectlanguage [0]{\@gobble}%
\providecommand \bibinfo  [0]{\@secondoftwo}%
\providecommand \bibfield  [0]{\@secondoftwo}%
\providecommand \translation [1]{[#1]}%
\providecommand \BibitemOpen [0]{}%
\providecommand \bibitemStop [0]{}%
\providecommand \bibitemNoStop [0]{.\EOS\space}%
\providecommand \EOS [0]{\spacefactor3000\relax}%
\providecommand \BibitemShut  [1]{\csname bibitem#1\endcsname}%
\let\auto@bib@innerbib\@empty
\bibitem [{\citenamefont {Zubova}\ and\ \citenamefont
  {Strelnikov}(2023)}]{2023-DNA-review-we}%
  \BibitemOpen
  \bibfield  {author} {\bibinfo {author} {\bibfnamefont {E.~A.}\ \bibnamefont
  {Zubova}}\ and\ \bibinfo {author} {\bibfnamefont {I.~A.}\ \bibnamefont
  {Strelnikov}},\ }\bibfield  {title} {\bibinfo {title} {{Experimental
  detection of conformational transitions between forms of DNA: problems and
  prospects}},\ }\href {https://doi.org/10.1007/s12551-023-01143-9} {\bibfield
  {journal} {\bibinfo  {journal} {Biophysical Reviews}\ }\textbf {\bibinfo
  {volume} {15}},\ \bibinfo {pages} {1053} (\bibinfo {year}
  {2023})}\BibitemShut {NoStop}%
\bibitem [{\citenamefont {Gao}\ \emph {et~al.}(1995)\citenamefont {Gao},
  \citenamefont {Robinson}, \citenamefont {van Boom},\ and\ \citenamefont
  {Wang}}]{1995-CoNH36-A-DNA-X-ray}%
  \BibitemOpen
  \bibfield  {author} {\bibinfo {author} {\bibfnamefont {Y.~G.}\ \bibnamefont
  {Gao}}, \bibinfo {author} {\bibfnamefont {H.}~\bibnamefont {Robinson}},
  \bibinfo {author} {\bibfnamefont {J.~H.}\ \bibnamefont {van Boom}},\ and\
  \bibinfo {author} {\bibfnamefont {A.~H.}\ \bibnamefont {Wang}},\ }\bibfield
  {title} {\bibinfo {title} {{Influence of counter-ions on the crystal
  structures of DNA decamers: binding of [Co(NH3)6]3+ and Ba2+ to A-DNA}},\
  }\href {https://doi.org/10.1016/S0006-3495(95)79929-5} {\bibfield  {journal}
  {\bibinfo  {journal} {Biophysical Journal}\ }\textbf {\bibinfo {volume}
  {69}},\ \bibinfo {pages} {559} (\bibinfo {year} {1995})}\BibitemShut
  {NoStop}%
\bibitem [{\citenamefont {Ahmad}\ \emph {et~al.}(1996)\citenamefont {Ahmad},
  \citenamefont {Naoui}, \citenamefont {Neault}, \citenamefont {Diamantoglou},\
  and\ \citenamefont {Tajmir-Riahi}}]{1996-Al-Ga-A-DNA}%
  \BibitemOpen
  \bibfield  {author} {\bibinfo {author} {\bibfnamefont {R.}~\bibnamefont
  {Ahmad}}, \bibinfo {author} {\bibfnamefont {M.}~\bibnamefont {Naoui}},
  \bibinfo {author} {\bibfnamefont {J.~F.}\ \bibnamefont {Neault}}, \bibinfo
  {author} {\bibfnamefont {S.}~\bibnamefont {Diamantoglou}},\ and\ \bibinfo
  {author} {\bibfnamefont {H.~A.}\ \bibnamefont {Tajmir-Riahi}},\ }\bibfield
  {title} {\bibinfo {title} {{An FTIR Spectroscopic Study of Calf-thymus DNA
  Complexation with Al(III) and Ga(III) Cations}},\ }\href
  {https://doi.org/10.1080/07391102.1996.10508892} {\bibfield  {journal}
  {\bibinfo  {journal} {J. of Biomol. Str. and Dyn.}\ }\textbf {\bibinfo
  {volume} {13}},\ \bibinfo {pages} {795} (\bibinfo {year} {1996})}\BibitemShut
  {NoStop}%
\bibitem [{\citenamefont {Ivanov}\ \emph {et~al.}(1973)\citenamefont {Ivanov},
  \citenamefont {Minchenkova}, \citenamefont {Schyolkina},\ and\ \citenamefont
  {Poletayev}}]{1973-Ivanov-CD-A-B-C}%
  \BibitemOpen
  \bibfield  {author} {\bibinfo {author} {\bibfnamefont {V.~I.}\ \bibnamefont
  {Ivanov}}, \bibinfo {author} {\bibfnamefont {L.~E.}\ \bibnamefont
  {Minchenkova}}, \bibinfo {author} {\bibfnamefont {A.~K.}\ \bibnamefont
  {Schyolkina}},\ and\ \bibinfo {author} {\bibfnamefont {A.~I.}\ \bibnamefont
  {Poletayev}},\ }\bibfield  {title} {\bibinfo {title} {{Different
  conformations of double-stranded nucleic acid in solution as revealed by
  circular dichroism}},\ }\href {https://doi.org/10.1002/bip.1973.360120109}
  {\bibfield  {journal} {\bibinfo  {journal} {Biopolymers}\ }\textbf {\bibinfo
  {volume} {12}},\ \bibinfo {pages} {89} (\bibinfo {year} {1973})}\BibitemShut
  {NoStop}%
\bibitem [{\citenamefont {Cheatham~III}\ and\ \citenamefont
  {Kollman}(1997)}]{1997-B-to-A-hexaaminecobalt-MD-Chetham}%
  \BibitemOpen
  \bibfield  {author} {\bibinfo {author} {\bibfnamefont {T.~E.}\ \bibnamefont
  {Cheatham~III}}\ and\ \bibinfo {author} {\bibfnamefont {P.~A.}\ \bibnamefont
  {Kollman}},\ }\bibfield  {title} {\bibinfo {title} {{Insight into the
  stabilization of A-DNA by specific ion association: spontaneous B-DNA to
  A-DNA transitions observed in molecular dynamics simulations of
  d[ACCCGCGGGT]2 in the presence of hexaamminecobalt(III)}},\ }\href
  {https://doi.org/10.1016/S0969-2126(97)00282-7} {\bibfield  {journal}
  {\bibinfo  {journal} {Structure}\ }\textbf {\bibinfo {volume} {5}},\ \bibinfo
  {pages} {1297} (\bibinfo {year} {1997})}\BibitemShut {NoStop}%
\bibitem [{\citenamefont
  {Mazur}(2008)}]{2008-Mazur-Electrostatic-Origin-DNA-Polymorphism}%
  \BibitemOpen
  \bibfield  {author} {\bibinfo {author} {\bibfnamefont {A.~K.}\ \bibnamefont
  {Mazur}},\ }\bibfield  {title} {\bibinfo {title} {{The Electrostatic Origin
  of Low-Hydration Polymorphism in DNA}},\ }\href
  {https://doi.org/10.1002/cphc.200800446} {\bibfield  {journal} {\bibinfo
  {journal} {ChemPhysChem}\ }\textbf {\bibinfo {volume} {9}},\ \bibinfo {pages}
  {2691} (\bibinfo {year} {2008})}\BibitemShut {NoStop}%
\bibitem [{\citenamefont {Manning}(2002)}]{2002-Manning-condesation}%
  \BibitemOpen
  \bibfield  {author} {\bibinfo {author} {\bibfnamefont {G.~S.}\ \bibnamefont
  {Manning}},\ }\bibfield  {title} {\bibinfo {title} {{Electrostatic free
  energy of the DNA double helix in counterion condensation theory}},\ }\href
  {https://doi.org/10.1016/S0301-4622(02)00162-X} {\bibfield  {journal}
  {\bibinfo  {journal} {Biophysical Chemistry}\ }\textbf {\bibinfo {volume}
  {101-102}},\ \bibinfo {pages} {461} (\bibinfo {year} {2002})}\BibitemShut
  {NoStop}%
\bibitem [{\citenamefont {Pasi}\ \emph {et~al.}(2015)\citenamefont {Pasi},
  \citenamefont {Maddocks},\ and\ \citenamefont
  {Lavery}}]{2015-MD-seq-dep-ions-distr-around-DNA}%
  \BibitemOpen
  \bibfield  {author} {\bibinfo {author} {\bibfnamefont {M.}~\bibnamefont
  {Pasi}}, \bibinfo {author} {\bibfnamefont {J.~H.}\ \bibnamefont {Maddocks}},\
  and\ \bibinfo {author} {\bibfnamefont {R.}~\bibnamefont {Lavery}},\
  }\bibfield  {title} {\bibinfo {title} {{Analyzing ion distributions around
  DNA: sequence-dependence of potassium ion distributions from microsecond
  molecular dynamics}},\ }\href {https://doi.org/10.1093/nar/gkv080} {\bibfield
   {journal} {\bibinfo  {journal} {Nucleic Acids Research}\ }\textbf {\bibinfo
  {volume} {43}},\ \bibinfo {pages} {2412} (\bibinfo {year}
  {2015})}\BibitemShut {NoStop}%
\bibitem [{\citenamefont {Hud}\ and\ \citenamefont
  {Polak}(2001)}]{2001-flexible-ionophors}%
  \BibitemOpen
  \bibfield  {author} {\bibinfo {author} {\bibfnamefont {N.~V.}\ \bibnamefont
  {Hud}}\ and\ \bibinfo {author} {\bibfnamefont {M.}~\bibnamefont {Polak}},\
  }\bibfield  {title} {\bibinfo {title} {{DNA–cation interactions: the major
  and minor grooves are flexible ionophores}},\ }\href
  {https://doi.org/10.1016/S0959-440X(00)00205-0} {\bibfield  {journal}
  {\bibinfo  {journal} {Current Opinion in Structural Biology}\ }\textbf
  {\bibinfo {volume} {11}},\ \bibinfo {pages} {293} (\bibinfo {year}
  {2001})}\BibitemShut {NoStop}%
\bibitem [{\citenamefont {Hsiao}\ \emph
  {et~al.}(2008{\natexlab{a}})\citenamefont {Hsiao}, \citenamefont
  {Tannenbaum}, \citenamefont {VanDeusen}, \citenamefont {Hershkovitz},
  \citenamefont {Perng}, \citenamefont {Tannenbaum},\ and\ \citenamefont
  {Williams}}]{2008-DNA-ions-compl-Williams}%
  \BibitemOpen
  \bibfield  {author} {\bibinfo {author} {\bibfnamefont {C.}~\bibnamefont
  {Hsiao}}, \bibinfo {author} {\bibfnamefont {E.}~\bibnamefont {Tannenbaum}},
  \bibinfo {author} {\bibfnamefont {H.}~\bibnamefont {VanDeusen}}, \bibinfo
  {author} {\bibfnamefont {E.}~\bibnamefont {Hershkovitz}}, \bibinfo {author}
  {\bibfnamefont {G.}~\bibnamefont {Perng}}, \bibinfo {author} {\bibfnamefont
  {A.~R.}\ \bibnamefont {Tannenbaum}},\ and\ \bibinfo {author} {\bibfnamefont
  {L.~D.}\ \bibnamefont {Williams}},\ }\bibfield  {title} {\bibinfo {title}
  {{Complexes of Nucleic Acids with Group I and II Cations}},\ }in\ \href
  {https://doi.org/10.1039/9781847558763-00001} {\emph {\bibinfo {booktitle}
  {Nucleic Acid–Metal Ion Interactions}}}\ (\bibinfo  {publisher} {The Royal
  Society of Chemistry},\ \bibinfo {year} {2008})\BibitemShut {NoStop}%
\bibitem [{\citenamefont {Howerton}\ \emph {et~al.}(2001)\citenamefont
  {Howerton}, \citenamefont {Sines}, \citenamefont {VanDerveer},\ and\
  \citenamefont {Williams}}]{2001-Xray-ions-in-grooves-B-DNA}%
  \BibitemOpen
  \bibfield  {author} {\bibinfo {author} {\bibfnamefont {S.~B.}\ \bibnamefont
  {Howerton}}, \bibinfo {author} {\bibfnamefont {C.~C.}\ \bibnamefont {Sines}},
  \bibinfo {author} {\bibfnamefont {D.}~\bibnamefont {VanDerveer}},\ and\
  \bibinfo {author} {\bibfnamefont {L.~D.}\ \bibnamefont {Williams}},\
  }\bibfield  {title} {\bibinfo {title} {{Locating Monovalent Cations in the
  Grooves of B-DNA}},\ }\href {https://doi.org/10.1021/bi010391+} {\bibfield
  {journal} {\bibinfo  {journal} {Biochemistry}\ }\textbf {\bibinfo {volume}
  {40}},\ \bibinfo {pages} {10023} (\bibinfo {year} {2001})}\BibitemShut
  {NoStop}%
\bibitem [{\citenamefont {Maehigashi}\ \emph {et~al.}(2011)\citenamefont
  {Maehigashi}, \citenamefont {Hsiao}, \citenamefont {Kruger~Woods},
  \citenamefont {Moulaei}, \citenamefont {Hud},\ and\ \citenamefont
  {Dean~Williams}}]{2011-Xray-Mg-many-places-B-DNA}%
  \BibitemOpen
  \bibfield  {author} {\bibinfo {author} {\bibfnamefont {T.}~\bibnamefont
  {Maehigashi}}, \bibinfo {author} {\bibfnamefont {C.}~\bibnamefont {Hsiao}},
  \bibinfo {author} {\bibfnamefont {K.}~\bibnamefont {Kruger~Woods}}, \bibinfo
  {author} {\bibfnamefont {T.}~\bibnamefont {Moulaei}}, \bibinfo {author}
  {\bibfnamefont {N.~V.}\ \bibnamefont {Hud}},\ and\ \bibinfo {author}
  {\bibfnamefont {L.}~\bibnamefont {Dean~Williams}},\ }\bibfield  {title}
  {\bibinfo {title} {{B-DNA structure is intrinsically polymorphic: even at the
  level of base pair positions}},\ }\href {https://doi.org/10.1093/nar/gkr1168}
  {\bibfield  {journal} {\bibinfo  {journal} {Nucleic Acids Research}\ }\textbf
  {\bibinfo {volume} {40}},\ \bibinfo {pages} {3714} (\bibinfo {year}
  {2011})}\BibitemShut {NoStop}%
\bibitem [{\citenamefont {Leonarski}\ \emph {et~al.}(2016)\citenamefont
  {Leonarski}, \citenamefont {D'Ascenzo},\ and\ \citenamefont
  {Auffinger}}]{2016-Xray-Mg-only-phosphates}%
  \BibitemOpen
  \bibfield  {author} {\bibinfo {author} {\bibfnamefont {F.}~\bibnamefont
  {Leonarski}}, \bibinfo {author} {\bibfnamefont {L.}~\bibnamefont
  {D'Ascenzo}},\ and\ \bibinfo {author} {\bibfnamefont {P.}~\bibnamefont
  {Auffinger}},\ }\bibfield  {title} {\bibinfo {title} {{Mg2+ ions: do they
  bind to nucleobase nitrogens?}},\ }\href
  {https://doi.org/10.1093/nar/gkw1175} {\bibfield  {journal} {\bibinfo
  {journal} {Nucleic Acids Research}\ }\textbf {\bibinfo {volume} {45}},\
  \bibinfo {pages} {987} (\bibinfo {year} {2016})}\BibitemShut {NoStop}%
\bibitem [{\citenamefont {Torrie}\ and\ \citenamefont
  {Valleau}(1977)}]{1977-Umbrella}%
  \BibitemOpen
  \bibfield  {author} {\bibinfo {author} {\bibfnamefont {G.}~\bibnamefont
  {Torrie}}\ and\ \bibinfo {author} {\bibfnamefont {J.}~\bibnamefont
  {Valleau}},\ }\bibfield  {title} {\bibinfo {title} {{Nonphysical sampling
  distributions in Monte Carlo free-energy estimation: Umbrella sampling}},\
  }\href {https://doi.org/10.1016/0021-9991(77)90121-8} {\bibfield  {journal}
  {\bibinfo  {journal} {Journal of Computational Physics}\ }\textbf {\bibinfo
  {volume} {23}},\ \bibinfo {pages} {187} (\bibinfo {year} {1977})}\BibitemShut
  {NoStop}%
\bibitem [{\citenamefont {Kumar}\ \emph {et~al.}(1992)\citenamefont {Kumar},
  \citenamefont {Rosenberg}, \citenamefont {Bouzida}, \citenamefont
  {Swendsen},\ and\ \citenamefont {Kollman}}]{1992-WHAM}%
  \BibitemOpen
  \bibfield  {author} {\bibinfo {author} {\bibfnamefont {S.}~\bibnamefont
  {Kumar}}, \bibinfo {author} {\bibfnamefont {J.~M.}\ \bibnamefont
  {Rosenberg}}, \bibinfo {author} {\bibfnamefont {D.}~\bibnamefont {Bouzida}},
  \bibinfo {author} {\bibfnamefont {R.~H.}\ \bibnamefont {Swendsen}},\ and\
  \bibinfo {author} {\bibfnamefont {P.~A.}\ \bibnamefont {Kollman}},\
  }\bibfield  {title} {\bibinfo {title} {{The weighted histogram analysis
  method for free-energy calculations on biomolecules. I. The method}},\ }\href
  {https://doi.org/10.1002/jcc.540130812} {\bibfield  {journal} {\bibinfo
  {journal} {Journal of Computational Chemistry}\ }\textbf {\bibinfo {volume}
  {13}},\ \bibinfo {pages} {1011} (\bibinfo {year} {1992})}\BibitemShut
  {NoStop}%
\bibitem [{\citenamefont {Hart}\ \emph {et~al.}(2012)\citenamefont {Hart},
  \citenamefont {Foloppe}, \citenamefont {Baker}, \citenamefont {Denning},
  \citenamefont {Nilsson},\ and\ \citenamefont
  {MacKerell}}]{2012-CHARMM-for-BII}%
  \BibitemOpen
  \bibfield  {author} {\bibinfo {author} {\bibfnamefont {K.}~\bibnamefont
  {Hart}}, \bibinfo {author} {\bibfnamefont {N.}~\bibnamefont {Foloppe}},
  \bibinfo {author} {\bibfnamefont {C.~M.}\ \bibnamefont {Baker}}, \bibinfo
  {author} {\bibfnamefont {E.~J.}\ \bibnamefont {Denning}}, \bibinfo {author}
  {\bibfnamefont {L.}~\bibnamefont {Nilsson}},\ and\ \bibinfo {author}
  {\bibfnamefont {A.~D.}\ \bibnamefont {MacKerell}},\ }\bibfield  {title}
  {\bibinfo {title} {Optimization of the charmm additive force field for dna:
  Improved treatment of the bi/bii conformational equilibrium},\ }\href
  {https://doi.org/10.1021/ct200723y} {\bibfield  {journal} {\bibinfo
  {journal} {JCTC}\ }\textbf {\bibinfo {volume} {8}},\ \bibinfo {pages} {348}
  (\bibinfo {year} {2012})}\BibitemShut {NoStop}%
\bibitem [{\citenamefont {Yoo}\ and\ \citenamefont
  {Aksimentiev}(2012)}]{2012-parametrization-DNA-ions}%
  \BibitemOpen
  \bibfield  {author} {\bibinfo {author} {\bibfnamefont {J.}~\bibnamefont
  {Yoo}}\ and\ \bibinfo {author} {\bibfnamefont {A.}~\bibnamefont
  {Aksimentiev}},\ }\bibfield  {title} {\bibinfo {title} {Improved
  parametrization of li+, na+, k+, and mg2+ ions for all-atom molecular
  dynamics simulations of nucleic acid systems},\ }\href
  {https://doi.org/10.1021/jz201501a} {\bibfield  {journal} {\bibinfo
  {journal} {The Journal of Physical Chemistry Letters}\ }\textbf {\bibinfo
  {volume} {3}},\ \bibinfo {pages} {45} (\bibinfo {year} {2012})}\BibitemShut
  {NoStop}%
\bibitem [{\citenamefont {Yoo}\ and\ \citenamefont
  {Aksimentiev}(2018)}]{2018-CUFIX-review}%
  \BibitemOpen
  \bibfield  {author} {\bibinfo {author} {\bibfnamefont {J.}~\bibnamefont
  {Yoo}}\ and\ \bibinfo {author} {\bibfnamefont {A.}~\bibnamefont
  {Aksimentiev}},\ }\bibfield  {title} {\bibinfo {title} {{New tricks for old
  dogs: improving the accuracy of biomolecular force fields by pair-specific
  corrections to non-bonded interactions}},\ }\href
  {https://doi.org/10.1039/C7CP08185E} {\bibfield  {journal} {\bibinfo
  {journal} {Phys. Chem. Chem. Phys.}\ }\textbf {\bibinfo {volume} {20}},\
  \bibinfo {pages} {8432} (\bibinfo {year} {2018})}\BibitemShut {NoStop}%
\bibitem [{\citenamefont {Ng}\ \emph {et~al.}(2000)\citenamefont {Ng},
  \citenamefont {Kopka},\ and\ \citenamefont
  {Dickerson}}]{2000-Dickerson-stable-intermediate-CAT}%
  \BibitemOpen
  \bibfield  {author} {\bibinfo {author} {\bibfnamefont {H.-L.}\ \bibnamefont
  {Ng}}, \bibinfo {author} {\bibfnamefont {M.~L.}\ \bibnamefont {Kopka}},\ and\
  \bibinfo {author} {\bibfnamefont {R.~E.}\ \bibnamefont {Dickerson}},\
  }\bibfield  {title} {\bibinfo {title} {{The structure of a stable
  intermediate in the A$\leftrightarrow$B DNA helix transition}},\ }\href
  {https://doi.org/10.1073/pnas.040571197} {\bibfield  {journal} {\bibinfo
  {journal} {PNAS}\ }\textbf {\bibinfo {volume} {97}},\ \bibinfo {pages} {2035}
  (\bibinfo {year} {2000})}\BibitemShut {NoStop}%
\bibitem [{\citenamefont {Humphrey}\ \emph {et~al.}(1996)\citenamefont
  {Humphrey}, \citenamefont {Dalke},\ and\ \citenamefont
  {Schulten}}]{1996-VMD}%
  \BibitemOpen
  \bibfield  {author} {\bibinfo {author} {\bibfnamefont {W.}~\bibnamefont
  {Humphrey}}, \bibinfo {author} {\bibfnamefont {A.}~\bibnamefont {Dalke}},\
  and\ \bibinfo {author} {\bibfnamefont {K.}~\bibnamefont {Schulten}},\
  }\bibfield  {title} {\bibinfo {title} {{VMD: Visual molecular dynamics}},\
  }\href {https://doi.org/10.1016/0263-7855(96)00018-5} {\bibfield  {journal}
  {\bibinfo  {journal} {Journal of Molecular Graphics}\ }\textbf {\bibinfo
  {volume} {14}},\ \bibinfo {pages} {33} (\bibinfo {year} {1996})}\BibitemShut
  {NoStop}%
\bibitem [{VMD()}]{VMD-site}%
  \BibitemOpen
  \href@noop {} {\emph {\bibinfo {title} {{\rm VMD
  (http://www.ks.uiuc.edu/Research/vmd/) is developed with NIH support by the
  Theoretical and Computational Biophysics group at the Beckman Institute,
  University of Illinois at Urbana-Champaign}}}}\BibitemShut {NoStop}%
\bibitem [{\citenamefont {Hsiao}\ \emph
  {et~al.}(2008{\natexlab{b}})\citenamefont {Hsiao}, \citenamefont
  {Tannenbaum}, \citenamefont {VanDeusen}, \citenamefont {Hershkovitz},
  \citenamefont {Perng}, \citenamefont {Tannenbaum},\ and\ \citenamefont
  {Williams}}]{2008-DNA-groupI-ions-X-ray-book1}%
  \BibitemOpen
  \bibfield  {author} {\bibinfo {author} {\bibfnamefont {C.}~\bibnamefont
  {Hsiao}}, \bibinfo {author} {\bibfnamefont {E.}~\bibnamefont {Tannenbaum}},
  \bibinfo {author} {\bibfnamefont {H.}~\bibnamefont {VanDeusen}}, \bibinfo
  {author} {\bibfnamefont {E.}~\bibnamefont {Hershkovitz}}, \bibinfo {author}
  {\bibfnamefont {G.}~\bibnamefont {Perng}}, \bibinfo {author} {\bibfnamefont
  {A.~R.}\ \bibnamefont {Tannenbaum}},\ and\ \bibinfo {author} {\bibfnamefont
  {L.~D.}\ \bibnamefont {Williams}},\ }\bibfield  {title} {\bibinfo {title}
  {{Complexes of Nucleic Acids with Group I and II Cations}},\ }in\ \href
  {https://doi.org/10.1039/9781847558763-00001} {\emph {\bibinfo {booktitle}
  {Nucleic Acid–Metal Ion Interactions}}}\ (\bibinfo  {publisher} {The Royal
  Society of Chemistry},\ \bibinfo {year} {2008})\BibitemShut {NoStop}%
\bibitem [{\citenamefont {Auffinger}\ \emph {et~al.}(2016)\citenamefont
  {Auffinger}, \citenamefont {D'Ascenzo},\ and\ \citenamefont
  {Ennifar}}]{2016-Na-K}%
  \BibitemOpen
  \bibfield  {author} {\bibinfo {author} {\bibfnamefont {P.}~\bibnamefont
  {Auffinger}}, \bibinfo {author} {\bibfnamefont {L.}~\bibnamefont
  {D'Ascenzo}},\ and\ \bibinfo {author} {\bibfnamefont {E.}~\bibnamefont
  {Ennifar}},\ }\bibinfo {title} {{Sodium and Potassium Interactions with
  Nucleic Acids}},\ in\ \href {https://doi.org/10.1007/978-3-319-21756-7_6}
  {\emph {\bibinfo {booktitle} {{The Alkali Metal Ions: Their Role for
  Life}}}},\ \bibinfo {editor} {edited by\ \bibinfo {editor} {\bibfnamefont
  {A.}~\bibnamefont {Sigel}}, \bibinfo {editor} {\bibfnamefont
  {H.}~\bibnamefont {Sigel}},\ and\ \bibinfo {editor} {\bibfnamefont
  {R.~K.~O.}\ \bibnamefont {Sigel}}}\ (\bibinfo  {publisher} {Springer
  International Publishing},\ \bibinfo {address} {Cham},\ \bibinfo {year}
  {2016})\ pp.\ \bibinfo {pages} {167--201}\BibitemShut {NoStop}%
\bibitem [{\citenamefont {Tajmir-Riahi}\ \emph {et~al.}(1993)\citenamefont
  {Tajmir-Riahi}, \citenamefont {Ahmad},\ and\ \citenamefont
  {Naoui}}]{1993-La-Eu-Tb-no-A-DNA}%
  \BibitemOpen
  \bibfield  {author} {\bibinfo {author} {\bibfnamefont {H.-A.}\ \bibnamefont
  {Tajmir-Riahi}}, \bibinfo {author} {\bibfnamefont {R.}~\bibnamefont
  {Ahmad}},\ and\ \bibinfo {author} {\bibfnamefont {M.}~\bibnamefont {Naoui}},\
  }\bibfield  {title} {\bibinfo {title} {{Interaction of calf-thymus DNA with
  trivalent La, Eu, and Tb ions. Metal ion binding, DNA condensation and
  structural features}},\ }\href
  {https://doi.org/10.1080/07391102.1993.10508680} {\bibfield  {journal}
  {\bibinfo  {journal} {J. of Biomol. Str. and Dyn.}\ }\textbf {\bibinfo
  {volume} {10}},\ \bibinfo {pages} {865} (\bibinfo {year} {1993})}\BibitemShut
  {NoStop}%
\bibitem [{\citenamefont {Robinson}\ and\ \citenamefont
  {Wang}(1996)}]{1996-B-to-A-in-solution-Hexaamminecobalt-exp}%
  \BibitemOpen
  \bibfield  {author} {\bibinfo {author} {\bibfnamefont {H.}~\bibnamefont
  {Robinson}}\ and\ \bibinfo {author} {\bibfnamefont {A.~H.-J.}\ \bibnamefont
  {Wang}},\ }\bibfield  {title} {\bibinfo {title} {{Neomycin, Spermine and
  Hexaamminecobalt(III) Share Common Structural Motifs in Converting B- to
  A-DNA}},\ }\href {https://doi.org/10.1093/nar/24.4.676} {\bibfield  {journal}
  {\bibinfo  {journal} {Nucleic Acids Research}\ }\textbf {\bibinfo {volume}
  {24}},\ \bibinfo {pages} {676} (\bibinfo {year} {1996})}\BibitemShut
  {NoStop}%
\bibitem [{\citenamefont {Hackl}\ \emph {et~al.}(2005)\citenamefont {Hackl},
  \citenamefont {Kornilova},\ and\ \citenamefont
  {Blagoi}}]{2005-DNA-Mg-phosphate-groups-compaction-Blagoi}%
  \BibitemOpen
  \bibfield  {author} {\bibinfo {author} {\bibfnamefont {E.~V.}\ \bibnamefont
  {Hackl}}, \bibinfo {author} {\bibfnamefont {S.~V.}\ \bibnamefont
  {Kornilova}},\ and\ \bibinfo {author} {\bibfnamefont {Y.~P.}\ \bibnamefont
  {Blagoi}},\ }\bibfield  {title} {\bibinfo {title} {{DNA structural
  transitions induced by divalent metal ions in aqueous solutions}},\ }\href
  {https://doi.org/10.1016/j.ijbiomac.2005.01.011} {\bibfield  {journal}
  {\bibinfo  {journal} {Int. J. of Biol. Macromol.}\ }\textbf {\bibinfo
  {volume} {35}},\ \bibinfo {pages} {175} (\bibinfo {year} {2005})}\BibitemShut
  {NoStop}%
\bibitem [{\citenamefont {Andrushchenko}\ and\ \citenamefont
  {Bou\v{r}}(2009)}]{2009-IR-DNA-ion-binding-Andrushchenko}%
  \BibitemOpen
  \bibfield  {author} {\bibinfo {author} {\bibfnamefont {V.}~\bibnamefont
  {Andrushchenko}}\ and\ \bibinfo {author} {\bibfnamefont {P.}~\bibnamefont
  {Bou\v{r}}},\ }\bibfield  {title} {\bibinfo {title} {{Infrared Absorption
  Detection of Metal Ion-Deoxyguanosine Monophosphate Binding: Experimental and
  Theoretical Study}},\ }\href {https://doi.org/10.1021/jp8058678} {\bibfield
  {journal} {\bibinfo  {journal} {The Journal of Physical Chemistry B}\
  }\textbf {\bibinfo {volume} {113}},\ \bibinfo {pages} {283} (\bibinfo {year}
  {2009})}\BibitemShut {NoStop}%
\bibitem [{\citenamefont {Robinson}\ \emph {et~al.}(2000)\citenamefont
  {Robinson}, \citenamefont {Gao}, \citenamefont {Sanishvili}, \citenamefont
  {Joachimiak},\ and\ \citenamefont {Wang}}]{2000-Mg-A-DNA-X-ray}%
  \BibitemOpen
  \bibfield  {author} {\bibinfo {author} {\bibfnamefont {H.}~\bibnamefont
  {Robinson}}, \bibinfo {author} {\bibfnamefont {Y.-G.}\ \bibnamefont {Gao}},
  \bibinfo {author} {\bibfnamefont {R.}~\bibnamefont {Sanishvili}}, \bibinfo
  {author} {\bibfnamefont {A.}~\bibnamefont {Joachimiak}},\ and\ \bibinfo
  {author} {\bibfnamefont {A.~H.-J.}\ \bibnamefont {Wang}},\ }\bibfield
  {title} {\bibinfo {title} {{Hexahydrated magnesium ions bind in the deep
  major groove and at the outer mouth of A-form nucleic acid duplexes}},\
  }\href {https://doi.org/10.1093/nar/28.8.1760} {\bibfield  {journal}
  {\bibinfo  {journal} {Nucleic Acids Research}\ }\textbf {\bibinfo {volume}
  {28}},\ \bibinfo {pages} {1760} (\bibinfo {year} {2000})}\BibitemShut
  {NoStop}%
\bibitem [{\citenamefont {Fingerhut}\ \emph {et~al.}(2021)\citenamefont
  {Fingerhut}, \citenamefont {Schauss}, \citenamefont {Kundu},\ and\
  \citenamefont {Elsaesser}}]{2021-RNA-P-cross}%
  \BibitemOpen
  \bibfield  {author} {\bibinfo {author} {\bibfnamefont {B.~P.}\ \bibnamefont
  {Fingerhut}}, \bibinfo {author} {\bibfnamefont {J.}~\bibnamefont {Schauss}},
  \bibinfo {author} {\bibfnamefont {A.}~\bibnamefont {Kundu}},\ and\ \bibinfo
  {author} {\bibfnamefont {T.}~\bibnamefont {Elsaesser}},\ }\bibfield  {title}
  {\bibinfo {title} {{Contact pairs of RNA with magnesium ions-electrostatics
  beyond the Poisson-Boltzmann equation}},\ }\href
  {https://doi.org/10.1016/j.bpj.2021.10.029} {\bibfield  {journal} {\bibinfo
  {journal} {Biophysical Journal}\ }\textbf {\bibinfo {volume} {120}},\
  \bibinfo {pages} {5322} (\bibinfo {year} {2021})}\BibitemShut {NoStop}%
\bibitem [{\citenamefont {Hays}\ \emph {et~al.}(2005)\citenamefont {Hays},
  \citenamefont {Teegarden}, \citenamefont {Jones}, \citenamefont {Harms},
  \citenamefont {Raup}, \citenamefont {Watson}, \citenamefont {Cavaliere},\
  and\ \citenamefont {Ho}}]{2005-A-sequences}%
  \BibitemOpen
  \bibfield  {author} {\bibinfo {author} {\bibfnamefont {F.~A.}\ \bibnamefont
  {Hays}}, \bibinfo {author} {\bibfnamefont {A.}~\bibnamefont {Teegarden}},
  \bibinfo {author} {\bibfnamefont {Z.~J.~R.}\ \bibnamefont {Jones}}, \bibinfo
  {author} {\bibfnamefont {M.}~\bibnamefont {Harms}}, \bibinfo {author}
  {\bibfnamefont {D.}~\bibnamefont {Raup}}, \bibinfo {author} {\bibfnamefont
  {J.}~\bibnamefont {Watson}}, \bibinfo {author} {\bibfnamefont
  {E.}~\bibnamefont {Cavaliere}},\ and\ \bibinfo {author} {\bibfnamefont
  {P.~S.}\ \bibnamefont {Ho}},\ }\bibfield  {title} {\bibinfo {title} {{How
  sequence defines structure: A crystallographic map of DNA structure and
  conformation}},\ }\href {https://doi.org/10.1073/pnas.0409455102} {\bibfield
  {journal} {\bibinfo  {journal} {PNAS}\ }\textbf {\bibinfo {volume} {102}},\
  \bibinfo {pages} {7157} (\bibinfo {year} {2005})}\BibitemShut {NoStop}%
\bibitem [{\citenamefont {Auffinger}\ \emph {et~al.}(2011)\citenamefont
  {Auffinger}, \citenamefont {Grover},\ and\ \citenamefont
  {Westhof}}]{2011-ions-binding-RNA-Auffinger}%
  \BibitemOpen
  \bibfield  {author} {\bibinfo {author} {\bibfnamefont {P.}~\bibnamefont
  {Auffinger}}, \bibinfo {author} {\bibfnamefont {N.}~\bibnamefont {Grover}},\
  and\ \bibinfo {author} {\bibfnamefont {E.}~\bibnamefont {Westhof}},\
  }\bibfield  {title} {\bibinfo {title} {{Metal Ion Binding to RNA}},\ }in\
  \href {https://doi.org/10.1039/9781849732512-00001} {\emph {\bibinfo
  {booktitle} {{Structural and Catalytic Roles of Metal Ions in RNA}}}}\
  (\bibinfo  {publisher} {The Royal Society of Chemistry},\ \bibinfo {year}
  {2011})\BibitemShut {NoStop}%
\bibitem [{\citenamefont {Egli}\ \emph {et~al.}(1998)\citenamefont {Egli},
  \citenamefont {Tereshko}, \citenamefont {Teplova}, \citenamefont {Minasov},
  \citenamefont {Joachimiak}, \citenamefont {Sanishvili}, \citenamefont
  {Weeks}, \citenamefont {Miller}, \citenamefont {Maier}, \citenamefont {An},
  \citenamefont {Dan~Cook},\ and\ \citenamefont
  {Manoharan}}]{1998-A-B-hydration-X-ray}%
  \BibitemOpen
  \bibfield  {author} {\bibinfo {author} {\bibfnamefont {M.}~\bibnamefont
  {Egli}}, \bibinfo {author} {\bibfnamefont {V.}~\bibnamefont {Tereshko}},
  \bibinfo {author} {\bibfnamefont {M.}~\bibnamefont {Teplova}}, \bibinfo
  {author} {\bibfnamefont {G.}~\bibnamefont {Minasov}}, \bibinfo {author}
  {\bibfnamefont {A.}~\bibnamefont {Joachimiak}}, \bibinfo {author}
  {\bibfnamefont {R.}~\bibnamefont {Sanishvili}}, \bibinfo {author}
  {\bibfnamefont {C.~M.}\ \bibnamefont {Weeks}}, \bibinfo {author}
  {\bibfnamefont {R.}~\bibnamefont {Miller}}, \bibinfo {author} {\bibfnamefont
  {M.~A.}\ \bibnamefont {Maier}}, \bibinfo {author} {\bibfnamefont
  {H.}~\bibnamefont {An}}, \bibinfo {author} {\bibfnamefont {P.}~\bibnamefont
  {Dan~Cook}},\ and\ \bibinfo {author} {\bibfnamefont {M.}~\bibnamefont
  {Manoharan}},\ }\bibfield  {title} {\bibinfo {title} {{X-ray crystallographic
  analysis of the hydration of A- and B-form DNA at atomic resolution}},\
  }\href
  {https://doi.org/10.1002/(SICI)1097-0282(1998)48:4<234::AID-BIP4>3.0.CO;2-H}
  {\bibfield  {journal} {\bibinfo  {journal} {Biopolymers}\ }\textbf {\bibinfo
  {volume} {48}},\ \bibinfo {pages} {234} (\bibinfo {year} {1998})}\BibitemShut
  {NoStop}%
\bibitem [{\citenamefont {Tereshko}\ \emph {et~al.}(1999)\citenamefont
  {Tereshko}, \citenamefont {Minasov},\ and\ \citenamefont
  {Egli}}]{1999-Rb-A-tract-X-ray}%
  \BibitemOpen
  \bibfield  {author} {\bibinfo {author} {\bibfnamefont {V.}~\bibnamefont
  {Tereshko}}, \bibinfo {author} {\bibfnamefont {G.}~\bibnamefont {Minasov}},\
  and\ \bibinfo {author} {\bibfnamefont {M.}~\bibnamefont {Egli}},\ }\bibfield
  {title} {\bibinfo {title} {{A 'Hydrat-Ion'{} Spine in a B-DNA Minor
  Groove}},\ }\href {https://doi.org/10.1021/ja984346+} {\bibfield  {journal}
  {\bibinfo  {journal} {JACS}\ }\textbf {\bibinfo {volume} {121}},\ \bibinfo
  {pages} {3590} (\bibinfo {year} {1999})}\BibitemShut {NoStop}%
\bibitem [{\citenamefont {Chiu}\ and\ \citenamefont
  {Dickerson}(2000)}]{2000-O4-cross-exp}%
  \BibitemOpen
  \bibfield  {author} {\bibinfo {author} {\bibfnamefont {T.~K.}\ \bibnamefont
  {Chiu}}\ and\ \bibinfo {author} {\bibfnamefont {R.~E.}\ \bibnamefont
  {Dickerson}},\ }\bibfield  {title} {\bibinfo {title} {{1A crystal structures
  of B-DNA reveal sequence-specific binding and groove-specific bending of DNA
  by magnesium and calcium}},\ }\href {https://doi.org/10.1006/jmbi.2000.4012}
  {\bibfield  {journal} {\bibinfo  {journal} {Journal of Molecular Biology}\
  }\textbf {\bibinfo {volume} {301}},\ \bibinfo {pages} {915} (\bibinfo {year}
  {2000})}\BibitemShut {NoStop}%
\bibitem [{\citenamefont {Fedeles}\ \emph {et~al.}(2015)\citenamefont
  {Fedeles}, \citenamefont {Freudenthal}, \citenamefont {Yau}, \citenamefont
  {Singh}, \citenamefont {chi Chang}, \citenamefont {Li}, \citenamefont
  {Delaney}, \citenamefont {Wilson},\ and\ \citenamefont
  {Essigmann}}]{2015-O4cross-polymerasa}%
  \BibitemOpen
  \bibfield  {author} {\bibinfo {author} {\bibfnamefont {B.~I.}\ \bibnamefont
  {Fedeles}}, \bibinfo {author} {\bibfnamefont {B.~D.}\ \bibnamefont
  {Freudenthal}}, \bibinfo {author} {\bibfnamefont {E.}~\bibnamefont {Yau}},
  \bibinfo {author} {\bibfnamefont {V.}~\bibnamefont {Singh}}, \bibinfo
  {author} {\bibfnamefont {S.}~\bibnamefont {chi Chang}}, \bibinfo {author}
  {\bibfnamefont {D.}~\bibnamefont {Li}}, \bibinfo {author} {\bibfnamefont
  {J.~C.}\ \bibnamefont {Delaney}}, \bibinfo {author} {\bibfnamefont {S.~H.}\
  \bibnamefont {Wilson}},\ and\ \bibinfo {author} {\bibfnamefont {J.~M.}\
  \bibnamefont {Essigmann}},\ }\bibfield  {title} {\bibinfo {title} {{Intrinsic
  mutagenic properties of 5-chlorocytosine: A mechanistic connection between
  chronic inflammation and cancer}},\ }\href
  {https://doi.org/10.1073/pnas.1507709112} {\bibfield  {journal} {\bibinfo
  {journal} {PNAS}\ }\textbf {\bibinfo {volume} {112}},\ \bibinfo {pages}
  {E4571} (\bibinfo {year} {2015})}\BibitemShut {NoStop}%
\bibitem [{\citenamefont {Kowal}\ \emph {et~al.}(2011)\citenamefont {Kowal},
  \citenamefont {Ganguly}, \citenamefont {Pallan}, \citenamefont {Marky},
  \citenamefont {Gold}, \citenamefont {Egli},\ and\ \citenamefont
  {Stone}}]{2011-Na-in-Min-Dir-accidently}%
  \BibitemOpen
  \bibfield  {author} {\bibinfo {author} {\bibfnamefont {E.~A.}\ \bibnamefont
  {Kowal}}, \bibinfo {author} {\bibfnamefont {M.}~\bibnamefont {Ganguly}},
  \bibinfo {author} {\bibfnamefont {P.~S.}\ \bibnamefont {Pallan}}, \bibinfo
  {author} {\bibfnamefont {L.~A.}\ \bibnamefont {Marky}}, \bibinfo {author}
  {\bibfnamefont {B.}~\bibnamefont {Gold}}, \bibinfo {author} {\bibfnamefont
  {M.}~\bibnamefont {Egli}},\ and\ \bibinfo {author} {\bibfnamefont {M.~P.}\
  \bibnamefont {Stone}},\ }\bibfield  {title} {\bibinfo {title} {{Altering the
  Electrostatic Potential in the Major Groove: Thermodynamic and Structural
  Characterization of 7-Deaza-2'-deoxyadenosine: dT Base Pairing in DNA}},\
  }\href {https://doi.org/10.1021/jp207104w} {\bibfield  {journal} {\bibinfo
  {journal} {The Journal of Physical Chemistry B}\ }\textbf {\bibinfo {volume}
  {115}},\ \bibinfo {pages} {13925} (\bibinfo {year} {2011})}\BibitemShut
  {NoStop}%
\bibitem [{\citenamefont {Feig}\ and\ \citenamefont
  {Pettitt}(1999)}]{1999-MD-A-B-Pettitt}%
  \BibitemOpen
  \bibfield  {author} {\bibinfo {author} {\bibfnamefont {M.}~\bibnamefont
  {Feig}}\ and\ \bibinfo {author} {\bibfnamefont {B.~M.}\ \bibnamefont
  {Pettitt}},\ }\bibfield  {title} {\bibinfo {title} {{Sodium and Chlorine Ions
  as Part of the DNA Solvation Shell}},\ }\href
  {https://doi.org/10.1016/S0006-3495(99)77023-2} {\bibfield  {journal}
  {\bibinfo  {journal} {Biophysical Journal}\ }\textbf {\bibinfo {volume}
  {77}},\ \bibinfo {pages} {1769} (\bibinfo {year} {1999})}\BibitemShut
  {NoStop}%
\bibitem [{\citenamefont {V\'{a}rnai}\ and\ \citenamefont
  {Zakrzewska}(2004)}]{2004-MD-Na-K-DNA}%
  \BibitemOpen
  \bibfield  {author} {\bibinfo {author} {\bibfnamefont {P.}~\bibnamefont
  {V\'{a}rnai}}\ and\ \bibinfo {author} {\bibfnamefont {K.}~\bibnamefont
  {Zakrzewska}},\ }\bibfield  {title} {\bibinfo {title} {{DNA and its
  counterions: a molecular dynamics study}},\ }\href
  {https://doi.org/10.1093/nar/gkh765} {\bibfield  {journal} {\bibinfo
  {journal} {Nucleic Acids Research}\ }\textbf {\bibinfo {volume} {32}},\
  \bibinfo {pages} {4269} (\bibinfo {year} {2004})}\BibitemShut {NoStop}%
\bibitem [{\citenamefont {Girod}\ \emph {et~al.}(1973)\citenamefont {Girod},
  \citenamefont {Johnson}, \citenamefont {Huntington},\ and\ \citenamefont
  {Maestre}}]{1973-methanol-ethanol-C-noA-condensation}%
  \BibitemOpen
  \bibfield  {author} {\bibinfo {author} {\bibfnamefont {J.~C.}\ \bibnamefont
  {Girod}}, \bibinfo {author} {\bibfnamefont {W.~C.~J.}\ \bibnamefont
  {Johnson}}, \bibinfo {author} {\bibfnamefont {S.~K.}\ \bibnamefont
  {Huntington}},\ and\ \bibinfo {author} {\bibfnamefont {M.~F.}\ \bibnamefont
  {Maestre}},\ }\bibfield  {title} {\bibinfo {title} {{Conformation of
  deoxyribonucleic acid in alcohol solutions}},\ }\href
  {https://doi.org/10.1021/bi00749a011} {\bibfield  {journal} {\bibinfo
  {journal} {Biochemistry}\ }\textbf {\bibinfo {volume} {12}},\ \bibinfo
  {pages} {5092} (\bibinfo {year} {1973})}\BibitemShut {NoStop}%
\bibitem [{\citenamefont {Kypr}\ \emph {et~al.}(2009)\citenamefont {Kypr},
  \citenamefont {Kejnovsk\'{a}}, \citenamefont {Ren\v{c}iuk},\ and\
  \citenamefont {Vorl\'{\i}\v{c}kov\'{a}}}]{2009-Kypr-CD-review}%
  \BibitemOpen
  \bibfield  {author} {\bibinfo {author} {\bibfnamefont {J.}~\bibnamefont
  {Kypr}}, \bibinfo {author} {\bibfnamefont {I.}~\bibnamefont {Kejnovsk\'{a}}},
  \bibinfo {author} {\bibfnamefont {D.}~\bibnamefont {Ren\v{c}iuk}},\ and\
  \bibinfo {author} {\bibfnamefont {M.}~\bibnamefont
  {Vorl\'{\i}\v{c}kov\'{a}}},\ }\bibfield  {title} {\bibinfo {title} {{Circular
  dichroism and conformational polymorphism of DNA}},\ }\href
  {https://doi.org/10.1093/nar/gkp026} {\bibfield  {journal} {\bibinfo
  {journal} {Nucleic Acids Research}\ }\textbf {\bibinfo {volume} {37}},\
  \bibinfo {pages} {1713} (\bibinfo {year} {2009})}\BibitemShut {NoStop}%
\bibitem [{\citenamefont {Lenton}\ \emph {et~al.}(2018)\citenamefont {Lenton},
  \citenamefont {Rhys}, \citenamefont {Towey}, \citenamefont {Soper},\ and\
  \citenamefont {Dougan}}]{2017-alcohol-water-clusters-bi-percolation}%
  \BibitemOpen
  \bibfield  {author} {\bibinfo {author} {\bibfnamefont {S.}~\bibnamefont
  {Lenton}}, \bibinfo {author} {\bibfnamefont {N.~H.}\ \bibnamefont {Rhys}},
  \bibinfo {author} {\bibfnamefont {J.~J.}\ \bibnamefont {Towey}}, \bibinfo
  {author} {\bibfnamefont {A.~K.}\ \bibnamefont {Soper}},\ and\ \bibinfo
  {author} {\bibfnamefont {L.}~\bibnamefont {Dougan}},\ }\bibfield  {title}
  {\bibinfo {title} {Temperature-dependent segregation in alcohol-water binary
  mixtures is driven by water clustering},\ }\href
  {https://doi.org/10.1021/acs.jpcb.8b03543} {\bibfield  {journal} {\bibinfo
  {journal} {The Journal of Physical Chemistry B}\ }\textbf {\bibinfo {volume}
  {122}},\ \bibinfo {pages} {7884} (\bibinfo {year} {2018})}\BibitemShut
  {NoStop}%
\bibitem [{\citenamefont {Nagasaka}\ \emph {et~al.}(2022)\citenamefont
  {Nagasaka}, \citenamefont {Bouvier}, \citenamefont {Yuzawa},\ and\
  \citenamefont {Kosugi}}]{2022-exp-ethanol-water-clusters}%
  \BibitemOpen
  \bibfield  {author} {\bibinfo {author} {\bibfnamefont {M.}~\bibnamefont
  {Nagasaka}}, \bibinfo {author} {\bibfnamefont {M.}~\bibnamefont {Bouvier}},
  \bibinfo {author} {\bibfnamefont {H.}~\bibnamefont {Yuzawa}},\ and\ \bibinfo
  {author} {\bibfnamefont {N.}~\bibnamefont {Kosugi}},\ }\bibfield  {title}
  {\bibinfo {title} {{Hydrophobic Cluster Formation in Aqueous Ethanol
  Solutions Probed by Soft X-ray Absorption Spectroscopy}},\ }\href
  {https://doi.org/10.1021/acs.jpcb.2c02990} {\bibfield  {journal} {\bibinfo
  {journal} {The Journal of Physical Chemistry B}\ }\textbf {\bibinfo {volume}
  {126}},\ \bibinfo {pages} {4948} (\bibinfo {year} {2022})}\BibitemShut
  {NoStop}%
\bibitem [{\citenamefont {Po\v{z}ar}\ \emph {et~al.}(2016)\citenamefont
  {Po\v{z}ar}, \citenamefont {Lovrin\v{c}evi\'{c}}, \citenamefont
  {Zorani\'{c}}, \citenamefont {Primora\'{c}}, \citenamefont {Sokoli\'{c}},\
  and\ \citenamefont {Perera}}]{2016-wat-alc-binary-microheterogeneity}%
  \BibitemOpen
  \bibfield  {author} {\bibinfo {author} {\bibfnamefont {M.}~\bibnamefont
  {Po\v{z}ar}}, \bibinfo {author} {\bibfnamefont {B.}~\bibnamefont
  {Lovrin\v{c}evi\'{c}}}, \bibinfo {author} {\bibfnamefont {L.}~\bibnamefont
  {Zorani\'{c}}}, \bibinfo {author} {\bibfnamefont {T.}~\bibnamefont
  {Primora\'{c}}}, \bibinfo {author} {\bibfnamefont {F.}~\bibnamefont
  {Sokoli\'{c}}},\ and\ \bibinfo {author} {\bibfnamefont {A.}~\bibnamefont
  {Perera}},\ }\bibfield  {title} {\bibinfo {title} {"micro-heterogeneity
  versus clustering in binary mixtures of ethanol with water or alkanes"},\
  }\href {https://doi.org/10.1039/C6CP04676B} {\bibfield  {journal} {\bibinfo
  {journal} {Phys. Chem. Chem. Phys.}\ }\textbf {\bibinfo {volume} {18}},\
  \bibinfo {pages} {23971} (\bibinfo {year} {2016})}\BibitemShut {NoStop}%
\bibitem [{\citenamefont {Hawlicka}\ and\ \citenamefont
  {Swiatla-Wojcik}(1998)}]{1998-NaCl-water-methanol-MD-Hawlicka}%
  \BibitemOpen
  \bibfield  {author} {\bibinfo {author} {\bibfnamefont {E.}~\bibnamefont
  {Hawlicka}}\ and\ \bibinfo {author} {\bibfnamefont {D.}~\bibnamefont
  {Swiatla-Wojcik}},\ }\bibfield  {title} {\bibinfo {title} {{Solvation of ions
  in binary solvents - experimental and MD simulation studies}},\ }\href
  {https://doi.org/10.1016/S0167-7322(98)00079-8} {\bibfield  {journal}
  {\bibinfo  {journal} {Journal of Molecular Liquids}\ }\textbf {\bibinfo
  {volume} {78}},\ \bibinfo {pages} {7} (\bibinfo {year} {1998})}\BibitemShut
  {NoStop}%
\bibitem [{\citenamefont {Chatterjee}\ \emph {et~al.}(2013)\citenamefont
  {Chatterjee}, \citenamefont {Dixit},\ and\ \citenamefont
  {Tembe}}]{2013-MgCl2-water-ethanol-Tembe}%
  \BibitemOpen
  \bibfield  {author} {\bibinfo {author} {\bibfnamefont {A.}~\bibnamefont
  {Chatterjee}}, \bibinfo {author} {\bibfnamefont {M.~K.}\ \bibnamefont
  {Dixit}},\ and\ \bibinfo {author} {\bibfnamefont {B.~L.}\ \bibnamefont
  {Tembe}},\ }\bibfield  {title} {\bibinfo {title} {Solvation structures and
  dynamics of the magnesium chloride (mg2+-cl-) ion pair in water-ethanol
  mixtures},\ }\href {https://doi.org/10.1021/jp4031706} {\bibfield  {journal}
  {\bibinfo  {journal} {The Journal of Physical Chemistry A}\ }\textbf
  {\bibinfo {volume} {117}},\ \bibinfo {pages} {8703} (\bibinfo {year}
  {2013})}\BibitemShut {NoStop}%
\bibitem [{\citenamefont {Keshri}\ \emph {et~al.}(2015)\citenamefont {Keshri},
  \citenamefont {Sarkar},\ and\ \citenamefont
  {Tembe}}]{2015-NaCl-water-methanol-MD-Tembe}%
  \BibitemOpen
  \bibfield  {author} {\bibinfo {author} {\bibfnamefont {S.}~\bibnamefont
  {Keshri}}, \bibinfo {author} {\bibfnamefont {A.}~\bibnamefont {Sarkar}},\
  and\ \bibinfo {author} {\bibfnamefont {B.~L.}\ \bibnamefont {Tembe}},\
  }\bibfield  {title} {\bibinfo {title} {{Molecular Dynamics Simulation of
  Na+-Cl- Ion-Pair in Water–Methanol Mixtures under Supercritical and Ambient
  Conditions}},\ }\href {https://doi.org/10.1021/acs.jpcb.5b05401} {\bibfield
  {journal} {\bibinfo  {journal} {The Journal of Physical Chemistry B}\
  }\textbf {\bibinfo {volume} {119}},\ \bibinfo {pages} {15471} (\bibinfo
  {year} {2015})}\BibitemShut {NoStop}%
\bibitem [{\citenamefont {Eastman}\ \emph {et~al.}(2017)\citenamefont
  {Eastman}, \citenamefont {Swails}, \citenamefont {Chodera}, \citenamefont
  {McGibbon}, \citenamefont {Zhao}, \citenamefont {Beauchamp}, \citenamefont
  {Wang}, \citenamefont {Simmonett}, \citenamefont {Harrigan}, \citenamefont
  {Stern} \emph {et~al.}}]{2017-openmm}%
  \BibitemOpen
  \bibfield  {author} {\bibinfo {author} {\bibfnamefont {P.}~\bibnamefont
  {Eastman}}, \bibinfo {author} {\bibfnamefont {J.}~\bibnamefont {Swails}},
  \bibinfo {author} {\bibfnamefont {J.~D.}\ \bibnamefont {Chodera}}, \bibinfo
  {author} {\bibfnamefont {R.~T.}\ \bibnamefont {McGibbon}}, \bibinfo {author}
  {\bibfnamefont {Y.}~\bibnamefont {Zhao}}, \bibinfo {author} {\bibfnamefont
  {K.~A.}\ \bibnamefont {Beauchamp}}, \bibinfo {author} {\bibfnamefont {L.-P.}\
  \bibnamefont {Wang}}, \bibinfo {author} {\bibfnamefont {A.~C.}\ \bibnamefont
  {Simmonett}}, \bibinfo {author} {\bibfnamefont {M.~P.}\ \bibnamefont
  {Harrigan}}, \bibinfo {author} {\bibfnamefont {C.~D.}\ \bibnamefont {Stern}},
  \emph {et~al.},\ }\bibfield  {title} {\bibinfo {title} {{OpenMM 7: Rapid
  development of high performance algorithms for molecular dynamics}},\ }\href
  {https://doi.org/10.1371/journal.pcbi.1005659} {\bibfield  {journal}
  {\bibinfo  {journal} {PLoS computational biology}\ }\textbf {\bibinfo
  {volume} {13}},\ \bibinfo {pages} {e1005659} (\bibinfo {year}
  {2017})}\BibitemShut {NoStop}%
\bibitem [{\citenamefont {Strelnikov}\ \emph {et~al.}(2023)\citenamefont
  {Strelnikov}, \citenamefont {Kovaleva}, \citenamefont {Klinov},\ and\
  \citenamefont {Zubova}}]{2023-CBA-test-we}%
  \BibitemOpen
  \bibfield  {author} {\bibinfo {author} {\bibfnamefont {I.~A.}\ \bibnamefont
  {Strelnikov}}, \bibinfo {author} {\bibfnamefont {N.~A.}\ \bibnamefont
  {Kovaleva}}, \bibinfo {author} {\bibfnamefont {A.~P.}\ \bibnamefont
  {Klinov}},\ and\ \bibinfo {author} {\bibfnamefont {E.~A.}\ \bibnamefont
  {Zubova}},\ }\bibfield  {title} {\bibinfo {title} {{C-B-A Test of DNA Force
  Fields}},\ }\href {https://doi.org/10.1021/acsomega.2c07781} {\bibfield
  {journal} {\bibinfo  {journal} {ACS Omega}\ }\textbf {\bibinfo {volume}
  {8}},\ \bibinfo {pages} {10253} (\bibinfo {year} {2023})}\BibitemShut
  {NoStop}%
\bibitem [{\citenamefont {Joung}\ and\ \citenamefont
  {Cheatham}(2008)}]{2008-Young-Cheatham-ions}%
  \BibitemOpen
  \bibfield  {author} {\bibinfo {author} {\bibfnamefont {I.~S.}\ \bibnamefont
  {Joung}}\ and\ \bibinfo {author} {\bibfnamefont {T.~E.}\ \bibnamefont
  {Cheatham}},\ }\bibfield  {title} {\bibinfo {title} {{Determination of Alkali
  and Halide Monovalent Ion Parameters for Use in Explicitly Solvated
  Biomolecular Simulations}},\ }\href {https://doi.org/10.1021/jp8001614}
  {\bibfield  {journal} {\bibinfo  {journal} {The Journal of Physical Chemistry
  B}\ }\textbf {\bibinfo {volume} {112}},\ \bibinfo {pages} {9020} (\bibinfo
  {year} {2008})}\BibitemShut {NoStop}%
\bibitem [{\citenamefont {Guvench}\ \emph {et~al.}(2008)\citenamefont
  {Guvench}, \citenamefont {Greene}, \citenamefont {Kamath}, \citenamefont
  {Brady}, \citenamefont {Venable}, \citenamefont {Pastor},\ and\ \citenamefont
  {Mackerell~Jr}}]{2008-alcohols-CHARMM}%
  \BibitemOpen
  \bibfield  {author} {\bibinfo {author} {\bibfnamefont {O.}~\bibnamefont
  {Guvench}}, \bibinfo {author} {\bibfnamefont {S.~N.}\ \bibnamefont {Greene}},
  \bibinfo {author} {\bibfnamefont {G.}~\bibnamefont {Kamath}}, \bibinfo
  {author} {\bibfnamefont {J.~W.}\ \bibnamefont {Brady}}, \bibinfo {author}
  {\bibfnamefont {R.~M.}\ \bibnamefont {Venable}}, \bibinfo {author}
  {\bibfnamefont {R.~W.}\ \bibnamefont {Pastor}},\ and\ \bibinfo {author}
  {\bibfnamefont {A.~D.}\ \bibnamefont {Mackerell~Jr}},\ }\bibfield  {title}
  {\bibinfo {title} {{Additive empirical force field for hexopyranose
  monosaccharides}},\ }\href {https://doi.org/10.1002/jcc.21004} {\bibfield
  {journal} {\bibinfo  {journal} {Journal of Computational Chemistry}\ }\textbf
  {\bibinfo {volume} {29}},\ \bibinfo {pages} {2543} (\bibinfo {year}
  {2008})}\BibitemShut {NoStop}%
\bibitem [{\citenamefont {M.H.~Kabir}\ and\ \citenamefont
  {Huque}(2004)}]{2004-MethEthDens}%
  \BibitemOpen
  \bibfield  {author} {\bibinfo {author} {\bibfnamefont {M.~M.}\ \bibnamefont
  {M.H.~Kabir}}\ and\ \bibinfo {author} {\bibfnamefont {M.}~\bibnamefont
  {Huque}},\ }\bibfield  {title} {\bibinfo {title} {Densities and excess molar
  volumes of methanol, ethanol and n-propanol in pure water and in water + surf
  excel solutions at different temperatures},\ }\href
  {https://doi.org/10.1080/0031910042000205346} {\bibfield  {journal} {\bibinfo
   {journal} {Physics and Chemistry of Liquids}\ }\textbf {\bibinfo {volume}
  {42}},\ \bibinfo {pages} {279} (\bibinfo {year} {2004})}\BibitemShut
  {NoStop}%
\bibitem [{\citenamefont {Akerlof}(1932)}]{1932-eth-meth-epsilon-exp}%
  \BibitemOpen
  \bibfield  {author} {\bibinfo {author} {\bibfnamefont {G.}~\bibnamefont
  {Akerlof}},\ }\bibfield  {title} {\bibinfo {title} {Dielectric constants of
  some organic solvent-water mixtures at various temperatures},\ }\href
  {https://doi.org/10.1021/ja01350a001} {\bibfield  {journal} {\bibinfo
  {journal} {Journal of the American Chemical Society}\ }\textbf {\bibinfo
  {volume} {54}},\ \bibinfo {pages} {4125} (\bibinfo {year}
  {1932})}\BibitemShut {NoStop}%
\bibitem [{\citenamefont {Jorge}\ \emph {et~al.}(2022)\citenamefont {Jorge},
  \citenamefont {Gomes},\ and\ \citenamefont
  {Barrera}}]{2022-eth-meth-dip-mom-NOT-exp}%
  \BibitemOpen
  \bibfield  {author} {\bibinfo {author} {\bibfnamefont {M.}~\bibnamefont
  {Jorge}}, \bibinfo {author} {\bibfnamefont {J.~R.}\ \bibnamefont {Gomes}},\
  and\ \bibinfo {author} {\bibfnamefont {M.~C.}\ \bibnamefont {Barrera}},\
  }\bibfield  {title} {\bibinfo {title} {The dipole moment of alcohols in the
  liquid phase and in solution},\ }\href
  {https://doi.org/10.1016/j.molliq.2022.119033} {\bibfield  {journal}
  {\bibinfo  {journal} {Journal of Molecular Liquids}\ }\textbf {\bibinfo
  {volume} {356}},\ \bibinfo {pages} {119033} (\bibinfo {year}
  {2022})}\BibitemShut {NoStop}%
\bibitem [{\citenamefont {Price}\ \emph {et~al.}(2003)\citenamefont {Price},
  \citenamefont {Ide},\ and\ \citenamefont {Arata}}]{2003-meth-eth-SD-exp}%
  \BibitemOpen
  \bibfield  {author} {\bibinfo {author} {\bibfnamefont {W.~S.}\ \bibnamefont
  {Price}}, \bibinfo {author} {\bibfnamefont {H.}~\bibnamefont {Ide}},\ and\
  \bibinfo {author} {\bibfnamefont {Y.}~\bibnamefont {Arata}},\ }\bibfield
  {title} {\bibinfo {title} {{Solution Dynamics in Aqueous Monohydric Alcohol
  Systems}},\ }\href {https://doi.org/10.1021/jp027257z} {\bibfield  {journal}
  {\bibinfo  {journal} {The Journal of Physical Chemistry A}\ }\textbf
  {\bibinfo {volume} {107}},\ \bibinfo {pages} {4784} (\bibinfo {year}
  {2003})}\BibitemShut {NoStop}%
\bibitem [{\citenamefont {Hawlicka}(1983)}]{1983-meth-SD-exp-alternative}%
  \BibitemOpen
  \bibfield  {author} {\bibinfo {author} {\bibfnamefont {E.}~\bibnamefont
  {Hawlicka}},\ }\bibfield  {title} {\bibinfo {title} {{Self-Diffusion in
  Water-Alcohol Mixtures. Part I. Water-Methanol Solution of NaI}},\ }\href
  {https://doi.org/10.1002/bbpc.19830870513} {\bibfield  {journal} {\bibinfo
  {journal} {Ber. der Bunsengesellschaft fur phys. Chem.}\ }\textbf {\bibinfo
  {volume} {87}},\ \bibinfo {pages} {425} (\bibinfo {year} {1983})}\BibitemShut
  {NoStop}%
\bibitem [{\citenamefont {Reiher~III}(1985)}]{1985-water-TIP3P-parameters}%
  \BibitemOpen
  \bibfield  {author} {\bibinfo {author} {\bibfnamefont {W.~E.}\ \bibnamefont
  {Reiher~III}},\ }\emph {\bibinfo {title} {Theoretical Studies of Hydrogen
  Bonding (Molecular Mechanics, Ab Initio Calculation, Potential Surfaces,
  Formamide, Water)}},\ \href@noop {} {Ph.D. thesis},\ \bibinfo  {school}
  {Harvard University} (\bibinfo {year} {1985})\BibitemShut {NoStop}%
\bibitem [{\citenamefont {Badyal}\ \emph
  {et~al.}(2000{\natexlab{a}})\citenamefont {Badyal}, \citenamefont {Saboungi},
  \citenamefont {Price}, \citenamefont {Shastri}, \citenamefont {Haeffner},\
  and\ \citenamefont {Soper}}]{2000-mu-liquid-water}%
  \BibitemOpen
  \bibfield  {author} {\bibinfo {author} {\bibfnamefont {Y.~S.}\ \bibnamefont
  {Badyal}}, \bibinfo {author} {\bibfnamefont {M.-L.}\ \bibnamefont
  {Saboungi}}, \bibinfo {author} {\bibfnamefont {D.~L.}\ \bibnamefont {Price}},
  \bibinfo {author} {\bibfnamefont {S.~D.}\ \bibnamefont {Shastri}}, \bibinfo
  {author} {\bibfnamefont {D.~R.}\ \bibnamefont {Haeffner}},\ and\ \bibinfo
  {author} {\bibfnamefont {A.~K.}\ \bibnamefont {Soper}},\ }\bibfield  {title}
  {\bibinfo {title} {{Electron distribution in water}},\ }\href
  {https://doi.org/10.1063/1.481541} {\bibfield  {journal} {\bibinfo  {journal}
  {The Journal of Chemical Physics}\ }\textbf {\bibinfo {volume} {112}},\
  \bibinfo {pages} {9206} (\bibinfo {year} {2000}{\natexlab{a}})}\BibitemShut
  {NoStop}%
\bibitem [{\citenamefont {Badyal}\ \emph
  {et~al.}(2000{\natexlab{b}})\citenamefont {Badyal}, \citenamefont {Saboungi},
  \citenamefont {Price}, \citenamefont {Shastri}, \citenamefont {Haeffner},\
  and\ \citenamefont {Soper}}]{2000-water-dip-moment-exp}%
  \BibitemOpen
  \bibfield  {author} {\bibinfo {author} {\bibfnamefont {Y.~S.}\ \bibnamefont
  {Badyal}}, \bibinfo {author} {\bibfnamefont {M.-L.}\ \bibnamefont
  {Saboungi}}, \bibinfo {author} {\bibfnamefont {D.~L.}\ \bibnamefont {Price}},
  \bibinfo {author} {\bibfnamefont {S.~D.}\ \bibnamefont {Shastri}}, \bibinfo
  {author} {\bibfnamefont {D.~R.}\ \bibnamefont {Haeffner}},\ and\ \bibinfo
  {author} {\bibfnamefont {A.~K.}\ \bibnamefont {Soper}},\ }\bibfield  {title}
  {\bibinfo {title} {{Electron distribution in water}},\ }\href
  {https://doi.org/10.1063/1.481541} {\bibfield  {journal} {\bibinfo  {journal}
  {The Journal of Chemical Physics}\ }\textbf {\bibinfo {volume} {112}},\
  \bibinfo {pages} {9206} (\bibinfo {year} {2000}{\natexlab{b}})}\BibitemShut
  {NoStop}%
\bibitem [{\citenamefont {Chow}\ and\ \citenamefont
  {Ferguson}(1995)}]{1995-MC-barostat1}%
  \BibitemOpen
  \bibfield  {author} {\bibinfo {author} {\bibfnamefont {K.-H.}\ \bibnamefont
  {Chow}}\ and\ \bibinfo {author} {\bibfnamefont {D.~M.}\ \bibnamefont
  {Ferguson}},\ }\bibfield  {title} {\bibinfo {title} {{Isothermal-isobaric
  molecular dynamics simulations with Monte Carlo volume sampling}},\ }\href
  {https://doi.org/10.1016/0010-4655(95)00059-O} {\bibfield  {journal}
  {\bibinfo  {journal} {Computer Physics Communications}\ }\textbf {\bibinfo
  {volume} {91}},\ \bibinfo {pages} {283} (\bibinfo {year} {1995})}\BibitemShut
  {NoStop}%
\bibitem [{\citenamefont {Aqvist}\ \emph {et~al.}(2004)\citenamefont {Aqvist},
  \citenamefont {Wennerstrom}, \citenamefont {Nervall}, \citenamefont
  {Bjelic},\ and\ \citenamefont {Brandsdal}}]{2004-MC-barostat2}%
  \BibitemOpen
  \bibfield  {author} {\bibinfo {author} {\bibfnamefont {J.}~\bibnamefont
  {Aqvist}}, \bibinfo {author} {\bibfnamefont {P.}~\bibnamefont {Wennerstrom}},
  \bibinfo {author} {\bibfnamefont {M.}~\bibnamefont {Nervall}}, \bibinfo
  {author} {\bibfnamefont {S.}~\bibnamefont {Bjelic}},\ and\ \bibinfo {author}
  {\bibfnamefont {B.}~\bibnamefont {Brandsdal}},\ }\bibfield  {title} {\bibinfo
  {title} {{Molecular dynamics simulations of water and biomolecules with a
  Monte Carlo constant pressure algorithm}},\ }\href
  {https://doi.org/10.1016/j.cplett.2003.12.039} {\bibfield  {journal}
  {\bibinfo  {journal} {Chemical Physics Letters}\ }\textbf {\bibinfo {volume}
  {384}},\ \bibinfo {pages} {288} (\bibinfo {year} {2004})}\BibitemShut
  {NoStop}%
\bibitem [{\citenamefont {Zgarbov\'{a}}\ \emph {et~al.}(2014)\citenamefont
  {Zgarbov\'{a}}, \citenamefont {Otyepka}, \citenamefont {\v{S}poner},
  \citenamefont {Lanka\v{s}},\ and\ \citenamefont
  {Jure\v{c}ka}}]{2014-modeling-fraying}%
  \BibitemOpen
  \bibfield  {author} {\bibinfo {author} {\bibfnamefont {M.}~\bibnamefont
  {Zgarbov\'{a}}}, \bibinfo {author} {\bibfnamefont {M.}~\bibnamefont
  {Otyepka}}, \bibinfo {author} {\bibfnamefont {J.}~\bibnamefont {\v{S}poner}},
  \bibinfo {author} {\bibfnamefont {F.}~\bibnamefont {Lanka\v{s}}},\ and\
  \bibinfo {author} {\bibfnamefont {P.}~\bibnamefont {Jure\v{c}ka}},\
  }\bibfield  {title} {\bibinfo {title} {{Base Pair Fraying in Molecular
  Dynamics Simulations of DNA and RNA}},\ }\href
  {https://doi.org/10.1021/ct500120v} {\bibfield  {journal} {\bibinfo
  {journal} {J. Chem. Theory Comput.}\ }\textbf {\bibinfo {volume} {10}},\
  \bibinfo {pages} {3177} (\bibinfo {year} {2014})}\BibitemShut {NoStop}%
\bibitem [{\citenamefont {Ben~Imeddourene}\ \emph {et~al.}(2015)\citenamefont
  {Ben~Imeddourene}, \citenamefont {Elbahnsi}, \citenamefont
  {Gu$\rm\acute{e}$roult}, \citenamefont {Oguey}, \citenamefont {Foloppe},\
  and\ \citenamefont {Hartmann}}]{2015-modeling-against-fraying}%
  \BibitemOpen
  \bibfield  {author} {\bibinfo {author} {\bibfnamefont {A.}~\bibnamefont
  {Ben~Imeddourene}}, \bibinfo {author} {\bibfnamefont {A.}~\bibnamefont
  {Elbahnsi}}, \bibinfo {author} {\bibfnamefont {M.}~\bibnamefont
  {Gu$\rm\acute{e}$roult}}, \bibinfo {author} {\bibfnamefont {C.}~\bibnamefont
  {Oguey}}, \bibinfo {author} {\bibfnamefont {N.}~\bibnamefont {Foloppe}},\
  and\ \bibinfo {author} {\bibfnamefont {B.}~\bibnamefont {Hartmann}},\
  }\bibfield  {title} {\bibinfo {title} {Simulations meet experiment to reveal
  new insights into dna intrinsic mechanics},\ }\href
  {https://doi.org/10.1371/journal.pcbi.1004631} {\bibfield  {journal}
  {\bibinfo  {journal} {PLOS Computational Biology}\ }\textbf {\bibinfo
  {volume} {11}},\ \bibinfo {pages} {e1004631} (\bibinfo {year}
  {2015})}\BibitemShut {NoStop}%
\bibitem [{\citenamefont {Kovaleva}\ \emph {et~al.}(2017)\citenamefont
  {Kovaleva}, \citenamefont {Koroleva~(Kikot)}, \citenamefont {Mazo},\ and\
  \citenamefont {Zubova}}]{2017-our-CG-DNA-model}%
  \BibitemOpen
  \bibfield  {author} {\bibinfo {author} {\bibfnamefont {N.~A.}\ \bibnamefont
  {Kovaleva}}, \bibinfo {author} {\bibfnamefont {I.~P.}\ \bibnamefont
  {Koroleva~(Kikot)}}, \bibinfo {author} {\bibfnamefont {M.~A.}\ \bibnamefont
  {Mazo}},\ and\ \bibinfo {author} {\bibfnamefont {E.~A.}\ \bibnamefont
  {Zubova}},\ }\bibfield  {title} {\bibinfo {title} {{The "sugar"
  coarse-grained DNA model}},\ }\href
  {https://doi.org/10.1007/s00894-017-3209-z} {\bibfield  {journal} {\bibinfo
  {journal} {J. Mol. Model.}\ }\textbf {\bibinfo {volume} {23}},\ \bibinfo
  {pages} {66} (\bibinfo {year} {2017})}\BibitemShut {NoStop}%
\bibitem [{\citenamefont {Zhang}\ \emph {et~al.}(2019)\citenamefont {Zhang},
  \citenamefont {Fu}, \citenamefont {Shao}, \citenamefont {Dehez},
  \citenamefont {Chipot},\ and\ \citenamefont
  {Cai}}]{2019-CCGGGCCCGG-MD-ethanol}%
  \BibitemOpen
  \bibfield  {author} {\bibinfo {author} {\bibfnamefont {H.}~\bibnamefont
  {Zhang}}, \bibinfo {author} {\bibfnamefont {H.}~\bibnamefont {Fu}}, \bibinfo
  {author} {\bibfnamefont {X.}~\bibnamefont {Shao}}, \bibinfo {author}
  {\bibfnamefont {F.}~\bibnamefont {Dehez}}, \bibinfo {author} {\bibfnamefont
  {C.}~\bibnamefont {Chipot}},\ and\ \bibinfo {author} {\bibfnamefont
  {W.}~\bibnamefont {Cai}},\ }\bibfield  {title} {\bibinfo {title} {{Changes in
  Microenvironment Modulate the B- to A-DNA Transition}},\ }\href
  {https://doi.org/10.1021/acs.jcim.8b00885} {\bibfield  {journal} {\bibinfo
  {journal} {J. Chem. Inf. Model.}\ }\textbf {\bibinfo {volume} {59}},\
  \bibinfo {pages} {2324} (\bibinfo {year} {2019})}\BibitemShut {NoStop}%
\bibitem [{\citenamefont {Vargason}\ \emph {et~al.}(2000)\citenamefont
  {Vargason}, \citenamefont {Eichman},\ and\ \citenamefont
  {Ho}}]{2000-E-DNA-Ho-1}%
  \BibitemOpen
  \bibfield  {author} {\bibinfo {author} {\bibfnamefont {J.~M.}\ \bibnamefont
  {Vargason}}, \bibinfo {author} {\bibfnamefont {B.~F.}\ \bibnamefont
  {Eichman}},\ and\ \bibinfo {author} {\bibfnamefont {P.~S.}\ \bibnamefont
  {Ho}},\ }\bibfield  {title} {\bibinfo {title} {{The extended and eccentric
  E-DNA structure induced by cytosine methylation or bromination}},\ }\href
  {https://doi.org/10.1038/78985} {\bibfield  {journal} {\bibinfo  {journal}
  {Nat. Struct. Biol.}\ }\textbf {\bibinfo {volume} {7}},\ \bibinfo {pages}
  {758} (\bibinfo {year} {2000})}\BibitemShut {NoStop}%
\bibitem [{\citenamefont {Chodera}\ \emph {et~al.}(2007)\citenamefont
  {Chodera}, \citenamefont {Swope}, \citenamefont {Pitera}, \citenamefont
  {Seok},\ and\ \citenamefont {Dill}}]{2007-statistical-inefficiency}%
  \BibitemOpen
  \bibfield  {author} {\bibinfo {author} {\bibfnamefont {J.~D.}\ \bibnamefont
  {Chodera}}, \bibinfo {author} {\bibfnamefont {W.~C.}\ \bibnamefont {Swope}},
  \bibinfo {author} {\bibfnamefont {J.~W.}\ \bibnamefont {Pitera}}, \bibinfo
  {author} {\bibfnamefont {C.}~\bibnamefont {Seok}},\ and\ \bibinfo {author}
  {\bibfnamefont {K.~A.}\ \bibnamefont {Dill}},\ }\bibfield  {title} {\bibinfo
  {title} {{Use of the Weighted Histogram Analysis Method for the Analysis of
  Simulated and Parallel Tempering Simulations}},\ }\href
  {https://doi.org/10.1021/ct0502864} {\bibfield  {journal} {\bibinfo
  {journal} {Journal of Chemical Theory and Computation}\ }\textbf {\bibinfo
  {volume} {3}},\ \bibinfo {pages} {26} (\bibinfo {year} {2007})}\BibitemShut
  {NoStop}%
\bibitem [{\citenamefont {Zhu}\ and\ \citenamefont
  {Hummer}(2012)}]{2012-systematic-error-in-WHAM}%
  \BibitemOpen
  \bibfield  {author} {\bibinfo {author} {\bibfnamefont {F.}~\bibnamefont
  {Zhu}}\ and\ \bibinfo {author} {\bibfnamefont {G.}~\bibnamefont {Hummer}},\
  }\bibfield  {title} {\bibinfo {title} {{Convergence and error estimation in
  free energy calculations using the weighted histogram analysis method}},\
  }\href {https://doi.org/10.1002/jcc.21989} {\bibfield  {journal} {\bibinfo
  {journal} {Journal of Computational Chemistry}\ }\textbf {\bibinfo {volume}
  {33}},\ \bibinfo {pages} {453} (\bibinfo {year} {2012})}\BibitemShut
  {NoStop}%
\bibitem [{Gro()}]{Grossfield-bootstrap}%
  \BibitemOpen
  \href {http://membrane.urmc.rochester.edu/wordpress/?page_id=126} {\emph
  {\bibinfo {title} {{\rm Grossfield, Alan, "WHAM: the weighted histogram
  \newline analysis method", version 2.1.0, \newline
  http://membrane.urmc.rochester.edu/wordpress/ \newline ?page\textunderscore
  id=126}}}}\BibitemShut {NoStop}%
\bibitem [{\citenamefont {Marvin}\ \emph {et~al.}(1958)\citenamefont {Marvin},
  \citenamefont {Spencer}, \citenamefont {Wilkins},\ and\ \citenamefont
  {Hamilton}}]{1958-MARVIN-Twist-C-DNA}%
  \BibitemOpen
  \bibfield  {author} {\bibinfo {author} {\bibfnamefont {D.~A.}\ \bibnamefont
  {Marvin}}, \bibinfo {author} {\bibfnamefont {M.}~\bibnamefont {Spencer}},
  \bibinfo {author} {\bibfnamefont {M.~H.~F.}\ \bibnamefont {Wilkins}},\ and\
  \bibinfo {author} {\bibfnamefont {L.~D.}\ \bibnamefont {Hamilton}},\
  }\bibfield  {title} {\bibinfo {title} {{A New Configuration of
  Deoxyribonucleic Acid}},\ }\href {https://doi.org/10.1038/182387b0}
  {\bibfield  {journal} {\bibinfo  {journal} {Nature}\ }\textbf {\bibinfo
  {volume} {182}},\ \bibinfo {pages} {387} (\bibinfo {year}
  {1958})}\BibitemShut {NoStop}%
\bibitem [{\citenamefont {{van Dam}}\ and\ \citenamefont
  {Levitt}(2000)}]{2000-C-DNA-Levitt}%
  \BibitemOpen
  \bibfield  {author} {\bibinfo {author} {\bibfnamefont {L.}~\bibnamefont {{van
  Dam}}}\ and\ \bibinfo {author} {\bibfnamefont {M.~H.}\ \bibnamefont
  {Levitt}},\ }\bibfield  {title} {\bibinfo {title} {Bii nucleotides in the b
  and c forms of natural-sequence polymeric dna: A new model for the c form of
  dna},\ }\href {https://doi.org/https://doi.org/10.1006/jmbi.2000.4194}
  {\bibfield  {journal} {\bibinfo  {journal} {Journal of Molecular Biology}\
  }\textbf {\bibinfo {volume} {304}},\ \bibinfo {pages} {541 } (\bibinfo {year}
  {2000})}\BibitemShut {NoStop}%
\bibitem [{\citenamefont {Plimpton}(1995)}]{1995-LAMMPS}%
  \BibitemOpen
  \bibfield  {author} {\bibinfo {author} {\bibfnamefont {S.}~\bibnamefont
  {Plimpton}},\ }\bibfield  {title} {\bibinfo {title} {{Fast Parallel
  Algorithms for Short-Range Molecular Dynamics}},\ }\href
  {https://doi.org/10.1006/jcph.1995.1039} {\bibfield  {journal} {\bibinfo
  {journal} {J. Comput. Phys.}\ }\textbf {\bibinfo {volume} {117}},\ \bibinfo
  {pages} {1} (\bibinfo {year} {1995})}\BibitemShut {NoStop}%
\bibitem [{LAM()}]{LAMMPS-site}%
  \BibitemOpen
  \href {http://lammps.sandia.gov/index.html} {\emph {\bibinfo {title} {\rm
  http://lammps.sandia.gov/index.html}}}\BibitemShut {NoStop}%
\bibitem [{\citenamefont {Hockney}\ and\ \citenamefont
  {Eastwood}(2021)}]{2021-computer-book-Hockney}%
  \BibitemOpen
  \bibfield  {author} {\bibinfo {author} {\bibfnamefont {R.~W.}\ \bibnamefont
  {Hockney}}\ and\ \bibinfo {author} {\bibfnamefont {J.~W.}\ \bibnamefont
  {Eastwood}},\ }\href {https://doi.org/10.1201/9780367806934} {\emph {\bibinfo
  {title} {Computer simulation using particles}}}\ (\bibinfo  {publisher} {CRC
  Press},\ \bibinfo {year} {2021})\BibitemShut {NoStop}%
\end{thebibliography}%
\end{document}